\documentclass[12pt]{article} 
\usepackage{latexsym,amsfonts}
\newcommand{\text}[1]{\mbox{\rm #1}}
\newenvironment{arablist}{\begin{list}{(\theenumi)}
{\usecounter{enumi}}}{\end{list}}

\newenvironment{romanlist}{\begin{list}{(\theenumiii)}
{\usecounter{enumiii}}}{\end{list}}

\newenvironment{letterlist}{\begin{list}{(\alph{enumiv})}
{\usecounter{enumiv}}}{\end{list}}
\newtheorem{theorem}{THEOREM}
\newtheorem{corollary}[theorem]{COROLLARY}
\newtheorem{lemma}[theorem]{LEMMA}
\newenvironment{definition}[1]
{\mbox{ } \\ \begin{bf} #1 \end{bf} 
    \begin{list}{}{\setlength{\leftmargin}{\parindent}}\item}
{\end{list}\begin{flushright} End of Dfn.\end{flushright}}
\newcounter{subeq}
\newenvironment{abc}
{\setcounter{subeq}{\value{equation}} \setcounter{equation}{0}
    
    \stepcounter{subeq}}
{
    \setcounter{equation}{\value{subeq}}
    \setcounter{subeq}{0}}
\newcommand{\proof}{\noindent{\em Proof:~}}
\newcommand{\proofs}{\noindent{\em Proofs:~}}
\newcommand{\cheers}{\hfill {\em End~of~proof.}}
\newcommand{\A}{{\mathcal A}}
\newcommand{\B}{{\mathcal B}}

\newcommand{\E}{{\mathcal E}}
\newcommand{\F}{{\mathcal F}}

\newcommand{\I}{{\mathcal I}}

\newcommand{\K}{{\mathcal K}}

\newcommand{\M}{{\mathcal M}}
\newcommand{\N}{{\mathcal N}}

\renewcommand{\S}{{\mathcal S}}

\newcommand{\Y}{{\mathcal Y}}

\newcommand{\bgreek}[1]{\mbox{\boldmath{$#1$}}}			
\newcommand{\x}{{\mathbf x}}

\newcommand{\bu}{{\mathbf u}}
\newcommand{\bv}{{\mathbf v}}
\newcommand{\bw}{{\mathbf w}}

\newcommand{\bC}{{\mathbf C}}

\newcommand{\bV}{{\mathbf V}}
\newcommand{\bsigma}{\bgreek{\sigma}}
\newcommand{\btheta}{\bgreek{\theta}}
\newcommand{\bvarphi}{\bgreek{\varphi}}

\newcommand{\nbd}[1]{{\mathrm nbd}#1}
\renewcommand{\Re}[1]{{\mathrm Re}~#1}
\renewcommand{\Im}[1]{{\mathrm Im}~#1}
\newcommand{\dom}[1]{{\mathrm dom}~#1}
\newcommand{\tr}[1]{{\mathrm tr}~#1}
\newcommand{\triple}{(x_{0},x_{1},x_{2})}
\newcommand{\s}{(\x_{0},\I^{(3)},\I^{(4)})}

\newcommand{\zip}[1]{{#1}$_{0}$}

\begin{document}
\title{Proof of a generalized Geroch conjecture 
for the hyperbolic Ernst equation}
\author{I.\ Hauser\thanks{Home address: 4500 19th Street, \#342,
	Boulder, CO 80304} and F.\ J.\ Ernst\thanks{E-mail: gravity@slic.com} \\
	FJE Enterprises, 511 County Route 59, Potsdam, NY 13676, 
	USA\thanks{Homepage URL: http://pages.slic.com/gravity}}
\date{August 18, 1999; Minor corrections October 14, 2000}
\maketitle
\begin{abstract}
We enunciate and prove here a generalization of Geroch's famous conjecture 
concerning analytic solutions of the elliptic Ernst equation.  Our 
generalization is stated for solutions of the hyperbolic Ernst equation
that are not necessarily analytic, although it can be formulated also for
solutions of the elliptic Ernst equation that are nowhere axis-accessible.
\end{abstract}


\section{A generalized Geroch conjecture\label{Sec_1}}

In terms of Weyl canonical coordinates $(z,\rho)$, the Ernst equation of 
general relativity can be expressed in the form
\begin{equation}
(\Re{\E}) \left\{ \frac{\partial^{2}\E}{\partial z^{2}}
\pm \frac{1}{\rho} \frac{\partial}{\partial\rho} 
\left( \rho \frac{\partial\E}{\partial\rho} \right) \right\}
= \left( \frac{\partial\E}{\partial z} \right)^{2}
\pm \left( \frac{\partial\E}{\partial\rho} \right)^{2},
\label{EE}
\end{equation}
where the upper signs correspond to the elliptic equation associated with
stationary axisymmetric (spinning body) gravitational fields and the lower
signs correspond to the hyperbolic equation associated with colliding 
gravitational plane wave pairs and cylindrical gravitational 
waves.\footnote{In the latter case, one of the Weyl coordinates has the 
character of a time coordinate.  In
practice a notation more appropriate for the physical problem being
treated would be in order.}  In 1972 R.\ Geroch asserted a 
conjecture\footnote{R.~Geroch, J.~Math.~Phys.\ {\bf 13}, 394-404 (1972).} 
concerning the solution manifold of the elliptic Ernst equation that was
eventually proved\footnote{I.~Hauser and F.~J.~Ernst, {\em A new proof
of an old conjecture}, in Gravitation and Geometry, Eds.\ Rindler and
Trautman, Bibliopolis, Naples (1987).} by the present authors, who used
their own homogeneous Hilbert problem version of the Kinnersley--Chitre
realization of the Geroch group.

In 1986, at the suggestion of S.\ Chandrasekhar, we turned our attention
from stationary axisymmetric fields to colliding gravitational plane wave
pairs.  While the Kinnersley--Chitre transformations 
could still be used to generate scores of exact analytic solutions of the
hyperbolic Ernst equation, we were aware of the fact that there might 
exist a significantly larger group, for, whereas any $\bC^{3}$ solution of the
axis-accessible elliptic Ernst equation can be shown to be automatically
an analytic solution, a solution of the hyperbolic Ernst equation can be
even $\bC^{\infty}$ without being analytic.\footnote{Even the elliptic
equation admits a larger group if solutions are considered that are
everywhere axis-inaccessible.}  Clearly, one should not expect a 
non-analytic solution of the hyperbolic Ernst equation to be related to
Minkowski space by a K--C transformation, for these transformations preserve
analyticity.

\setcounter{equation}{0}
\subsection{Linear systems for the Ernst equation}

Any discussion of the Geroch group or its extensions requires a knowledge 
of at least one linear system\footnote{Such linear systems have been found
by many authors, including Chinea, Harrison, Kinnersley and Chitre, Maison,
Neugebauer and Papanicolaou.  A more complicated type of linear system was
found by Belinskii and Zakharov.}
\begin{equation}
dF(\x,\tau) = \Gamma(\x,\tau) F(\x,\tau)
\end{equation}
for the Ernst equation.  Here $\x$ is shorthand for the nonignorable 
spacetime coordinates (e.g., $z$ and $\rho$), $\tau$ is a 
spacetime-independent complex-valued parameter, and the $1$-form 
$2 \times 2$ matrix $\Gamma(\x,\tau)$ satisfies the integrability condition
\begin{equation}
d\Gamma(\x,\tau) - \Gamma(\x,\tau) \Gamma(\x,\tau) = 0 
\label{zerocurv}
\end{equation}
if and only if the Ernst equation is satisfied.  The symbol $\Gamma(\x,\tau)$
was chosen because of the resemblance of the last equation to a zero-curvature
condition for a connection $1$-form.

If there exists one such $\Gamma(\x,\tau)$ for the Ernst equation, then
there are infinitely many, for if
\begin{equation}
\Gamma'(\x,\tau) := p(\x,\tau) \Gamma(\x,\tau) p(\x,\tau)^{-1}
+ dp(\x,\tau) p(\x,\tau)^{-1}, 
\label{gauge0}
\end{equation}
where $p(\x,\tau)$ is an invertible matrix, then
\begin{equation}
d\Gamma'(\x,\tau) - \Gamma'(\x,\tau) \Gamma'(\x,\tau)
= p(\x,\tau) \left\{ d\Gamma(\x,\tau) - \Gamma(\x,\tau) \Gamma(\x,\tau)
\right\} p(\x,\tau)^{-1}.
\end{equation}
This transformation is nothing but a gauge transformation, the analog 
of the effect that a mere change of basis has upon a connection $1$-form. 
Under such a gauge transformation, the matrix $F(\x,\tau)$ transforms
into the matrix
\begin{equation}
F'(\x,\tau) := p(\x,\tau) F(\x,\tau).
\label{gauge1}
\end{equation}
While, in one sense, the various possible representations of the linear
system may be regarded as equivalent, in another sense they may be quite
different, with the matrices $F(\x,\tau)$ and $F'(\x,\tau)$ possibly
having very different domains in the space $R^{2} \times C$, as well as
different continuity and/or differentiability properties.  Often one
representation is more useful for one part of the analysis, while another
representation is more useful for another part.

Different formalisms may also differ with respect to the number of columns 
that the matrix $F$ has.  Here we shall follow an approach that we described
long ago that effectively sidesteps the question of number of columns by
introducing an auxilliary $2 \times 2$ matrix potential $\F(\x,\tau)$ such
that
\begin{equation}
F(\x,\tau) = \F(\x,\tau) F(\x_{0},\tau),
\end{equation}
\begin{equation}
d\F(\x,\tau) = \Gamma(\x,\tau) \F(\x,\tau)
\label{linsys}
\end{equation}
and
\begin{equation}
\F(\x_{0},\tau) = I,
\label{init}
\end{equation}
where $I$ is a unit matrix, and $\x_{0}$ is a selected spacetime point
within the domain of $\E(\x)$.  Clearly, under a gauge transformation
(\ref{gauge1}), $\F(\x,\tau)$ transforms into
\begin{equation}
\F'(\x,\tau) := p(\x,\tau) \F(\x,\tau) p(\x_{0},\tau)^{-1}.
\label{gauge}
\end{equation}

One of the simplest formulations of the linear system is that of G.\
Neugebauer,\footnote{G.~Neugebauer, {\em B\"{a}cklund transformations 
of axially symmetric stationary gravitational fields}, Phys.\ Lett.~A
{\bf 12}, L67 (1979).} 
in which $\Gamma(\x,\tau) = \Gamma_{N}(\x,\tau)$, where
\begin{equation}
\Gamma_{N}(\x,\tau) := \left( \frac{\tau-z\pm\rho\star}{\tau-z\mp\rho\star}
\right)^{\frac{1}{2}} \left( \begin{array}{cc}
0 & \frac{d\E(\x)}{2f(\x)} \\
\frac{d\E^{*}(\x)}{2f(\x)} & 0
\end{array} \right) + \left( \begin{array}{cc}
\frac{d\E(\x)}{2f(\x)} & 0 \\
0 & \frac{d\E^{*}(\x)}{2f(\x)}
\end{array} \right),
\label{GammaN}
\end{equation}
where $\star$ is a $2$-dimensional duality operator such that
\begin{equation}
\star d\rho = \pm dz, \quad \star dz = - d\rho,
\end{equation}
the upper signs applying in the stationary axisymmetric (elliptic) case, 
and the lower signs applying in the gravitational wave (hyperbolic) case.  
Here $\Gamma(\x,\tau)$ is expressed directly in terms of the Ernst potential
$\E(\x)$ and its complex conjugate, with $f(\x) := \Re{\E(\x)}$.  
Using these notations, the Ernst equation (\ref{EE}) can be expressed as
\begin{equation}
(\Re{\E}) d (\rho \star d\E) = \rho d\E \star d\E.
\label{EE1}
\end{equation}

A slightly different linear system that is due to the authors and is more
suited to our purpose employs $\Gamma = \Gamma_{HE}$, where
\begin{equation} 
\Gamma_{HE}(\x,\tau) := -\left(\frac{\tau-z\pm\rho\star}
{\tau-z\mp\rho\star}\right)^{\frac{1}{2}} \left(\frac{I df(\x) \mp J d\chi(\x)}
{2f(\x)}\right) \sigma_{3} \mp J \frac{d\chi(\x)}{2f(\x)}
\label{GammaHE}
\end{equation}
and
\begin{equation}
\chi := \Im{\E}, \quad J := \left( \begin{array}{cc}
0 & 1 \\ -1 & 0
\end{array} \right), \quad \sigma_{3} := \left( \begin{array}{cc}
1 & 0 \\ 0 & -1
\end{array} \right).
\end{equation}
The 1-form $\Gamma_{HE}$ can be obtained from $\Gamma_{N}$ by the gauge
transformation (\ref{gauge0}) corresponding to $p = p_{N \rightarrow HE}$, 
where
\begin{equation} 
p_{N \rightarrow HE}(\x,\tau) = \frac{1}{2\sqrt{|f(\x)|}} 
\left( \begin{array}{cc}
1 \pm i & -1 \mp i \\ 1 \mp i & 1 \mp i
\end{array} \right).
\end{equation}

On the other hand, the Kinnersley--Chitre formulation of the linear 
system\footnote{W.~Kinnersley and D.~M.~Chitre, {\em Symmetries of the 
stationary Einstein-Maxwell field equations, III}, J.\ Math.\ Phys.\ 
{\bf 19}, 1926--1931 (1978).} 
corresponds to the choice $\Gamma(\x,\tau) = \Gamma_{KC}(\x,\tau)$, where 
\begin{equation}
\Gamma_{KC}(\x,\tau) := \frac{1}{2} \Lambda(\x,\tau)^{-1} dH(\x) \Omega,
\quad \Omega := \left( \begin{array}{cc}
0 & i \\ -i & 0
\end{array} \right),
\label{GammaKC}
\end{equation}
with
\begin{equation}
\Lambda(\x,\tau) := \tau - (z \pm \rho \star)
\end{equation}
and $H(\x)$ a $2 \times 2$ matrix generalization of the Ernst potential 
$\E(\x)$ that can be introduced in the following manner.

It is well-known that any vacuum spacetime possessing two commuting 
Killing vector fields can be described in terms of a $2 \times 2$ real
symmetric matrix $h(\x)$ (a $2 \times 2$ block of the metric tensor) 
that depends exclusively on the nonignorable coordinates, and that this 
matrix satisfies the equation
\begin{equation}
d[\rho \star dh h^{-1}] = d[\rho \star h^{-1} dh] = 0,
\label{h_eqn}
\end{equation}
where
\begin{equation}
\rho := \sqrt{|\det{h}|}.
\end{equation}
Equation (\ref{h_eqn}) can be used to justify the introduction of a 
complex $H$-potential that satisfies the equations
\begin{equation}
\rho d(\Im{H}) = i h \Omega \star dh \text{ and } \Re{H} = -h,
\end{equation}
or, equivalently,
\begin{equation}
2 (z \pm \rho \star) dH = \left( H + H^{\dagger} \right) \Omega dH, 
\label{sdr}
\end{equation}
where
\begin{equation}
H - H^{T} = 2z \Omega \text{ and } \Re{H} = -h.
\end{equation}
Then it is not difficult to establish that $\Gamma(\x,\tau)$ as given by 
Eq.\ (\ref{GammaKC}) satisfies the zero-curvature condition (\ref{zerocurv}) 
if and only if $\E := H_{22}$ satisfies the Ernst equation.

The reader can verify that the K--C connection (\ref{GammaKC}) is related 
to the H--E connection (\ref{GammaHE}) by 
\begin{equation}
\Gamma_{KC}(\x,\tau) := p(\x,\tau) \Gamma_{HE}(\x,\tau) p(\x,\tau)^{-1}
+ dp(\x,\tau) p(\x,\tau)^{-1}, 
\end{equation}
and $\F_{KC}(\x,\tau)$ is related to $\F_{HE}(\x,\tau)$ by
\begin{equation}
\F_{KC}(\x,\tau) = p(\x,\tau) \F_{HE}(\x,\tau) p(\x_{0},\tau)^{-1},
\label{G2.14a}
\end{equation}
where
\begin{equation} 
p(\x,\tau) = \frac{1}{\sqrt{|h_{22}(\x)|}}\left( \begin{array}{cc}
1 & \mp h_{12}(\x) \\ 0 & |h_{22}(\x)|
\end{array} \right) P^{M}(\x,\tau),
\end{equation}
\begin{equation}
P^{M}(\x,\tau) :=
\left( \begin{array}{cc}
1 & \pm i(\tau-z) \\ 0 & 1
\end{array} \right) \left( \begin{array}{cc}
1 & 0 \\ 0 & \mu(\x,\tau)^{-1}
\end{array} \right) \frac{1}{\sqrt{2}} (\mp\sigma_{3}-\sigma_{2})
\label{Uref}
\end{equation}
and
\begin{equation}
\mu(\x,\tau) := \sqrt{(\tau-z)^{2} \pm \rho^{2}}, \quad
\lim_{\tau\rightarrow\infty}\frac{\mu(\x,\tau)}{\tau} := 1.
\label{mu}
\end{equation}
Note that, for fixed $\x$, $\mu(\x,\tau)$ is a holomorphic function of $\tau$
throughout a cut complex plane.  It has branch points of index $1/2$ at the
zeroes of $\mu(\x,\tau)$, which are at the end points of the branch cut, and
a simple pole at $\tau = \infty$.  

We shall assume that this brief review of the three formulations of the linear
system for the Ernst equation and the relationships among these formulations
will suffice.  In the rest
of this paper we shall suppress the subscript $KC$ on $\Gamma_{KC}(\x,\tau)$
and $\F_{KC}(\x,\tau)$ as we proceed to discuss how a group $\K$ such as the
Geroch group can be described in terms of its action upon the potentials
$\F_{KC}(\x,\tau)$ associated with the spacetimes in question.

\setcounter{equation}{0}
\subsection{The set $\S_{\F}$ of Kinnersley--Chitre $\F$-potentials}

In order to discuss in a meaningful way the action of the group $\K$ upon 
the potentials $\F(\x,\tau)$, we must first identify the set 
$\S_{\F}$ of $\F$-potentials being considered, and this requires,
in particular, the specification of the domain of $\F(\x,\tau)$.  This
can best be done by first specifying the domain of $H(\x)$ [and $\E(\x)$], 
and then choosing the gauge of $\F(\x,\tau)$ so as to minimize its 
singularities in the complex $\tau$-plane.  Throughout the rest of this 
paper we shall be concerned exclusively with the hyperbolic Ernst 
equation, where we find it convenient to introduce null coordinates 
$r := z - \rho$ and $s := z + \rho$ and to adopt the $\E$-potential 
domain (see Fig.~1)
\begin{equation}
D := \dom{\E} := \{(r,s): r_{1} < r < r_{2}, s_{2} < s < s_{1}, r < s\}.
\end{equation}
It is to be understood that $r_{1}$ may be $-\infty$ and/or $s_{1}$ may be
$+\infty$.  Moreover, we restrict attention to domains $D$ such that
$r_{1} < s_{2}$ and $r_{2} < s_{1}$; i.e., $\rho > 0$ at both the
lower left vertex $(r_{1},s_{2})$ and the upper right vertex $(r_{2},s_{1})$,
while $\rho$ may be greater than, less than or equal to zero at the lower 
right vertex $\x_{2} := (r_{2},s_{2})$.  Finally, we select one point 
$\x_{0} := (r_{0},s_{0}) \in D$ such that the null line segments 
$\{(r,s_{0}):r_{1}<r<r_{2}\}$ and $\{(r_{0},s):s_{2}<s<s_{1}\}$ lie
entirely within $D$; and at this point we assign the Minkowski space
value $\E(\x_{0})=-1$ to the complex $\E$-potential.\footnote{We have 
also considered more general domains and a more general choice for
$\x_{0}$, but to include discussion of these extensions here would
unnecessarily complicate our exposition.}  It is our intention to solve
an initial value problem in which $\E(\x)$ is determined throughout $D$
from its values specified on the two null line segments through the point
$\x_{0}$.

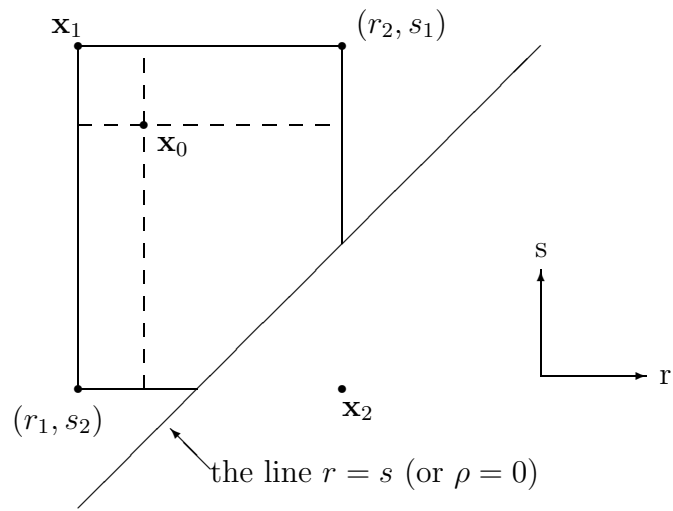
\begin{figure}[htbp] 
\begin{picture}(300,300)(-60,60)
\put(75,75){\line(1,1){175}}
\put(125,85){the line $r=s$ (or $\rho=0$)}
\put(125,90){\vector(-1,1){15}}
\put(250,125){\vector(0,1){40}}
\put(250,125){\vector(1,0){40}}
\put(248,170){s}
\put(295,122){r}
\put(75,250){\line(1,0){100}}
\put(75,120){\line(1,0){45}}
\put(75,120){\line(0,1){130}}
\put(175,175){\line(0,1){75}}
\put(75,250){\circle*{3}}
\put(75,120){\circle*{3}}
\put(175,250){\circle*{3}}
\put(175,120){\circle*{3}}
\put(65,255){${\mathbf x}_{1}$}
\put(175,110){${\mathbf x}_{2}$}
\put(50,105){$(r_{1},s_{2})$}
\put(180,255){$(r_{2},s_{1})$}
\put(100,220){\circle*{3}}
\put(105,210){${\mathbf x}_{0}$}
\multiput(75,220)(10,0){10}{\line(1,0){5}}
\multiput(100,120)(0,10){13}{\line(0,1){5}}
\end{picture}
\caption{An $\E$-potential domain $D$ for which $s_{2} < r_{2}$ is
illustrated.  The null line segments through $\x_{0}$ are represented by 
the vertical and horizontal dashed lines.}
\end{figure}

For a given choice of the triple $(\x_{0},\x_{1},\x_{2})$, we shall define
\begin{eqnarray}
\S_{\E} & := & \text{the set of all complex-valued functions $\E$
such that} \nonumber \\ & & 
\text{$\dom{\E} = D$, the derivatives $\E_{r}(\x)$, $\E_{s}(\x)$ and}
\nonumber \\ & &
\text{$\E_{rs}(\x)$ exist and are continuous at all $\x \in D$, }
\\ & &
\text{$f := \Re{\E} > 0$ and $\E$ satisfies Eq.\ (\ref{EE1}) thoughout}
\nonumber \\ & &
\text{$D$, and $\E(\x_{0})=-1$.}
\nonumber
\end{eqnarray}
The metric components $h_{ab}$ corresponding to each given $\E \in \S_{\E}$
are defined by $h_{22} := -f$, $d\omega := \rho f^{-2} \star d\chi$ such
that $\omega(\x_{0}) := 0$, $h_{12} := \omega h_{22}$ and $h_{11} := 
[(h_{12})^{2} + \rho^{2}]/h_{22}$.

Naturally, we shall let $\dom{H} = D$ and assign the value
\begin{equation}
H(\x_{0}) = H^{M}(\x_{0}),
\label{1.12c}
\end{equation}
where $H^{M}$ is the Minkowski space $H$-potential with values
\begin{equation}
H^{M}(\x) = - \left( \begin{array}{cc}
\rho^{2} & 0 \\ 2iz & 1
\end{array} \right).
\label{1.14c}
\end{equation}
For a given choice of the triple $(\x_{0},\x_{1},\x_{2})$, we shall define
\begin{eqnarray}
\S_{H} & := & \text{the set of all complex-valued $2 \times 2$ matrix
functions } \nonumber
\\ & & \text{$H$ with $\dom{H} := D$ such that there exists $\E \in \S_{\E}$}
\\ & & \text{for which $\Re{H} = -h$, $d(\Im{H})$ exists and satisfies }
\nonumber
\\ & & \rho d(\Im{H}) = i h \Omega \star dh \text{ and the gauge condition 
(\ref{1.12c}) holds.}
\nonumber
\end{eqnarray}

Let $\I^{(3)}(\x)$ denote the open interval with end points $r,r_{0}$
and $\I^{(4)}(\x)$ denote the open interval with end points $s,s_{0}$,
and let $\bar{\I}^{(3)}(\x)$ and $\bar{\I}^{(4)}(\x)$ denote, respectively,
the closures of these two intervals.  Furthermore, let
\begin{eqnarray}
\I(\x) & := & \I^{(3)}(\x) \cup \I^{(4)}(\x), \text{ and } \\
\bar{\I}(\x) & := & \bar{\I}^{(3)}(\x) \cup \bar{\I}^{(4)}(\x).
\end{eqnarray}
Note that $\bar{\I}^{(3)}(\x)$ is empty if $r=r_{0}$ and 
$\bar{\I}^{(4)}(\x)$ is empty if $s=s_{0}$.  When neither $r=r_{0}$
nor $s=s_{0}$, the set $\bar{\I}(\x)$ comprises two disjoint closed
sets (for $\x_{0}$ chosen as indicated earlier).  The gauge of the 
$\F$-potential can be chosen so that 
\begin{equation}
\dom{\F} := \{(\x,\tau): \x \in D, \tau \in C - \bar{\I}(\x)\}.
\label{Fdom}
\end{equation}
For a given choice of the triple $(\x_{0},\x_{1},\x_{2})$, we shall define
\begin{eqnarray}
\S_{\F} & := & \text{the set of all complex-valued $2 \times 2$
matrix functions } 
\nonumber
\\ & & \text{$\F$ with domain (\ref{Fdom}) such that there exists 
$H \in \S_{H}$}
\nonumber
\\ & & \text{such that, for all $\x \in D$ and
$\tau \in [C - \bar{\I}(\x)]-\{r_{0},s_{0}\}$,}
\nonumber
\\ & & \text{$d\F(\x,\tau)$ exists and Eq.\ (\ref{linsys}) holds,
subject to the}
\\ & & \text{condition (\ref{init}), and, for each $(r,s) \in D$, 
$\F((r,s_{0}),\tau)$}
\nonumber
\\ & & \text{and $\F((r_{0},s),\tau)$ are continuous functions of $\tau$ at}
\nonumber
\\ & & \text{$\tau = s_{0}$ and at $\tau = r_{0}$, respectively.}
\nonumber
\end{eqnarray}
Remember that at $\x = \x_{0}$, $\F(\x,\tau)$ reduces to the
$2 \times 2$ unit matrix.

With these definitions one can establish the properties enumerated in the
following theorem, the proof of which is (except for conventions and 
notations and the choice of the domain $D$) essentially the same as that 
given in two earlier papers\footnote{I.\ Hauser and F.\ J.\ Ernst, {\em
Initial value problem for colliding gravitational plane waves-III/IV\/},
J.~Math.~Phys.\ {\bf 31}, 871--881 (1990), {\bf 32}, 198--209 (1991).
In these papers we used `$P$' in place of `$\F$'.} on 
the IVP (initial value problem) for colliding gravitional plane wave pairs
by the present authors.  The complex-valued functions $\E^{(3)}$ and
$\E^{(4)}$ with respective domains $\I^{(3)} := \{r: r_{1} < r < r_{2}\}$
and $\I^{(4)} := \{s: s_{2} < s < s_{1}\}$ serve as initial value data
for the $\E$-potential on the null line segments through the point $\x_{0}$.

\begin{theorem}[Initial Value Problem]
\label{Thm_3}
\mbox{ } \\ \vspace{-3ex}
\begin{romanlist}
\item 
For each $H \in \S_{H}$, the corresponding $\F \in 
\S_{\F}$ exists and is unique; and, for each $\x \in D$,
$\F(\x,\tau)$ is a holomorphic function of $\tau$ throughout 
$C - \bar{\I}(\x)$ and, in at least one neighborhood of $\tau = \infty$,
\begin{equation}
\F(\x,\tau) = I + (2\tau)^{-1} \left[ H(\x) - H(\x_{0}) \right] \Omega
+ O(\tau^{-2}).
\end{equation}
\item 
For each $\F \in \S_{\F}$, there is only one 
$H \in \S_{H}$ for which $d\F(\x,\tau) = \Gamma(\x,\tau)
\F(\x,\tau)$.
\item 
With the understanding that
\begin{equation}
\dom{\nu} := \{ (\x,\tau): \x \in D \text{ and } 
\tau \in C - \bar{\I}(\x)\}
\end{equation}
and that $\nu(\x,\infty) = 1$, we have
\begin{equation}
\det{\F(\x,\tau)} = \nu(\x,\tau) := \frac{\mu(\x_{0},\tau)}{\mu(\x,\tau)} = 
\left( \frac{\tau-r_{0}}{\tau-r} \right)^{1/2}
\left( \frac{\tau-s_{0}}{\tau-s} \right)^{1/2}.
\label{barnu}
\end{equation}
\item 
The member of $\S_{\F}$ that corresponds to $\E^{M}$ is given by
\begin{equation}
\F^{M}(\x,\tau) = \left( \begin{array}{cc}
1 & -i(\tau-z) \\ 0 & 1
\end{array} \right) \left( \begin{array}{cc}
1 & 0 \\ 0 & \nu(\x,\tau)
\end{array} \right) \left( \begin{array}{cc}
1 & i(\tau-z_{0}) \\ 0 & 1
\end{array} \right).
\label{G2.10c}
\end{equation}
\item 
For each $\E \in \S_{\E}$, there is exactly one $H \in \S_{H}$
such that $\E = H_{22}$.
\item 
If, for each $i \in \{3,4\}$, $\E^{(i)}$ is $\bC^{n_{i}}$ ($n_{i} \ge
1$), then, for all $0 \le k < n_{3}$ and $0 \le m \le n_{4}$,
the partial derivatives $\partial^{k+m}H(\x)/\partial r^{k} \partial
s^{m}$ exist and are continuous throughout $D$.  If, for each $i \in
\{3,4\}$, $\E^{(i)}$ is analytic, then $H$ is analytic.
\item 
For each choice of complex valued functions $\E^{(3)}$ and $\E^{(4)}$ 
for which (for $i \in \{3,4\}$) $\dom{\E^{(i)}} = \I^{(i)}$, $\E^{(i)}$ 
is $\bC^{1}$, $f^{(i)} := \Re{\E^{(i)}} < 0$ throughout $\I^{(i)}$, and
$\E^{(3)}(r_{0}) = -1 = \E^{(4)}(s_{0})$, there exists exactly one 
$\E \in \S_{\E}$ such that
$$
\E^{(3)}(r) = \E(r,s_{0}) \text{ and } \E^{(4)}(s) = \E(r_{0},s)
$$
for all $r \in \I^{(3)}$ and $s \in \I^{(4)}$, respectively.
\end{romanlist}
\end{theorem}

\setcounter{equation}{0}
\subsection{Homogeneous Hilbert problem}

The HHP that we developed for effecting K--C transformations\footnote{I.~Hauser 
and F.~J.~Ernst, {\em A homogeneous Hilbert problem for the
Kinnersley--Chitre transformations}, J.\ Math.\ Phys.\ {\bf 21}, 1126-1140
(1980).} (adapted to the hyperbolic case) involved
a closed contour in the complex $\tau$-plane surrounding the arcs that 
comprise $\bar{\I}(\x)$.  This was fine as long as we were dealing with
the analytic case, but now we must instead formulate an HHP on those
arcs themselves, and this will involve the limiting values of 
$\F(\x,\tau)$ as $\tau$ approaches points on those arcs.  What we
discovered concerning these limiting values is contained in the 
following theorems, the proofs of which are based upon a
classic method of reducing the solving of a total differential equation
to the solving of a pair of ordinary linear differential equations along 
characteristic lines in $D$.  The Picard method of successive approximations
and certain well known theorems of infinite sequences of functions are used
to demonstrate existence, continuity and differentiability properties of the
solution.\footnote{For each $\sigma \in R^{1}$ and for fixed $\x \in D$, the
limits of $\F_{HE}(\x,\sigma\pm\zeta)$ as $\zeta \rightarrow 0 (\Im{\zeta}
> 0)$ exist.  Moreover, $\F_{HE}(\x,\tau^{*}) = \F_{HE}(\x,\tau)^{*}$ and
$\det{\F(\x,\tau)}=1$.  For these reasons, we found it convenient to use
the H--E representation of the linear system in developing this proof,
translating the results into corresponding results for the K--C
representation.}

\begin{theorem}[Limits of $\F$]
\label{3.7G}
\mbox{ } \\ \vspace{-3ex}
\begin{romanlist}
\item
For each $\x \in D$ and and $\sigma \in \I(\x)$
the limits $\F^{\pm}(\x,\sigma) := \lim_{\zeta \rightarrow 0} 
\F(\x,\sigma \pm \zeta) (\Im{\zeta} > 0)$ exist.
\item
Further, let $\alpha$ and $\beta$ be points of $\bar{\I}(\x)$ such that
$\alpha \in \{r_{0},s_{0}\}$ and $\beta \in \{r,s\}$, and let $\tau \in 
C - \bar{\I}(\x)$.  Then the following limits all exist and are equal
as indicated:
\begin{eqnarray}
\lim_{\sigma \rightarrow \alpha} \F^{\pm}(\x,\sigma) & = &
\lim_{\tau \rightarrow \alpha} \F(\x,\tau),
\label{3.46a} \\
\lim_{\sigma \rightarrow \beta} [\F^{\pm}(\x,\sigma)^{-1}] & = &
\lim_{\tau \rightarrow \beta} [\F(\x,\tau)^{-1}].
\label{3.46b}
\end{eqnarray}
\end{romanlist}
\end{theorem}

We shall employ $\Box$ as a generic superscript that stands for $n$, $n+$,
$\infty$ or `an' (analytic).  The symbols $\bC^{n}$ and $\bC^{\infty}$ are
self explanatory.  We shall say that $f$ is $\bC^{n+}$ if its $n$th 
derivative $D^{n}f$ exists throughout $\dom{f}$ and $D^{n}f$ obeys a 
H\"{o}lder condition of arbitrary index on each closed subinterval of 
$\dom{f}$.\footnote{The index may be different for different closed 
subintervals of $\dom{f}$.}  

If $f$ is a real- or complex-valued function, the domain of which 
is a union of disjoint intervals of $R^{1}$, and $[a,b]$ is a given
closed subinterval of $\dom{f}$, then $f$ is said {\em to obey a
H\"{o}lder condition of index $0 < \gamma \le 1$ on $[a,b]$; i.e.,
to be $H(\gamma)$ on $[a,b]$}, if there exists $M(a,b,\gamma) > 0$
such that $|f(x')-f(x)| \le M(a,b,\gamma) |x'-x|^{\gamma}$ for all
$x, x' \in [a,b]$.  The same terminology is used if $f(x)$ is a matrix
with real or complex elements, and $|f(x)|$ is its norm.

\begin{definition}{Dfn.\ of the groups $K^{\Box}$ and $K$}
In order to describe our extensions $\K^{\Box}$ of the Geroch group, we 
shall introduce groups $K^{\Box}$ of $2 \times 2$ matrix pairs; namely, the 
multiplicative groups of all ordered pairs $\bv = (v^{(3)},v^{(4)})$ 
of $2 \times 2$ matrix functions such that, for both $i=3$ and 
$i=4$,
\begin{equation}
\dom{v^{(i)}} = \I^{(i)}, \quad \det{v^{(i)}} = 1, \quad v^{(i)} 
\text{ is } \bC^{\Box} 
\label{gKbox}
\end{equation}
and the condition 
\begin{equation}
v^{(i)}(\sigma)^{\dagger} \A^{M}(\x_{0},\sigma)
v^{(i)}(\sigma) = \A^{M}(\x_{0},\sigma)
\text{ for all } \sigma \in \I^{(i)}
\label{4.2}
\end{equation}
holds, where
\begin{equation}
\A^{M}(\x_{0},\sigma) := (\sigma-z_{0}) \Omega + \Omega h^{M}(\x_{0})
\Omega, \quad h^{M}(\x_{0}) := \left( \begin{array}{cc}
\rho_{0}^{2} & 0 \\ 0 & 1
\end{array} \right).
\end{equation}
Moreover, the symbol $K$ will denote the multiplicative group of all 
ordered pairs $\bv = (v^{(3)},v^{(4)})$ of $2 \times 2$ matrix functions
such that, for both $i=3$ and $i=4$,
\begin{equation}
\dom{v^{(i)}} = \I^{(i)}, \quad \det{v^{(i)}} = 1, \quad
v^{(i)} \text{ is } H(1/2) \text{ on each closed subinterval of }
\I^{(i)}
\label{gK}
\end{equation}
and the condition (\ref{4.2}) holds.
\end{definition}

\begin{definition}{Dfn.\ of the HHP corresponding to $(\bv,\F_{0})$}
For each $\bv \in K^{\Box}$ and $\F_{0} \in \S_{\F}$,
{\em the\/} HHP {\em corresponding to} $(\bv,\F_{0})$ will mean 
the set of all functions $\F$ [which are not presumed to be members 
of $\S_{\F}$] such that $\dom{\F} = \{(\x,\tau):\x \in
D, \tau \in C - \bar{\I}(\x)\}$ and such that, for each $\x \in D$, 
the functions $\F(\x)$ whose domains are $C - \bar{\I}(\x)$ and whose 
values are $\F(\x,\tau)$ is a solution of {\em the\/} HHP {\em 
corresponding to} $(\bv,\F_{0},\x)$, i.e., a member of the set of all
$2 \times 2$ matrix functions $\F(x)$ such that
\begin{arablist}
\item 
$\F(\x)$ is holomorphic throughout $\dom{\F(\x)} := 
C - \bar{\I}(\x)$, 
\item 
$\F(\x,\infty) = I$, 
\item 
$\F^{\pm}(\x)$ exist, and 
\begin{eqnarray}
Y^{(i)}(\x,\sigma) & := & \F^{+}(\x,\sigma) v^{(i)}(\sigma)
[\F_{0}^{+}(\x,\sigma)]^{-1} \nonumber \\
& = & \F^{-}(\x,\sigma) v^{(i)}(\sigma) [\F_{0}^{-}(\x,\sigma)]^{-1} 
\label{G3.17} \\
& & \text{ for each $i \in \{3,4\}$ and $\sigma \in \I^{(i)}(\x)$,}
\nonumber
\end{eqnarray}
\item 
$\F(\x)$ is bounded at $\x_{0}$ and $\nu(\x)^{-1} \F(\x)$
is bounded at $\x$, and the function $Y(\x)$ whose domain is $\I(\x)$ 
and whose values are given by $Y(\x,\sigma) := Y^{(i)}(\x,\sigma)$ for each
$\sigma \in \I^{(i)}(\x)$ is bounded at $\x_{0}$ and at $\x$. 
\end{arablist}
The members of the HHP corresponding to $(\bv,\F_{0})$ will be 
called its {\em solutions}.
\end{definition}

Notes:  
\begin{itemize}
\item
$\F^{+}(\x)$ and $\F^{-}(\x)$ denote the functions that have 
the common domain $\I(\x)$ and the values ($\Im{\zeta} > 0$)
\begin{equation}
\F^{\pm}(\x,\sigma) := \lim_{\zeta \rightarrow 0} \F(\x,\sigma\pm\zeta).
\label{G3.19}
\end{equation}
It is understood that $\F^{+}(\x)$ and $\F^{-}(\x)$ exist if and only if
the above limits exist for every $\sigma \in \I(\x)$.  $\nu^{+}(\x)$ and
$\nu^{-}(\x)$ are similarly defined.
\item
$\nu(\x)$ denotes the function whose domain is $C - \bar{\I}(\x)$ 
and whose values $\nu(\x,\tau)$ are defined in Eq.\ (\ref{barnu}).
\item
It is to be understood that $\F(\x)$, with domain $C - \bar{\I}(\x)$,
is {\em bounded at $\x_{0}$} if there exists a neighborhood $\nbd(\x_{0})$ 
of the set $\{r_{0},s_{0}\}$ in the space $C$ such that
\begin{equation}
\{ \F(\x,\tau):\tau \in \nbd(\x_{0})-\bar{\I}(\x) \}
\end{equation}
is bounded.  Likewise, $\F(\x)$ is said to be {\em bounded at $\x$}
if there exists a neighborhood $\nbd(\x)$ of the set $\{r,s\}$ in the space
$C$ such that
\begin{equation}
\{ \F(\x,\tau):\tau \in \nbd(\x)-\bar{\I}(\x) \}
\end{equation}
is bounded.
\item
We say that $Y(\x)$, with domain $\I(\x)$, is {\em bounded at $\x_{0}$} 
if there exists a neighborhood $\nbd(\x_{0})$ of the set $\{r_{0},s_{0}\}$
in the space $R^{1}$ such that
\begin{equation}
\{ Y(\x,\sigma):\sigma \in \nbd(\x_{0})\cap\I(\x) \}
\end{equation}
is bounded.  Likewise, $Y(\x)$ is {\em bounded at $\x$} if there
exists a neighborhood $\nbd(\x)$ of the set $\{r,s\}$ in the space
$R^{1}$ such that
\begin{equation}
\{ Y(\x,\sigma):\sigma \in \nbd(\x)\cap\I(\x) \}
\end{equation}
is bounded.
\end{itemize}

\begin{theorem}[Properties of HHP solution]
\label{4.3D}
\mbox{ } \\
Suppose that $\bv \in K^{\Box}$, $\F_{0} \in \S_{\F}$ and 
$\x \in D$ exist such that a solution $\F(\x)$ of the HHP 
corresponding to $(\bv,\F_{0},\x)$ exists.  Then
\begin{romanlist}
\item
${\F}^{+}(\x)$, ${\F}^{-}(\x)$ and $Y(\x)$ are continuous throughout 
$\I(\x)$,
\item
${\F}^{\pm}(\x)$ are bounded at $\x_{0}$, and 
$[\nu^{\pm}(\x)]^{-1} {\F}^{\pm}(\x)$ are bounded at $\x$, 
\item
$\det{\F(\x)} = \nu(\x), \quad \det{Y(\x)} = 1$, 
\item
the solution $\F(\x)$ is unique, and
\item
the solution of the HHP corresponding to $(\bv,\F_{0},\x_{0})$
is given by
\begin{equation}
\F(\x_{0},\tau) = I 
\label{4.30}
\end{equation}
for all $\tau \in C$.
\end{romanlist}
\end{theorem}

\proofs
\begin{romanlist}
\item 
The statement that $\F^{+}(\x)$ and $\F^{-}(\x)$ are continuous is a
direct consequence of a theorem by P.~Painlev\'{e} which is stated and
proved by N.\ I.\ Muskhelishvili.\footnote{N.~I.~Muskhelishvili, {\em
Singular Integral Equations}, Ch.\ 2, Sec.~14, pp.\ 33-34 (Dover, 1992).
\label{Musk}} The continuity of $Y(\x)$ then follows from its definition
by Eq.\ (\ref{G3.17}), the fact that $v^{(i)}$ is continuous and the fact 
that $\F_{0}^{+}(\x)$ and $\F_{0}^{-}(\x)$ are continuous.
\cheers

\item 
From Eq.\ (\ref{G3.17}),
\begin{equation}
\F^{\pm}(\x) = Y^{(i)}(\x) \F_{0}^{\pm}(\x) [v^{(i)}]^{-1}
\label{4.31}
\end{equation}
for each $i \in \{3,4\}$.  The function $Y(\x)$ is bounded at $\x$ and
at $\x_{0}$ according to condition (4) in the definition of the HHP,
and $v^{(i)}$ and its inverse are continuous throughout $\I^{(i)}$.
Finally, $\F_{0}^{\pm}(\x)$ is bounded at $\x_{0}$ and 
$[\nu^{\pm}(\x)]^{-1} \F_{0}^{\pm}(\x)$ is bounded at $\x$, so, from 
Eq.\ (\ref{4.31}), $\F^{\pm}(\x)$ is bounded at $\x_{0}$, and
$[\nu(\x)]^{-1} \F^{\pm}(\x)$ is bounded at $\x$.
\cheers

\item 
Conditions (1), (2), (3) and (4) of the definition of the HHP imply that 
\begin{equation}
Z_{1}(\x) := \det{\F(\x)}/\nu(\x) \text{ is holomorphic throughout } C - \bar{\I}(\x),
\label{4.33b}
\end{equation}
\begin{equation}
Z_{1}(\x,\infty) = 1,
\label{4.33c}
\end{equation}
\begin{equation}
\begin{array}{l}
\text{the limits $Z_{1}^{\pm}(\x)$ exist and } \\
\det{Y(\x,\sigma)} = Z_{1}^{+}(\x,\sigma) = Z_{1}^{-}(\x,\sigma)
\text{ for all } \sigma \in \I(\x),
\end{array}
\label{4.33d}
\end{equation}
\begin{equation}
\begin{array}{l}
\nu(\x) Z_{1}(\x) \text{ is bounded at } \x_{0} \text{ and } \\
\nu(\x)^{-1} Z_{1}(\x) \text{ is bounded at } \x,
\end{array}
\label{4.33e}
\end{equation}
and
\begin{equation}
\det{Y(\x)} = Z_{1}^{\pm}(\x) \text{ is bounded at $\x$ and at $\x_{0}$.}
\label{4.33f}
\end{equation}

From the above statements (\ref{4.33b}) and (\ref{4.33d}) together with
the theorem of Riemann\footnote{See Sec.~24, Ch.~1, of {\em A Course of
Higher Mathematics}, Vol.~III, Part Two, by V.\ I.\ Smirnov (Addison-Wesley,
1964).\label{acaa}} on analytic continuation across an arc, $Z_{1}(\x)$ has a 
holomorphic extension to the domain $C - \{r,s,r_{0},s_{0}\}$; and, from 
the statements (\ref{4.33e}) and (\ref{4.33f}), together with the theorem 
of Riemann\footnote{See Sec.~133 of {\em Theory of Functions of a Complex
Variable}, Vol.~1, by C.\ Caratheodory, 2nd English edition (Chelsea
Publishing Company, 1983).\label{isohf}} on isolated singularities of holomorphic 
functions, $Z_{1}(\x)$ has a further holomorphic extension $Z_{1}^{ex}(\x)$
to $C$.  Finally, the theorem of Liouville\footnote{See Secs.~167-168 of
the text by Caratheodory cited above.\label{Liouville}} on entire functions
that do not have an essential singularity at $\tau=\infty$, together with 
Eq.\ (\ref{4.33c}), then yields 
\begin{equation}
Z_{1}^{ex}(\x,\tau) = 1 \text{ for all } C.
\end{equation}
Thus, we have shown that $\det{\F(\x)} = \nu(\x)$, whereupon
Eq.\ (\ref{4.33d}) yields $\det{Y(\x)} = 1$.
\cheers

\item 
Suppose that $\F'(\x)$ is also a solution of the HHP corresponding
to $(\bv,\F_{0},\x)$.  Since $\det{\F(\x)} = \nu(\x)$, $\F(\x)$ is 
invertible.  Conditions (1), (2), (3) and (4) in the definition of
the HHP imply that
\begin{equation}
Z_{2}(\x) := \F'(\x) \F(\x)^{-1} \text{ is holomorphic throughout }
C - \bar{\I}(\x),
\end{equation}
\begin{equation}
Z_{2}(\x,\infty) = I,
\end{equation}
\begin{equation}
\begin{array}{l}
\text{the limits $Z_{2}^{\pm}(\x)$ exist and } \\
Y'(\x) Y(\x)^{-1} = Z_{2}^{+}(\x) = Z_{2}^{-}(\x) 
\text{ throughout } \I(\x),
\end{array}
\end{equation}
\begin{equation}
Z_{2}(\x) \text{ is bounded at $\x$ and at $\x_{0}$,}
\end{equation}
and
\begin{equation}
Y'(\x) Y(\x)^{-1} = Z_{2}^{\pm}(\x) 
\text{ is bounded at $\x$ and at $\x_{0}$.}
\end{equation}
The same kind of reasoning that was used in the proof of part (iii) of
the theorem nets $Z(\x) = I$.  So $\F'(\x) = \F(\x)$.
\cheers

\item 
When $\x = \x_{0}$, $\I(x)$ and its closure $\bar{\I}(\x)$ are empty.
So, condition (1) of the HHP definition implies that $\F(\x_{0})$
is holomorphic throughout $C$, whereupon condition (2) tells us that
$\F(\x_{0})$ has the value $I$ throughout $C$.  [$\F^{\pm}(\x)$
are empty sets when $\x = \x_{0}$; and conditions (3) and (4) hold
trivially when $\x = \x_{0}$.]
\cheers
\end{romanlist}

\setcounter{equation}{0}
\subsection{The generalized Geroch conjecture}

At this point we shall conjecture that for each $\Box$, where $\Box$
may be $n$ or $n+$, where $n \ge 3$, $\infty$ or `an' (analytic),
the following theorems hold:
\begin{itemize}
\item 
There exists a subset $\S_{\F}^{\Box}$ of $\S_{\F}$ such that, for
each $\F_{0} \in \S_{\F}^{\Box}$ and each $\bv \in K^{\Box}$, there 
exists exactly one solution $\F \in \S_{\F}^{\Box}$ of the HHP 
corresponding to $(\bv,\F_{0})$, enabling us to define a mapping
\begin{equation}
[\bv]:\S_{\F}^{\Box} \rightarrow \S_{\F}^{\Box}
\end{equation}
such that, for each $\F_{0} \in \S_{\F}^{\Box}$,
\begin{equation}
[\bv](\F_{0}) = \F
\end{equation}
is that unique solution of the HHP corresponding to $(\bv,\F_{0})$.
We then define our extension $\K^{\Box}$ of the K--C group by
\begin{equation}
\K^{\Box} := \{[\bv]:\bv\in K^{\Box}\}.
\end{equation}

\item 
The mapping $[\bv]$ is the identity map on $\S_{\F}^{\Box}$ iff
$\bv \in Z^{(3)} \times Z^{(4)}$, where
\begin{equation}
Z^{(i)} := \{ \delta^{(i)},-\delta^{(i)} \}
\end{equation}
and
\begin{equation}
\delta^{(i)}(\sigma) = I \text{ for all $\sigma \in \I^{(i)}$.}
\end{equation}

\item 
The set $\K^{\Box}$ is a group of permutations of $\S_{\F}^{\Box}$
such that the mapping $\bv \rightarrow [\bv]$ is a homomorphism of
$K^{\Box}$ onto $\K^{\Box}$; and the mapping $\{\bv\bw: \bw \in
Z^{(3)} \times Z^{(4)}\} \rightarrow [\bv]$ is an isomorphism [viz,
the isomorphism of $K^{\Box}/(Z^{(3)} \times Z^{(4)})$ onto $\K^{\Box}$].

\item 
The group $\K^{\Box}$ is transitive [i.e., for each $\F_{0},
\F \in \S_{\F}^{\Box}$ there exists at least one element of
$\K^{\Box}$ that transforms $\F_{0}$ into $\F$]. 
\end{itemize}

It will later be seen when we come to Thm.~\ref{4.2D} that to prove the
first part of the above generalized Geroch conjecture it is sufficient
to prove that, for each $\bv \in \S_{\Y}^{\Box}$ with $\Box = n$,
$n+ (n \ge 3)$, $\infty$ or `an', the solution $\F$ of the HHP
corresponding to $(\bv,\F^{M})$ exists, and $\F \in \S_{\Y}^{\Box}$.
For this reason, we shall now focus on the HHP corresponding to
$(\bv,\F^{M})$.

We shall begin with
a study of an Alekseev-type singular integral equation and a Fredholm
integral equation of the second kind that are, under suitable 
circumstances, equivalent to the HHP corresponding to $(\bv,\F^{M})$.  
Ultimately we shall have
to return to the identification of the sets $\S_{\F}^{\Box}$ for
$\Box = n$, $n+$, $\infty$ and `an' (analytic), which will require us
to introduce the concept of generalized Abel transforms of the initial
data functions $\E^{(3)}$ and $\E^{(4)}$.

\setcounter{equation}{0}
\section{An Alekseev-type singular integral equation that is equivalent
to the HHP corresponding to $(\bv,\F^{M})$ when $\bv \in K^{1+}$}

Using an ingenious argument G.\ A.\ Alekseev\footnote{G.~A.\ Alekseev, 
{\em The method of the inverse scattering problem and singular integral
equation for interacting massless fields}, Dokl.\ Akad.\ Nauk SSSR
{\bf 283}, 577--582 (1985) [Sov.\ Phys.\ Dokl.\ (USA) {\bf 30}, 565
(1985)], {\em Exact solutions in the general theory of relativity}, 
Trudy Matem.\ Inst.\ Steklova {\bf 176}, 215--262 (1987). \label{Alekseev}}
derived a singular integral equation, supposing that $\F(\tau)$ was
analytic in a neighborhood of $\{r,s\}$ except for branch points of index
$1/2$ at $\tau=r$ and $\tau=s$.  We shall now show that the same type integral
equation arises in connection with solutions of our new HHP that need not be 
analytic.

\setcounter{equation}{0}
\subsection{A preliminary theorem}

Henceforth, whenever there is no danger of ambiguity, the arguments `$\x$'
and `$\x_{0}$' will be suppressed.  For example, `$\F(\tau)$'
and `$\F^{\pm}(\sigma)$' will generally be used as abbreviations for
`$\F(\x,\tau)$' and `$\F^{\pm}(\x,\sigma)$', respectively; and 
`$\nu(\tau)$', `$\nu^{\pm}(\sigma)$' and `$\bar{\I}$' will 
generally stand for `$\nu(\x,\tau)$',
`$\nu^{\pm}(\x,\sigma)$' and `$\bar{\I}(\x)$', respectively.

\begin{theorem}[Alekseev preliminaries]
\label{5.4A} \mbox{ } \\ \vspace{-3ex}
\begin{romanlist}
\item
Suppose that the solution $\F(\x)$ of the HHP corresponding to
$(\bv,\F_{0},\x)$ exists.  Then, for each $\tau \in C - \bar{\I}(\x)$,
\begin{eqnarray}
[\nu^{+}(\sigma')]^{-1} \frac{{\F}^{+}(\sigma') + {\F}^{-}(\sigma')}
{\sigma' - \tau} 
\text{ is summable over $\sigma' \in \bar{\I}(\x)$}, \nonumber \\
\text{with assigned orientation in the direction of increasing $\sigma'$},
\label{5.13a}
\end{eqnarray}
and
\begin{equation}
\left[\nu(\tau)\right]^{-1} \F(\tau) = 
I + \frac{1}{2\pi i} \int_{\bar{\I}} d\sigma'
[\nu^{+}(\sigma')]^{-1} \frac{{\F}^{+}(\sigma') + {\F}^{-}(\sigma')}
{\sigma'-\tau}, 
\label{5.13c}
\end{equation}
where the meaning we attribute to the symbol $\int_{\bar{\I}}$ should 
be obvious. 
\item
Moreover, for each $\sigma \in \I(\x)$, 
\begin{eqnarray}
[\nu^{+}(\sigma')]^{-1} \frac{{\F}^{+}(\sigma') + {\F}^{-}(\sigma')}
{\sigma' - \sigma} 
\text{ is summable over $\sigma' \in \bar{\I}(\x)$} \nonumber \\
\text{in the principal value (PV) sense,}
\label{5.14a}
\end{eqnarray}
and
\begin{equation}
\frac{1}{2} [\nu^{+}(\sigma)]^{-1} \left\{ {\F}^{+}(\sigma) - {\F}^{-}(\sigma)
\right\} = I + \frac{1}{2\pi i} \int_{\bar{\I}} d\sigma' 
[\nu^{+}(\sigma')]^{-1} \frac{{\F}^{+}(\sigma') + {\F}^{-}(\sigma')}
{\sigma' - \sigma} . 
\label{5.14c}
\end{equation}
\end{romanlist}
\end{theorem}

\proofs
\begin{romanlist}
\item 
From Thms.~\ref{4.3D}(i) and (ii), the function of $\sigma'$ given by
$\nu^{\pm}(\sigma')^{-1} \F^{\pm}(\sigma') (\sigma'-\tau)^{-1}$ is
continuous throughout $\I$ and is bounded at $\x$, while
$\F^{\pm}(\sigma')(\sigma'-\tau)^{-1}$ is bounded at $\x_{0}$.
Moreover, it is clear that $\nu^{\pm}(\sigma')$ and $\nu^{\pm}(\sigma')^{-1}$
are summable on $\bar{\I}$, and $\nu^{-}(\sigma') = - \nu^{+}(\sigma')$
throughout $\I$.  Statement (\ref{5.13a}) can now be obtained by employing
the well-known theorem\footnote{See {\em Integration}, by Edward J.\ McShane
(Princeton University Press, 1944).\label{McShane}} that the product of any
complex-valued
function which is summable on $[a,b] \subset R^{1}$ by a function which is
continuous and bounded on $[a,b]-$(any given finite set) is also summable
on $[a,b]$.

To obtain the conclusion (\ref{5.13c}), one employs Cauchy's integral
formula and the HHP condition $\F(\infty)=I$ to infer that
\begin{equation}
\nu(\tau)^{-1}\F(\tau) = I - \frac{1}{2\pi i}\int_{\Lambda}d\tau' 
\frac{[\nu(\tau')]^{-1}\F(\tau')}{\tau'-\tau},
\end{equation}
where $\Lambda$ is a closed positively oriented contour enclosing $\bar{\I}$
but not the point $\tau$, which we may assume to be rectangular.  This 
equation can be expressed in the form
\begin{equation}
\nu(\tau)^{-1}\F(\tau) = I - \frac{1}{2\pi i}\int_{\Lambda^{+}}d\tau' 
\frac{[\nu^{+}(\tau')]^{-1}\F^{+}(\tau')}{\tau'-\tau} 
- \frac{1}{2\pi i}\int_{\Lambda^{-}}d\tau' 
\frac{[\nu^{-}(\tau')]^{-1}\F^{-}(\tau')}{\tau'-\tau},
\end{equation}
where $\Lambda^{\pm} := \Lambda \cap \bar{C}^{\pm}$ denote the parts of
the contour $\Lambda$ that lie respectively in the upper and lower half
planes, $\bar{C}^{\pm}$.

To evaluate each of the integrals, one applies a well known 
generalization\footnote{See Remark 2 in Sec.~2, Ch.~II, of 
{\em Analytic Functions} by M.~A.~Evgrafov (Dover Publications,
1978).} of Cauchy's integral theorem which asserts that the integral of a 
function about a simple piecewise smooth contour $\K$ is zero if the given 
function is holomorphic throughout $\K_{int}$ and is continuous throughout 
$\K \cup \K_{int}$.  In the case of the first integral, we select the 
contour as in Fig.~2.  The other integral is evaluated in a similar way,
using a contour in $\bar{C}^{-}$.

\begin{figure}[htbp] 
\begin{picture}(300,30)(-130,-10)
\put(145,22){$\tau$}
\put(153,25){\circle*{2}}	
\put(5,-12){$a^{3}$}
\put(85,-12){$b^{3}$}
\put(10,0){\circle*{2}}		
\put(90,0){\circle*{2}} 	
\put(10,0){\oval(20,20)[t]}
\put(90,0){\oval(20,20)[t]}
\put(20,0){\line(1,0){60}}
\put(100,14){\vector(-1,0){50}}
\put(50,14){\line(-1,0){50}}
\put(0,0){\line(0,1){14}}
\put(100,0){\line(0,1){14}}
\put(125,-12){$a^{4}$}
\put(205,-12){$b^{4}$}
\put(130,0){\circle*{2}}	
\put(210,0){\circle*{2}}	
\put(130,0){\oval(20,20)[t]}
\put(210,0){\oval(20,20)[t]}
\put(140,0){\line(1,0){60}}
\put(220,14){\vector(-1,0){50}}
\put(170,14){\line(-1,0){50}}
\put(120,0){\line(0,1){14}}
\put(220,0){\line(0,1){14}}
\end{picture}
\caption{\mbox{}}
\end{figure}
Here $a^{i}$ and $b^{i}$ are the left and right endpoints, respectively,
of the arc $\bar{{\mathcal I}}^{(i)}$.  The radius of each semicircular 
arc is $\alpha$ and each of the vertical segments of the closed contours
has length $\sqrt{2}\alpha$.  One ultimately takes the limit as $\alpha
\rightarrow 0$.

From a well known theorem\footnote{See Cor.~27.7 in Ref.~\ref{McShane}.}
on Lebesgue integrals,
\begin{equation}
\int_{a^{i}+\alpha}^{b^{i}-\alpha} d\sigma'
\frac{[\nu^{\pm}(\sigma')]^{-1}\F^{\pm}(\sigma')}{\sigma'-\tau} \rightarrow
\int_{a^{i}}^{b^{i}} d\sigma'
\frac{[\nu^{\pm}(\sigma')]^{-1}\F^{\pm}(\sigma')}{\sigma'-\tau}
\text{ as } \alpha \rightarrow 0.
\label{5.17d}
\end{equation}
Upon applying the above statement (\ref{5.17d}) and the easily proved 
statement that the integral on each semicircular arc $\rightarrow 0$ as
$\alpha \rightarrow 0$, and using the fact that
$\nu^{-}(\sigma') = - \nu^{+}(\sigma')$ for all $\sigma' \in \I(\x)$,
one obtains the conclusion (\ref{5.13c}).
\cheers

\item 
To obtain statement (\ref{5.14a}) and Eq.\ (\ref{5.14c}) when 
$\sigma \in \I^{(3)}(\x)$, we again 
employ the Cauchy integral formula and the generalized Cauchy integral 
theorem, this time using (for the integral over $\Lambda^{+}$) the positively
oriented closed contours depicted in Fig.~3.  The case $\sigma \in \I^{(4)}(\x)$
is treated similarly.

\begin{figure}[htbp] 
\begin{picture}(300,30)(-130,-10)
\put(5,-12){$a^{3}$}
\put(85,-12){$b^{3}$}
\put(10,0){\circle*{2}}		
\put(90,0){\circle*{2}} 	
\put(10,0){\oval(20,20)[t]}
\put(90,0){\oval(20,20)[t]}
\put(20,0){\line(1,0){15}}
\put(45,-12){$\sigma$}
\put(50,0){\circle*{2}}		
\put(50,0){\oval(30,30)[t]}
\put(65,0){\line(1,0){15}}
\put(100,20){\vector(-1,0){50}}
\put(50,20){\line(-1,0){50}}
\put(0,0){\line(0,1){20}}
\put(100,0){\line(0,1){20}}
\put(125,-12){$a^{4}$}
\put(205,-12){$b^{4}$}
\put(130,0){\circle*{2}}	
\put(210,0){\circle*{2}}	
\put(130,0){\oval(20,20)[t]}
\put(210,0){\oval(20,20)[t]}
\put(140,0){\line(1,0){60}}
\put(220,14){\vector(-1,0){50}}
\put(170,14){\line(-1,0){50}}
\put(120,0){\line(0,1){14}}
\put(220,0){\line(0,1){14}}
\end{picture}
\caption{\mbox{}}
\end{figure}
Here the radius of the semicircular arc about $\sigma$ is $\beta$
and each of the vertical segments of the left closed contour has
length $\sqrt{2}\beta$.  The radius of each of the other semicircular 
arcs is $\alpha$, and each of the vertical segments of the right 
closed contour has length $\sqrt{2}\alpha$.  One ultimately takes the 
limit as $\alpha \rightarrow 0$ followed by the limit as $\beta 
\rightarrow 0$.  It is clear that the integral on the semicircular
arc with center $\sigma$ has the limit $\frac{1}{2} \nu^{+}(\sigma)
\F^{+}(\sigma)$ as $\beta \rightarrow 0$.  
\cheers
\end{romanlist}

\setcounter{equation}{0}
\subsection{Derivation of an Alekseev-type singular integral equation}

Proceeding from equations (\ref{5.13c}) and (\ref{5.14c}), one can construct
a singular integral equation of the Alekseev type and, if $\bv \in K^{1+}$,
a Fredholm equation of the second kind. 

We begin by observing that Eq.\ (\ref{G2.10c}) implies that, for each
$\sigma \in \I(\x) \cup \{r_{0},s_{0}\}$, 
\begin{eqnarray}
\frac{1}{2} \left\{ {\F}^{M+}(\sigma) + {\F}^{M-}(\sigma) \right\} & = &
\left( \begin{array}{cc}
1 & -i(\sigma-z) \\ 0 & 1
\end{array} \right) \left( \begin{array}{cc}
1 & 0 \\ 0 & 0
\end{array} \right) \left( \begin{array}{cc}
1 & i(\sigma-z_{0}) \\ 0 & 1
\end{array} \right),
\end{eqnarray}
and
\begin{eqnarray}
\frac{1}{2} [\nu^{+}(\sigma)]^{-1}
\lefteqn{
\left\{ {\F}^{M+}(\sigma) - {\F}^{M-}(\sigma) \right\} = } 
\nonumber \\ & &
\left( \begin{array}{cc}
1 & -i(\sigma-z) \\ 0 & 1
\end{array} \right) \left( \begin{array}{cc}
0 & 0 \\ 0 & 1
\end{array} \right) \left( \begin{array}{cc}
1 & i(\sigma-z_{0}) \\ 0 & 1
\end{array} \right).
\end{eqnarray}
If $\F$ is a solution of the HHP corresponding to $(\bv,\F^{M})$,
Eq.\ (\ref{G3.17}) tells us that, for any $\sigma \in \I(\x)$,
\begin{equation}
{\F}^{\pm}(\sigma) v^{(i)}(\sigma) = Y^{(i)}(\sigma) {\F}^{M\pm}(\sigma),
\end{equation}
and, therefore,
\begin{eqnarray}
\lefteqn{\frac{1}{2} \left\{ {\F}^{+}(\sigma) 
+ {\F}^{-}(\sigma) \right\} v^{(i)}(\sigma) = } \nonumber \\
& & Y^{(i)}(\sigma) \left( \begin{array}{cc}
1 & -i(\sigma-z) \\ 0 & 1
\end{array} \right) \left( \begin{array}{cc}
1 & 0 \\ 0 & 0
\end{array} \right) \left( \begin{array}{cc}
1 & i(\sigma-z_{0}) \\ 0 & 1
\end{array} \right), 
\end{eqnarray}
and
\begin{eqnarray}
\lefteqn{
\frac{1}{2} [\nu^{+}(\sigma)]^{-1}
\left\{ {\F}^{+}(\sigma) - {\F}^{-}(\sigma) \right\} v^{(i)}(\sigma) = }
\nonumber \\ & &
Y^{(i)}(\sigma) \left( \begin{array}{cc}
1 & -i(\sigma-z) \\ 0 & 1
\end{array} \right) \left( \begin{array}{cc}
0 & 0 \\ 0 & 1
\end{array} \right) \left( \begin{array}{cc}
1 & i(\sigma-z_{0}) \\ 0 & 1
\end{array} \right).
\end{eqnarray}
This motivates the introduction of two new $2 \times 2$ matrices.

\begin{definition}{Dfn.\ of functions $W^{(i)}(\x)$ and $\Y^{(i)}(\x)$}
For each $\bv \in K$, we let $W^{(i)}(\x)$ denote the function whose domain
is $\I^{(i)}$ and whose value for each $\sigma \in \I^{(i)}$ is 
\begin{eqnarray}
W^{(i)}(\x,\sigma) & := & W^{(i)}(\x)(\sigma) := 
v^{(i)}(\sigma) \left( \begin{array}{cc}
1 & -i(\sigma-z_{0}) \\ 0 & 1
\end{array} \right),
\label{5.22a}
\end{eqnarray}
and, for each solution $\F(\x)$ of the HHP corresponding to
$(\bv,\F^{M},\x)$, we let $\Y^{(i)}(\x)$ denote the function whose
domain is $\I^{(i)}(\x)$ and whose value for each $\sigma \in \I^{(i)}(\x)$
is
\begin{eqnarray}
\Y^{(i)}(\x,\sigma) & := & \Y^{(i)}(\x)(\sigma) := Y^{(i)}(\x,\sigma) 
\left( \begin{array}{cc}
1 & -i(\sigma-z) \\ 0 & 1
\end{array} \right). 
\label{5.22b}
\end{eqnarray}
\end{definition}

In terms of these matrices we may write [suppressing `$\x$']
\begin{equation}
{\F}^{\pm}(\sigma) W^{(i)}(\sigma) = \Y^{(i)}(\sigma) \left( \begin{array}{cc}
1 & 0 \\ 0 & \nu^{\pm}(\sigma)
\end{array} \right) 
\label{G4.8}
\end{equation}
as well as
\begin{eqnarray}
\frac{1}{2} \left\{ {\F}^{+}(\sigma) + {\F}^{-}(\sigma) \right\} W^{(i)}(\sigma) 
& = & \Y^{(i)}(\sigma) \left( \begin{array}{cc}
1 & 0 \\ 0 & 0 
\end{array} \right), 
\label{5.25a}
\end{eqnarray}
and
\begin{eqnarray}
\frac{1}{2} [\nu^{+}(\sigma)]^{-1}
\left\{ {\F}^{+}(\sigma) - {\F}^{-}(\sigma) \right\} W^{(i)}(\sigma) & = &
\Y^{(i)}(\sigma) \left( \begin{array}{cc}
0 & 0 \\ 0 & 1
\end{array} \right). 
\label{5.25b}
\end{eqnarray}

\begin{definition}{Dfns.\ of $W(\x)$, $\Y(\x)$, $W_{a}(\x)$ and $\Y_{a}(\x)$}
Let $W(\x)$ and $\Y(\x)$ denote the functions\footnote{We shall frequently
suppress $\x$.} with domain $\I(\x)$ and values
\begin{equation}
\begin{array}{l}
W(\x,\sigma) := W(\x)(\sigma) := W^{(i)}(\x,\sigma) \text{ and } \\
\Y(\x,\sigma) := \Y(\x)(\sigma) := \Y^{(i)}(\x,\sigma) \\
\text{for each $i \in \{3,4\}$ and $\sigma \in \I^{(i)}(\x)$.}
\end{array}
\label{5.22c}
\end{equation}
Moreover, let
\begin{equation}
\begin{array}{l}
W_{a}(\x,\sigma) := a^{th} \text{ column of } W(\x,\sigma) \text{ and } \\ 
\Y_{a}(\x,\sigma) := a^{th} \text{ column of } \Y(\x,\sigma), \text{ where }
a \in \{1,2\}.
\end{array}
\end{equation}
\end{definition}

\begin{theorem}[Alekseev-type equation]
\label{5.1B} \mbox{ } \\
For each $\bv \in K$, $\x \in D$, solution $\F(\x)$ of the HHP
corresponding to $(\bv,\F^{M},\x)$, $\tau \in C - \bar{\I}(\x)$
and $\sigma \in \I(\x)$, the following statement holds:
\begin{equation}
[\nu^{+}(\sigma')]^{-1} \Y_{1}(\sigma') W_{2}^{T}(\sigma') (\sigma'-\tau)^{-1} 
\text{ is summable over } \sigma' \in \bar{\I}(\x),
\label{5.24a}
\end{equation}
\begin{equation}
\nu(\tau)^{-1}\F(\tau) = I + \frac{1}{\pi i} \int_{\bar{\I}} d\sigma'
[\nu^{+}(\sigma')]^{-1} \Y_{1}(\sigma') 
\frac{W_{2}^{T}(\sigma') J}{\sigma'-\tau},
\label{5.24b}
\end{equation}
\begin{equation}
\begin{array}{r}
[\nu^{+}(\sigma')]^{-1} \Y_{1}(\sigma') W_{2}^{T}(\sigma') (\sigma'-\sigma)^{-1}
\text{ is summable over } \sigma' \in \bar{\I}(\x) \\
\text{in the PV sense,}
\end{array}
\label{5.24c}
\end{equation}
\begin{eqnarray}
\Y_{2}(\sigma) & = & W_{2}(\sigma) - \frac{1}{\pi i} \int_{\bar{\I}}
d\sigma' [\nu^{+}(\sigma')]^{-1} \Y_{1}(\sigma') 
\frac{W_{2}^{T}(\sigma') J W_{2}(\sigma)}{\sigma'-\sigma},
\label{5.24d} \\
\text{and} & & \nonumber \\
0 & = & W_{1}(\sigma) + \frac{1}{\pi i} \int_{\bar{\I}} d\sigma' 
[\nu^{+}(\sigma')]^{-1} \Y_{1}(\sigma') 
\frac{W_{2}^{T}(\sigma') J W_{1}(\sigma)}{\sigma'-\sigma}.
\label{5.24e}
\end{eqnarray}
\end{theorem}
Here we have employed the symbol $J := -i\Omega = \left( \begin{array}{cc}
0 & 1 \\ -1 & 0
\end{array} \right)$.

\proof
The statements (\ref{5.24a}) to (\ref{5.24c}) are obtained by using Eqs.\ 
(\ref{5.25a}) and (\ref{5.25b}) together with the relation
$$
W(\sigma)^{-1} = - J W^{T}(\sigma) J
$$
to replace $\F^{+}(\sigma') - \F^{-}(\sigma')$ and $[\nu^{+}(\sigma')]^{-1}
[\F^{+}(\sigma') + \F^{-}(\sigma)]$ in statements (\ref{5.13a}) to
(\ref{5.14a}) of Thm.~\ref{5.4A}.  The same replacements are to be made in
the integrands on the right side of Eq.\ (\ref{5.14c}) in Thm.~\ref{5.4A}.  
Equation (\ref{5.24d}) is obtained by multiplying both sides of Eq.\
(\ref{5.14c}) by $W_{2}(\sigma)$ and replacing the product on the left
side with the second column of (\ref{5.25b}) multiplied by
$[\nu^{+}(\sigma)]^{-1}$.  Equation (\ref{5.24e}) is obtained by
multiplying both sides of Eq.\ (\ref{5.14c}) by $W_{1}(\sigma)$ and
replacing the product on the left side with the first column of (\ref{5.25b}).
\cheers

Equation (\ref{5.24e}) has the form of the singular integral equation
which Alekseev obtained in the analytic case.

\setcounter{equation}{0}
\subsection{Extension of the function $\Y(\x)$ from $\I(\x)$ to $\bar{\I}(\x)$}

Since $C^{T} J C = 0$ (the zero matrix) for any $2 \times 1$ matrix $C$,
Eq.\  (\ref{5.24d}) is expressible in the following 
form for each $i \in \{3,4\}$:
\begin{eqnarray}
\lefteqn{\Y_{2}^{(i)}(\sigma) = W_{2}^{(i)}(\sigma)} \nonumber \\
& & \mbox{ } - \frac{1}{\pi i} \int_{a^{i}}^{b^{i}} d\sigma'
[\nu^{+}(\sigma')]^{-1} \Y_{1}^{(i)}(\sigma') W_{2}^{(i)}(\sigma')^{T} J
\left[ \frac{W_{2}^{(i)}(\sigma) - W_{2}^{(i)}(\sigma')}{\sigma'-\sigma}
\right] \nonumber \\
& & \mbox{ } - \frac{1}{\pi i} \int_{a^{7-i}}^{b^{7-i}} d\sigma'
[\nu^{+}(\sigma')]^{-1} \Y_{1}^{(7-i)}(\sigma') W_{2}^{(7-i)}(\sigma')^{T} J
\left[ \frac{W_{2}^{(i)}(\sigma)}{\sigma'-\sigma} \right],
\label{5.30b}
\end{eqnarray}
for all $\sigma \in \I^{(i)}(\x)$, where recall that $a^{i} :=
\inf{\{x^{i},x^{i}_{0}\}}$ and $b^{i} := \sup{\{x^{i},x^{i}_{0}\}}$.
Without indicating the parallel proof, we simply remark that one can also
show that
\begin{eqnarray}
\lefteqn{\Y_{1}^{(i)}(\sigma) = W_{1}^{(i)}(\sigma)} \nonumber \\
& & \mbox{ } + \frac{1}{\pi i} \int_{a^{i}}^{b^{i}} d\sigma'
\nu^{+}(\sigma') \Y_{2}^{(i)}(\sigma') W_{1}^{(i)}(\sigma')^{T} J
\left[ \frac{W_{1}^{(i)}(\sigma) - W_{1}^{(i)}(\sigma')}{\sigma'-\sigma}
\right] \nonumber \\
& & \mbox{ } + \frac{1}{\pi i} \int_{a^{7-i}}^{b^{7-i}} d\sigma'
\nu^{+}(\sigma') \Y_{2}^{(7-i)}(\sigma') W_{1}^{(7-i)}(\sigma')^{T} J
\left[ \frac{W_{1}^{(i)}(\sigma)}{\sigma'-\sigma} \right],
\label{5.30a}
\end{eqnarray}

Now, from Thms.~\ref{4.3D}(i) and~(ii), Eq.\ (\ref{5.25a}) and Eq.\ (\ref{5.25b}),
\begin{equation}
\begin{array}{l}
\nu^{+}(\sigma') \Y_{2}(\sigma') W_{1}(\sigma')^{T} J \\
\text{and } [\nu^{+}(\sigma')]^{-1} \Y_{1}(\sigma') W_{2}(\sigma')^{T} J \\
\text{are summable over } \sigma' \in \bar{\I}(\x).
\end{array}
\label{5.30c}
\end{equation}
From the definition of $W^{(i)}$ by Eq.\ (\ref{5.22a}) and the definition
of $K^{\Box}$,
the following statement holds for each $\x \in D$ and $i \in \{3,4\}$:
\begin{equation}
\begin{array}{l}
\text{If $\bv \in K^{1}$, then $W^{(i)}$ is $\bC^{1}$ throughout $\I^{(i)}$,}
\\[0pt]
[W^{(i)}(\sigma') - W^{(i)}(\sigma)](\sigma'-\sigma)^{-1}
\text{ is a continuous } \\
\text{function of } (\sigma',\sigma) \text{ throughout } \I^{(i)} \times \I^{(i)},
\text{ and} \\
W^{(i)}(\sigma)(\sigma'-\sigma)^{-1} \text{ is a $\bC^{1}$ function of }
(\sigma',\sigma) \\
\text{throughout } \bar{\I}^{(7-i)}(\x) \times \check{\I}^{(i)}(x^{7-i}),
\end{array}
\label{5.30d}
\end{equation}
where
\begin{equation}
\begin{array}{rcl}
\check{\I}^{(3)}(s) & := & \{\sigma \in \I^{(3)}:\sigma < s\}, \text{ and } \\
\check{\I}^{(4)}(r) & := & \{\sigma \in \I^{(4)}:r < \sigma\}. 
\end{array}
\end{equation}
Note that (See Fig.~4)
\begin{equation}
\I^{(i)}(\x) \subset \check{\I}^{(i)}(x^{7-i}) \subset \I^{(i)}.
\end{equation}
From the above statements (\ref{5.30c}) and (\ref{5.30d}), and from the 
theorem that asserts the summability over a finite interval of the product
of a summable function by a continuous function, the extension of
$\Y^{(i)}(\x)$ that we shall define below exists.  Note that
$\Box$ is $n \ge 1$, $n+$ (with $n \ge 1$), $\infty$ or `an'.  
\vspace{3ex}

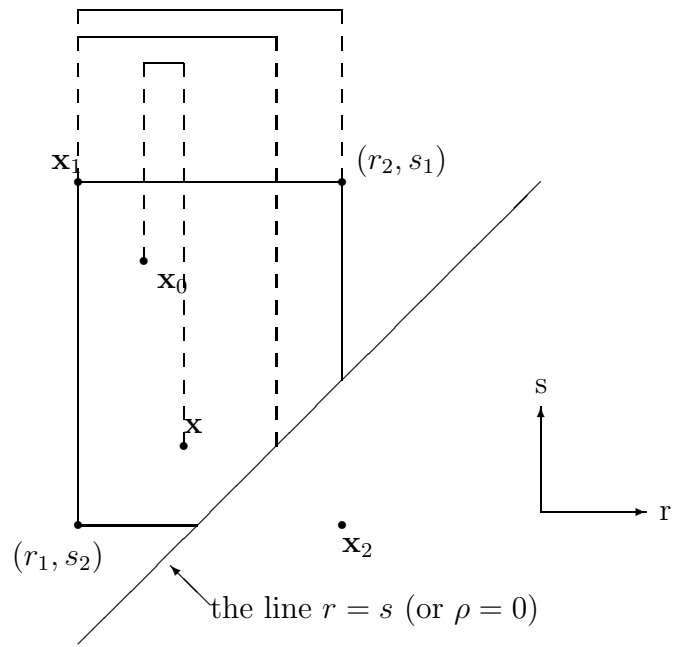
\begin{figure}[htbp] 
\begin{picture}(300,300)(-60,60)
\put(75,75){\line(1,1){175}}
\put(125,85){the line $r=s$ (or $\rho=0$)}
\put(125,90){\vector(-1,1){15}}
\put(250,125){\vector(0,1){40}}
\put(250,125){\vector(1,0){40}}
\put(248,170){s}
\put(295,122){r}
\put(75,250){\line(1,0){100}}
\put(75,120){\line(1,0){45}}
\put(75,120){\line(0,1){130}}
\multiput(75,250)(0,10){7}{\line(0,1){5}}
\put(175,175){\line(0,1){75}}
\multiput(175,250)(0,10){7}{\line(0,1){5}}
\put(75,250){\circle*{3}}
\put(75,120){\circle*{3}}
\put(175,250){\circle*{3}}
\put(175,120){\circle*{3}}
\put(65,255){${\mathbf x}_{1}$}
\put(175,110){${\mathbf x}_{2}$}
\put(50,105){$(r_{1},s_{2})$}
\put(180,255){$(r_{2},s_{1})$}
\put(115,150){\circle*{3}}
\put(115,155){${\mathbf x}$}
\multiput(115,150)(0,10){15}{\line(0,1){5}}
\put(100,220){\circle*{3}}
\put(105,210){${\mathbf x}_{0}$}
\multiput(100,220)(0,10){8}{\line(0,1){5}}
\multiput(150,150)(0,10){16}{\line(0,1){5}}
\put(75,315){\line(1,0){100}}
\put(75,305){\line(1,0){75}}
\put(100,295){\line(1,0){15}}
\end{picture}
\caption{Illustrating the relation
$\I^{(3)}(\x) \subset \check{\I}^{(3)}(x^{4}) \subset \I^{(3)}$.
In this example, $\check{\I}^{(4)}(x^{3}) = \I^{(4)}$.}
\end{figure}

\begin{definition}{Dfn.\ of an extension of $\Y^{(i)}(\x)$ when 
$\bv \in K^{\Box}$}
For each $\bv \in K^{\Box}$, $\x \in D$, solution $\F(\x)$ of the
HHP corresponding to $(\bv,\F^{M},\x)$ and $i \in \{3,4\}$, let
$\Y^{(i)}(\x)$ denote the function whose extended domain is
$\check{\I}(x^{7-i})$ and whose value for each $\sigma \in
\check{\I}^{(i)}(\x^{7-i})$ is given by [suppressing `$\x$']
\begin{eqnarray}
\Y_{1}^{(i)}(\sigma) & := & \text{ right side of Eq.\ (\ref{5.30a}),}
\label{5.31a} \\
\Y_{2}^{(i)}(\sigma) & := & \text{ right side of Eq.\ (\ref{5.30b}).}
\label{5.31b}
\end{eqnarray}
\end{definition}

\begin{lemma}[Continuity and differentiability of $W^{(i)}$]
\label{5.1C}
\mbox{ } \\ \vspace{-3ex}
\begin{romanlist}
\item 
If $\bv \in K^{\Box}$, then $W^{(i)}$ is $\bC^{\Box}$ throughout its domain
$\I^{(i)}$, and the function whose domain is $\bar{\I}^{(7-i)}(\x) \times
\check{\I}^{(i)}(x^{7-i})$ and whose values for each $(\sigma',\sigma)$ in
this domain is $W^{(i)}(\sigma)(\sigma'-\sigma)^{-1}$ is also $\bC^{\Box}$.
\item 
If $\bv \in K^{\Box}$, then the function of $(\sigma',\sigma)$ whose domain
is $\I^{(i)} \times \I^{(i)}$ and whose value for each $(\sigma',\sigma)$ in
this domain is $[W^{(i)}(\sigma)-W^{(i)}(\sigma')]/(\sigma'-\sigma)$ is
$\bC^{n-1}$ if $\Box$ is $n \ge 1$, is $\bC^{(n-1)+}$ if $\Box$ is 
$n^{+}$ ($n \ge 1$), is $\bC^{\infty}$ if $\Box$ is $\infty$, and is
$\bC^{an}$ if $\Box$ is `an'.
\end{romanlist}
\end{lemma}

\proofs
\begin{romanlist}
\item 
The conclusion follows by using the definition of $W^{(i)}$ by Eq.\ 
(\ref{5.22a}) together with the definition of $K^{\Box}$.
\cheers
\item 
The conclusions when $\Box$ is $n$, $\infty$ or `an' are well known.
As regards the case when $\Box$ is $n^{+}$ ($n \ge 1$), one can construct
a simple proof (which we shall not reproduce here) using the relation
\begin{equation}
\frac{W^{(i)}(\sigma) - W^{(i)}(\sigma')}{\sigma-\sigma'} =
\int_{0}^{1} dt (DW^{(i)})(t\sigma+(1-t)\sigma'),
\label{5.32a}
\end{equation}
where $D^{p}W^{(i)}$ ($1 \le p \le n$) denotes the function whose domain
is $\I^{(i)}$ and whose value for each $\sigma \in \I^{(i)}$ is
\begin{equation}
(D^{p}W^{(i)})(\sigma) :=
\frac{\partial^{p}W^{(i)}(\sigma)}{\partial\sigma^{p}};
\end{equation}
and $DW^{(i)} := D^{1}W^{(i)}$.  
\cheers
\end{romanlist}

We shall leave the proof of the following basic lemma to the reader.
\begin{lemma}[Integral of product]
\label{5.2C} \mbox{ } \\
Suppose $[a,b] \subset R^{1}$, $S$ is a connected open subset of
$R^{m}$ ($m \ge 1$), $f$ is a real-valued function defined almost
everywhere on and summable over $[a,b]$, and $g$ is a real-valued
function whose domain is $[a,b] \times S$ and which is continuous.
Let $\sigma := (\sigma^{1},\ldots,\sigma^{m})$ denote any member
of $S$, and let $F$ denote the function whose domain is $S$ and
whose value at each $\sigma \in S$ is
\begin{equation}
F(\sigma) := \int_{a}^{b} d\sigma' f(\sigma') g(\sigma',\sigma).
\label{5.33a}
\end{equation}
Then the following statements hold:
\begin{romanlist}
\item 
$F$ is continuous.
\item 
If $\partial g(\sigma',\sigma)/\partial\sigma^{k}$ exists for each
$k \in \{1,\ldots,m\}$ and is a continuous function of $(\sigma',\sigma)$
throughout $[a,b] \times S$, then $F$ is $\bC^{1}$ and
\begin{equation}
\frac{\partial F(\sigma)}{\partial\sigma^{k}} = \int_{a}^{b} d\sigma'
f(\sigma') \frac{\partial g(\sigma',\sigma)}{\partial\sigma^{k}}
\label{5.33b}
\end{equation}
for all $k \in \{1,\ldots,m\}$ and $\sigma \in S$.  In particular,
$F$ is $\bC^{1}$ if $g$ is $\bC^{1}$.
\item 
Assume (for simplicity) that $m=1$.  
If there exists a positive integer $p$ such that 
$\partial^{p}g(\sigma',\sigma)/(\partial\sigma)^{p}$ exists and is a
continuous function of $(\sigma',\sigma)$ throughout $[a,b] \times S$,
then $F$ is $bC^{p}$ and
\begin{equation}
\frac{\partial^{p}F(\sigma)}{\partial\sigma^{p}} = \int_{a}^{b} d\sigma'
f(\sigma') \frac{\partial^{p}g(\sigma',\sigma)}{\partial\sigma^{p}}
\end{equation}
for all $\sigma \in S$.  In particular, $F$ is $\bC^{p}$ if $g$ is
$\bC^{p}$.
\item 
Assume (for simplicity) that $m=1$.  If $[c,d] \subset S$ and $g$ obeys
a H\"{o}lder condition on $[a,b] \times [c,d]$, then $F$ obeys a H\"{o}lder
condition on $[c,d]$.
\item 
Assume (for simplicity) that $m=1$.  If $g$ is analytic [i.e., if $g$ has
an analytic extension to an open subset $[a-\epsilon,b+\epsilon] \times S$
of $R^{2}$, then $F$ is analytic.
\end{romanlist}
\end{lemma}

\begin{lemma}[Generalization of Lem.~\ref{5.2C}]
\label{5.3C} \mbox{ } \\
All of the conclusions of the preceding lemma remain valid when the only
alteration in the premises is to replace the statement that $f$ and $g$
are real valued by the statement that they are complex valued or are
finite matrices (such that the product $fg$ exists) with complex-valued
elements.
\end{lemma}

\proof
Use the definition
$$
\int_{a}^{b} d\sigma' h(\sigma') := \int_{a}^{b} d\sigma' \Re{h(\sigma')}
+ i \int_{a}^{b} d\sigma' \Im{h(\sigma')}
$$
for any complex-valued function $h$ whose real and imaginary parts are
summable over $[a,b]$.  The rest is obvious.
\cheers

\begin{theorem}[Continuity and differentiability of extended $\Y^{(i)}(\x)$]
\label{5.4C} \mbox{ } \\
For each $\bv \in K^{\Box}$, $\x \in D$, solution $\F(\x)$ of
the HHP corresponding to $(\bv,\F^{M},\x)$ and $i \in \{3,4\}$,
$\Y^{(i)}(\x)$ [see Eqs.\ (\ref{5.31a}) and (\ref{5.31b})] is
$\bC^{n-1}$ if $\Box$ is $n$, is $\bC^{(n-1)+}$ if $\Box$ is $n+$,
is $\bC^{\infty}$ if $\Box$ is $\infty$ and is $\bC^{an}$ if $\Box$
is `an'.
\end{theorem}

\proof
Apply Lemmas~\ref{5.1C}, \ref{5.2C} and~\ref{5.3C} to the definitions
(\ref{5.31a}) and (\ref{5.31b}) of $\Y_{1}^{(i)}(\x)$ and
$\Y_{2}^{(i)}(\x)$.  It is then easily shown that the second
term on the right side of each of the Eqs.\ (\ref{5.30a}) and (\ref{5.30b})
[with $\sigma \in \check{\I}^{(i)}(x^{7-i})$] is $\bC^{n-1}$ if $\Box$
is $n$, is $\bC^{(n-1)+}$ if $\Box$ is $n+$, is $\bC^{\infty}$ if
$\Box$ is $\infty$ and is $\bC^{an}$ if $\Box$ is `an'.  The first and
third terms on the right sides of each of the Eqs.\ (\ref{5.30a}) and
(\ref{5.30b}) are, on the other hand, both $\bC^{\Box}$ even when $\Box$
is $n$ or is $n+$.  However, a $\bC^{n}$ function is also a $\bC^{n-1}$
function; and a $\bC^{n+}$ function is also a $\bC^{(n-1)+}$ function.
\cheers

\begin{definition}{Dfns.\ of $\Y^{(i)}$, $\Y$ and the partial derivatives
of $\Y$}
Henceforth, $\Y^{(i)} \, (i \in \{3,4\})$ will denote the function whose
domain is
\begin{equation}
\dom{\Y^{(i)}} := \{(\x,\sigma):\x \in D, \sigma \in \check{\I}(x^{7-i})\}
\end{equation}
and whose values are given by
\begin{equation}
\Y^{(i)}(\x,\sigma) := \Y^{(i)}(\x)(\sigma),
\end{equation}
where $\Y^{(i)}(\x)$ is the extension of the original $\Y^{(i)}(\x)$ that
is defined by Eqs.\ (\ref{5.31a}) and (\ref{5.31b}).  We shall let $\Y$
denote the function whose domain is
$$
\dom{\Y} := \{(\x,\sigma):\x \in D, \sigma \in \bar{\I}(\x)\}
$$
and whose values are given by
$$
\Y(\x,\sigma) := \Y^{(i)}(\x,\sigma) \text{ whenever } \sigma \in
\bar{\I}^{(i)}(\x).
$$
[Thus, $\Y(\x,\sigma) = \Y(\x)(\sigma)$.]  Also, for each $\x \in D$,
$i \in \{3,4\}$ and $\sigma \in \bar{\I}^{(i)}(\x)$, we shall let
$$
\frac{\partial^{l+m+n}\Y(\x,\sigma)}{\partial r^{l} \partial s^{m} 
\partial \sigma^{n}} :=
\frac{\partial^{l+m+n}\Y^{(i)}(\x,\sigma)}{\partial r^{l} \partial s^{m} 
\partial \sigma^{n}},
$$
if the above partial derivative of $\Y^{(i)}$ exists.
\end{definition}

The domain of $\Y^{(i)}$, as defined above, is an open subset of $R^{3}$;
and (though the domain of $\Y$ is not an open subset of $R^{3}$) the 
partial derivatives of $\Y$ are defined in terms of partial derivatives
of $\Y^{(i)}$ and, therefore, employ sequences of points in $R^{3}$ which
may converge to a given point along any direction in $R^{3}$.  This has 
formal advantages when one employs the derivatives of $\Y$ at the boundary
of its domain.

\begin{corollary}[The extension $\Y(\x)$ when $\bv \in K^{\Box}$]
\label{5.5C} \mbox \\
\begin{romanlist}
\item 
Suppose $\bv \in K^{1}$, $\x \in D$ and the solution $\F(\x)$
of the HHP corresponding to $(\bv,\F^{M},\x)$ exists.  Then $\Y(\x)$
has a unique continuous extension to $\bar{\I}(\x)$.  
\item 
If $\bv \in K^{\Box}$, then the extension $\Y(\x)$ is $\bC^{n-1}$ if
$\Box$ is $n$, is $\bC^{(n-1)+}$ if $\Box$ is $n+$, is $\bC^{\infty}$
if $\Box$ is $\infty$ and is $\bC^{an}$ if $\Box$ is `an'.
\end{romanlist}
\end{corollary}

\proof
Statement (ii) of this corollary follows from Thm.~\ref{5.4C}.  The 
uniqueness follows, of course, from the fact that a function defined
and continuous on an open subset of $R^{1}$ has no more than one 
continuous extension to the closure of that subset.
\cheers

\setcounter{equation}{0}
\subsection{Equivalence of the HHP to an Alekseev-type equation when 
$\bv \in K^{1+}$}

\begin{theorem}[HHP-Alekseev equivalence theorem]
\label{5.1D} \mbox{ } \\
Suppose $\bv \in K^{1+}$ and $\x \in D$, and suppose that $\F(\x)$
and $\Y_{1}(\x)$ are $2 \times 2$ and $2 \times 1$ matrix functions,
respectively, such that
\begin{equation}
\dom{\F(\x)} = C - \bar{\I}(\x), \quad \dom{\Y_{1}(\x)} = \bar{\I}(\x)
\text{ and } \Y_{1}(\x) \text{ is } \bC^{0+}.
\label{5.39}
\end{equation}
Then the following two statements are equivalent to one another:
\begin{romanlist}
\item 
The function $\F(\x)$ is a solution of the HHP corresponding to
$(\bv,\F^{M},\x)$, and $\Y_{1}(\x)$ is the function whose 
restriction to $\I(\x)$ is defined in terms of $\F^{+}(\x) + \F^{-}(\x)$
by Eq.\ (\ref{5.25a}) [where $\x$ is suppressed] and whose existence and
uniqueness [for the given $\F(\x)$] is asserted by Cor.~\ref{5.5C}
when $\Box$ is $1+$.
\item 
The restriction of $\Y_{1}(\x)$ to $\I(\x)$ is a solution of the singular
integral equation (\ref{5.24e}), and $\F(\x)$ is defined in terms
of $\Y_{1}(\x)$ by Eq.\ (\ref{5.24b}) [where $\x$ is suppressed].
\end{romanlist}
\end{theorem}

\proof
That (i) implies (ii) has already been proved. [See Thm.~\ref{5.1B} and
Cor.~\ref{5.5C}.]  The proof that (ii) implies (i) will be given in four
parts:
\begin{arablist}
\item 
Assume that statement (ii) is true.  From the definition of $\F(\x)$
by Eq.\ (\ref{5.24b}),
\begin{equation}
\F(\x) \text{ is holomorphic;} 
\label{5.40a}
\end{equation}
and, from two theorems of Plemelj\footnote{See Secs.~16 and~17 of Ch.~II
of Ref.~\ref{Musk} (pp.\ 37-43).} 
\begin{equation}
\F^{+}(\x) \text{ and } \F^{-}(\x) \text{ exist }
\label{5.40b}
\end{equation}
and, since $\nu^{-}(\sigma) = -\nu^{+}(\sigma)$ for all $\sigma \in \I(\x)$,
\begin{equation}
\frac{1}{2}\left[\F^{+}(\sigma)+\F^{-}(\sigma)\right] =
- \Y_{1}(\sigma) W_{2}^{T}(\sigma) J
\label{5.41a}
\end{equation}
and
\begin{equation}
\frac{1}{2}[\nu^{+}(\sigma)]^{-1}\left[\F^{+}(\sigma)-\F^{-}(\sigma)\right] =
I - \frac{1}{\pi i} \int_{\bar{\I}} d\sigma' [\nu^{+}(\sigma')]^{-1}
\Y_{1}(\sigma') \frac{W_{2}^{T}(\sigma') J}{\sigma'-\sigma}
\label{5.41b}
\end{equation}
for all $\sigma \in \I(\x)$.  Upon multiplying Eqs.\ (\ref{5.41a}) and
(\ref{5.41b}) through by $W_{2}(\sigma)$ on the right, one obtains,
for all $\sigma \in \I(\x)$,
\begin{eqnarray}
\frac{1}{2}\left[\F^{+}(\sigma)+\F^{-}(\sigma)\right] W_{2}(\sigma) & = & \left(
\begin{array}{c}
0 \\ 0
\end{array} \right),
\label{5.42a} \\
\text{and} & & \nonumber \\
\frac{1}{2}[\nu^{+}(\sigma)]^{-1} \left[\F^{+}(\sigma)-\F^{-}(\sigma)\right]
W_{2}(\sigma) & = & \Y_{2}(\sigma),
\label{5.42b}
\end{eqnarray}
where $\Y_{2}(\x)$ has the domain $\bar{\I}(\x)$ and the values
\begin{eqnarray}
\lefteqn{\Y_{2}(\sigma) := W_{2}(\sigma) } \nonumber \\
& & \mbox{ }-\frac{1}{\pi I} \int_{\bar{\I}} d\sigma' [\nu^{+}(\sigma')]^{-1}
\Y_{1}(\sigma') \frac{W_{2}^{T}(\sigma') J [W_{2}(\sigma)-W_{2}(\sigma')]}
{\sigma'-\sigma}
\nonumber \\
& & \text{for all } \sigma \in \bar{\I}(\x).
\end{eqnarray}
From Lemmas~\ref{5.1C}(ii), \ref{5.2C}(iv) and \ref{5.3C},
\begin{equation}
\Y_{2}(\x) \text{ is } \bC^{0+}.
\label{5.43}
\end{equation}
Upon multiplying Eqs.\ (\ref{5.41a}) and (\ref{5.41b}) through by
$W_{1}(\sigma)$ on the right, upon using the fact that $\det{W(\sigma)}=1$ 
is equivalent to the equation
\begin{equation}
W_{2}^{T}(\sigma) J W_{1}(\sigma) = - (1), 
\end{equation}
and, upon using Eq.\ (\ref{5.24e}), one obtains, for all $\sigma \in \I(\x)$,
\begin{eqnarray}
\frac{1}{2}\left[\F^{+}(\sigma)+\F^{-}(\sigma)\right] W_{1}(\sigma) & = &
\Y_{1}(\sigma) 
\label{5.45a} \\
\text{and} & & \nonumber \\
\frac{1}{2}\left[\F^{+}(\sigma)-\F^{-}(\sigma)\right] W_{1}(\sigma) & = &
\left( \begin{array}{c}
0 \\ 0
\end{array} \right).
\label{5.45b}
\end{eqnarray}

\item 
We next note that the four equations (\ref{5.42a}), (\ref{5.42b}),
(\ref{5.45a}) and (\ref{5.45b}) are collectively equivalent to the
single equation
\begin{equation}
\F^{\pm}(\sigma) W(\sigma) = \Y(\sigma) \left( \begin{array}{cc}
1 & 0 \\ 0 & \nu^{\pm}(\sigma)
\end{array} \right) \text{ for all } \sigma \in \I(\x),
\label{5.46}
\end{equation}
where $\Y(\sigma)$ is defined to be the $2 \times 2$ matrix whose first
and second columns are $\Y_{1}(\sigma)$ and $\Y_{2}(\sigma)$, respectively.
From the definition of $W(\sigma)$ by Eqs.\ (\ref{5.22a}) and (\ref{5.22c}),
and from the expression for $\F^{M}(\tau)$ that is given by 
Eq.\ (\ref{G2.10c}), Eq.\ (\ref{5.46}) is equivalent to the statement
\begin{eqnarray}
\F^{+}(\sigma) v^{(i)}(\sigma) [\F^{M+}(\sigma)]^{-1} & = &
\F^{-}(\sigma) v^{(i)}(\sigma) [\F^{M-}(\sigma)]^{-1} 
\nonumber \\
& = & Y(\sigma) \text{ for all } \sigma \in \I^{(i)}(\x),
\label{5.47}
\end{eqnarray}
where
\begin{equation}
Y(\sigma) := \Y(\sigma) \left( \begin{array}{cc}
1 & i(\sigma-z) \\ 0 & 1
\end{array} \right) \text{ for all } \sigma \in \bar{\I}(\x).
\label{5.48}
\end{equation}
From the above Eq.\ (\ref{5.48}) and from statements (\ref{5.39}) and
(\ref{5.43}),
\begin{equation}
\begin{array}{l}
\text{the function } Y(\x) \text{ whose domain is } \bar{\I}(\x)
\text{ and whose value for each } 
\\ \sigma \in \bar{\I}(\x) \text{ is }
Y(\x)(\sigma) := Y(\x,\sigma) \text{ is } \bC^{0+}
\text{ and is, therefore, continuous.}
\end{array}
\end{equation}
So, 
\begin{equation}
Y(\x) \text{ is bounded at $\x$ and at $\x_{0}$.}
\label{5.49b}
\end{equation}

\item 
We now return to the definition of $\F(\x)$ in terms of $\Y_{1}(\x)$
by Eq.\ (\ref{5.24b}).  From Lemma~\ref{5.1C}(i) when $\Box$ is $1$, and
from statement (\ref{5.39}) concerning $\Y_{1}(\x)$ being $\bC^{0+}$ on
its domain $\bar{\I}(\x)$, note that the factors in the numerator of the
integrand in Eq.\ (\ref{5.24b}) have the following properties:
\begin{equation}
\begin{array}{l}
\Y_{1}(\sigma') [W_{2}(\sigma')]^{T} J \text{ is defined for all }
\sigma' \in \bar{\I}(\x) \\
\text{and obeys a H\"{o}lder condition on } \bar{\I}(\x);
\end{array}
\label{5.50a}
\end{equation}
and
\begin{equation}
\begin{array}{l}
[\nu^{\pm}(\sigma')]^{-1} \text{ is } H(1/2) \text{ on each closed subinterval
of } \I(\x) \\
\text{and converges to zero as } \sigma' \rightarrow r \text{ and as }
\sigma' \rightarrow s.
\end{array}
\label{5.50b}
\end{equation}
Also, recall that
\begin{equation}
\begin{array}{l}
\nu(\tau) \text{ is that branch of } (\tau-r_{0})^{1/2}
(\tau-s_{0})^{1/2} (\tau-r)^{-1/2} (\tau-s)^{-1/2} \\
\text{which has the cut } \bar{\I}(\x) \text{ and the value $1$ at
$\tau=\infty$.}
\end{array}
\label{5.50c}
\end{equation}
Several theorems on Cauchy intgrals near the end points of the lines of
inegration are given in Ref.~\ref{Musk}, Sec.~29, Ch.~4.  In particular,
by applying Muskelishvili's Eq.\ (29.4) to our Eq.\ (\ref{5.24b}), one
obtains the following conclusion from the above statements (\ref{5.50a})
and (\ref{5.50b}):
\begin{equation}
\nu(\tau)^{-1} \F(\tau) \text{ converges as } \tau \rightarrow
r \text{ and as } \tau \rightarrow s.
\label{5.51a}
\end{equation}
Moreover, by applying Muskelishvili's Eqs.\ (29.5) and (29.6) to our
Eq.\ (\ref{5.24b}), one obtains the following conclusion from the above
statements (\ref{5.50a}) to (\ref{5.50c}):
\begin{equation}
\F(\tau) \text{ converges as } \tau \rightarrow r_{0} \text{ and as }
\tau \rightarrow s_{0}.
\label{5.51b}
\end{equation}

\item 
From the above statements (\ref{5.40a}), (\ref{5.40b}), (\ref{5.47}),
(\ref{5.49b}), (\ref{5.51a}) and (\ref{5.51b}), all of the
defining conditions for a solution of the HHP corresponding to 
$(\bv,\F^{M},\x)$ are satisfied by $\F(\x)$ as defined in terms
of $\Y_{1}(\x)$ by Eq.\ (\ref{5.24b}).
\end{arablist}
\cheers

We already know from Thm.~\ref{4.3D}(iv) that there is not more than one
solution of the HHP corresponding to $(\bv,\F^{M},\x)$.

\begin{corollary}[Uniqueness of $\Y_{1}(\x)$]
\label{5.2D} \mbox{ } \\
For each $\bv \in K^{1+}$ and $\x \in D$, there is not more than one
$2 \times 1$ matrix function $\Y_{1}(\x)$ such that
\begin{eqnarray}
\dom{\Y_{1}(\x)} & = & \bar{\I}(\x), 
\label{5.52a} \\
\Y_{1}(\x) & \text{is} & \bC^{0+}
\label{5.52b}
\end{eqnarray}
and $\Y_{1}(\x,\sigma) := \Y_{1}(\x)(\sigma)$ satisfies the singular
integral equation (\ref{5.24e}) for all $\sigma \in \I(\x)$.
\end{corollary}

\proof
Suppose that $\Y_{1}(\x)$ and $\Y'_{1}(\x)$ are $2 \times 1$ matrix
functions, both of which have domain $\bar{\I}(\x)$, are $\bC^{0+}$
and satisfy Eq.\ (\ref{5.24e}) for all $\sigma \in \I(\x)$ and for
the same given $\bv \in K^{1+}$.  Let $\F(\x)$ and $\F'(\x)$
be the $2 \times 2$ matrix functions with domain $C - \bar{\I}(\x)$
that are defined in terms of $\Y_{1}(\x)$ and $\Y'_{1}(\x)$, respectively,
by Eq.\ (\ref{5.24b}).  Then, from the preceding Thm.~\ref{5.1D},
$\F(\x)$ and $\F'(\x)$ are both solutions of the HHP
corresponding to $(\bv,\F^{M},\x)$; and, therefore, from
Thm.~\ref{4.3D}(iv),
\begin{equation}
\F(\x) = \F'(\x);
\end{equation}
and, from Eq.\ (\ref{5.45a}) in the proof of Thm.~\ref{5.1D} and from
statements (\ref{5.52a}) and (\ref{5.52b}),
\begin{equation}
\Y_{1}(\x) = \Y'_{1}(\x).
\end{equation}
\cheers

\setcounter{equation}{0}
\section{A Fredholm integral equation of the second kind that is
equivalent to the Alekseev-type singular integral equation when 
$\bv \in K^{2+}$}

If $\bv \in K^{1+}$ and the particular solution $\Y_{1}(\x)$ of Eq.\ 
(\ref{5.24e}) that has a $\bC^{0+}$ extension to $\bar{\I}(\x)$ exists, 
then it can be shown that $\Y_{1}(\x)$ is also a solution of a Fredholm
integral equation of the second kind.


\setcounter{equation}{0}
\subsection{Derivation of Fredholm equation from Alekseev-type equation}

We shall employ a variant of the Poincar\'{e}-Bertrand commutator theorem.
Suppose that $L$ is a smooth oriented line or contour in $C-\{\infty\}$ 
and $\phi$ is a complex-valued function whose domain is $L \times L$ and 
which obeys a H\"{o}lder condition on $L \times L$.  Then the conventional
Poincar\'{e}-Bertrand theorem asserts
\begin{eqnarray}
\left[ \frac{1}{\pi i} \int_{L} d\tau'', \frac{1}{\pi i} \int_{L} d\tau'
\right] \frac{\phi(\tau',\tau'')}{(\tau''-\tau)(\tau'-\tau'')}
& = & \phi(\tau,\tau) \text{ for all } \tau \in L \nonumber \\
& & \text{minus its end points,}
\label{6.8a}
\end{eqnarray}
where the above bracketed expression is the commutator of the path integral
operators.  We are, of course, concerned here only with the case
$L=\bar{\I}(\x)$; and our variant asserts that, for any function $\phi$
which is $\bC^{0+}$ on $\bar{\I}(\x)^{2}$,
\begin{eqnarray}
\left[ \frac{1}{\pi i} \int_{\bar{\I}} d\sigma'' \nu^{+}(\sigma''),
\frac{1}{\pi i} \int_{\bar{\I}} d\sigma' \nu^{+}(\sigma')^{-1} \right]
\frac{\phi(\sigma',\sigma'')}{(\sigma''-\sigma)(\sigma'-\sigma'')}
& = & \phi(\sigma,\sigma) \text{ for all } \nonumber \\
& & \sigma \in \I(\x),
\label{6.8b}
\end{eqnarray}
or, alternatively,
\begin{eqnarray}
\left[ \frac{1}{\pi i} \int_{\bar{\I}} d\sigma'' \nu^{+}(\sigma'')^{-1},
\frac{1}{\pi i} \int_{\bar{\I}} d\sigma' \nu^{+}(\sigma') \right]
\frac{\phi(\sigma',\sigma'')}{(\sigma''-\sigma)(\sigma'-\sigma'')}
& = & \phi(\sigma,\sigma) \text{ for all } \nonumber \\
& & \sigma \in \I(\x),
\label{6.10c}
\end{eqnarray}

We shall not supply the proof here, as an elegant and thorough proof of the 
Poincar\'{e}-Bertrand theorem (\ref{6.8a}) is given by Sec.~23 of 
Muskhelishvili's treatise, and what we have done is to construct proofs of 
(\ref{6.8b}) and (\ref{6.10c}) that parallel his proof step by step.  

We shall now apply Eq.\ (\ref{6.8b}) to the Alekseev-type equation
(\ref{5.24e}), which we express in the form
\begin{equation}
\frac{1}{\pi i} \int_{\bar{\I}} d\sigma' \nu^{+}(\sigma')^{-1}
\frac{\Y_{1}(\sigma') d_{21}(\sigma',\sigma'')}{\sigma'-\sigma''}
= -W_{1}(\sigma'') \text{ for all } \sigma'' \in \I(\x),
\label{6.9a}
\end{equation}
where, for all $\sigma \in \bar{\I}(\x)$ and $\sigma' \in \bar{\I}(\x)$,
\begin{equation}
d_{21}(\sigma',\sigma) := W_{22}(\sigma')W_{11}(\sigma)
- W_{12}(\sigma')W_{21}(\sigma).
\label{6.1a}
\end{equation}
We suppose that, for a given $\bv \in K^{1+}$ and $\x \in D$, a solution
$\Y_{1}(\x)$ of the Alekseev-type equation (\ref{5.24e}) exists and is
$\bC^{0+}$ on $\bar{\I(\x)}$.  Then the product $\Y_{1}(\x)d_{21}$ is 
$\bC^{0+}$ on $\bar{\I}(\x)^{2}$.  Also, $\det{W(\sigma)} = 
d_{21}(\sigma,\sigma) = 1$.  Therefore, upon multiplying both sides of
Eq.\ (\ref{6.9a}) by $(\sigma''-\sigma)^{-1}$ and then applying the PV
integral operator
$$
\frac{1}{\pi i} \int_{\bar{\I}} d\sigma'' \nu^{+}(\sigma''),
$$
Eq.\ (\ref{6.8b}) gives us
\begin{equation}
\Y_{1}(\sigma) - \frac{1}{\pi i} \int_{\bar{\I}} d\sigma'
\nu^{+}(\sigma')^{-1} \Y_{1}(\sigma') K_{21}(\sigma',\sigma) 
= U_{1}(\sigma), 
\label{6.7}
\end{equation}
where, for each $\sigma \in \bar{\I}(\x)-\{r,s\}$,
\begin{equation}
U_{1}(\sigma) := - \frac{1}{\pi i} \int_{\bar{\I}} d\sigma'
\nu^{+}(\sigma') \frac{W_{1}(\sigma')}{\sigma'-\sigma};
\label{6.3}
\end{equation}
and, for each $(\sigma',\sigma) \in \bar{\I}(\x) \times 
[\bar{\I}(\x)-\{r,s\}]$,
\begin{equation}
K_{21}(\sigma',\sigma) := - \frac{1}{\pi i} \int_{\bar{\I}} d\sigma''
\nu^{+}(\sigma'') \frac{d_{21}(\sigma',\sigma'')}{(\sigma''-\sigma)
(\sigma'-\sigma'')}.
\label{6.4}
\end{equation}

So far, we have only established that Eq.\ (\ref{6.7}) holds for all 
$\sigma \in \I(\x)$.  However, using the expressions
\begin{eqnarray}
U_{1}(\sigma) & = & W_{1}(\sigma) - \frac{1}{\pi i} \int_{\bar{\I}} d\sigma' 
\nu^{+}(\sigma') \frac{W_{1}(\sigma')-W_{1}(\sigma)}{\sigma'-\sigma}, 
\label{6.2d} \\
K_{21}(\sigma',\sigma) & = & k_{21}(\sigma',\sigma)
- \frac{1}{\pi i} \int_{\bar{\I}} d\sigma'' \nu^{+}(\sigma'')
\left[\frac{k_{21}(\sigma',\sigma'') 
- k_{21}(\sigma',\sigma)}{\sigma''-\sigma}\right],
\label{6.2e}
\end{eqnarray}
where
\begin{equation}
k_{21}(\sigma',\sigma) := \frac{d_{21}(\sigma',\sigma)-1}{\sigma'-\sigma},
\label{6.2c}
\end{equation}
it is not difficult to prove the following lemma.
\begin{lemma}[Properties of $U_{1}(\x)$ and $K_{21}(\x)$]
\label{6.2A} \mbox{ } \\
For each $\x \in D$ and $\bv \in K^{\Box}$, 
$U_{1}(\x)$ is $\bC^{n-1}$ if $\Box$ is $n$ and $n \ge 2$, is 
$\bC^{(n-1)+}$ if $\Box$ is $n+$ and $n \ge 2$, are $\bC^{\infty}$ if
$\Box$ is $\infty$ and is $\bC^{an}$ if $\Box$ is `an'; and $K_{21}(\x)$
is $\bC^{n-2}$ if $\Box$ is $n$ and $n \ge 2$, is $\bC^{(n-2)+}$ if
$\Box$ is $n+$ and $n \ge 2$, is $\bC^{\infty}$ if $\Box$ is $\infty$,
and is $\bC^{an}$ if $\Box$ is `an'.

If $\bv \in K^{1+}$, then $U_{1}(\x)$ is $\bC^{0+}$
and $K_{21}(\x)$ is also $\bC^{0+}$ \{but, as we recall,
its domain is only $\bar{\I}(\x) \times [\bar{\I}(\x)-\{r,s\}]$\}.
\end{lemma}
From this it follows that
\begin{equation}
U_{1}(\x) \text{ is continuous on } \bar{\I}(\x)
\label{6.9d}
\end{equation}
and
\begin{equation}
K_{21}(\x) \text{ is continuous on } \dom{K_{21}(\x)}.
\label{6.9e}
\end{equation}
Moreover, Lem.~\ref{5.2C} remains valid if $S$ is a
closed or a semi-closed subinterval of $R^{1}$.  Therefore, from 
(\ref{6.9e}) and Lem.~\ref{5.2C}(i) [with a closed or a semi-closed 
$S \subset R^{1}$], the integral in Eq.\ (\ref{6.7}) is a continuous
function of $\sigma$ throughout $\bar{\I}(\x)$ if $\bv \in K^{2}$,
and throughout $\bar{\I}(\x)-\{r,s\}$ if $\bv \notin K^{2}$; and it then
follows from the fact that
\begin{equation}
\Y_{1}(\x) \text{ is continuous on } \bar{\I}(\x) 
\end{equation}
and from (\ref{6.9d}) that Eq.\ (\ref{6.7}) holds for all $\sigma \in
\bar{\I}(\x)$ if $\bv \in K^{2}$, and for all $\sigma \in 
\bar{\I}(\x)-\{r,s\}$ if $\bv \notin K^{2}$.  Thus, we have the
following theorem.
\begin{theorem}[Fredholm equation]
\label{6.3A} \mbox{ } \\
Suppose that, for a given $\bv \in K^{1+}$ and $\x \in D$, a solution
$\Y_{1}(\x)$ of the Alekseev-type equation (\ref{5.24e}) exists and is
$\bC^{0+}$ on $\bar{\I}(\x)$.  Then the Fredholm equation (\ref{6.7})
holds for all $\sigma \in \bar{\I}(\x)$ if $\bv \in K^{2}$ and for all
$\sigma \in \bar{I}(\x)-\{r,s\}$ if $\bv \notin K^{2}$.
\end{theorem}

\setcounter{equation}{0}
\subsection{Equivalence of Alekseev-type equation and Fredholm equation
when $\bv \in K^{2+}$}

The Fredholm equation (\ref{6.7}) generally has a singular kernel and is 
generally {\em not\/} equivalent to the Alekseev-type equation (\ref{5.24e}).
In this section we shall restrict our attention to the case $\bv \in K^{2+}$.

\begin{theorem}[Alekseev-Fredholm equivalence theorem]
\label{6.1B} \mbox{ } \\ 
Suppose $\bv \in K^{2+}$, $\x \in D$ and $\Y_{1}(\x)$ is a $2 \times 1$
column matrix function with domain $\bar{\I}(\x)$.  Then $U_{1}(\x)$ is
$\bC^{1+}$ and $K_{21}(\x)$ is $\bC^{0+}$.  Also, the following two
statements are equivalent to one another:
\begin{romanlist}
\item 
$\Y_{1}(\x)$ is $\bC^{0+}$ and is the solution of Eq.\ (\ref{5.24e}) for
all $\sigma \in \I(\x)$.
\item 
$\Y_{1}(\x)$ is summable over $\bar{\I}(\x)$ and is a solution of Eq.\
(\ref{6.7}) for all $\sigma \in \bar{\I}(\x)$.
\end{romanlist}
\end{theorem}

\proof
From Lem.~\ref{6.2A}, $U_{1}(\x)$ is $\bC^{1+}$ and $K_{21}(\x)$ is
$\bC^{0+}$; and Thm.~\ref{6.3A} already asserts that statement (i)
implies statement (ii).  It remains only to prove that statement (ii)
implies statement (i).

Grant statement (ii).  Since $U_{1}(\x)$ is $\bC^{1+}$ and $K_{21}(\x)$
is $\bC^{0+}$ on $\bar{\I}(\x)$ and since $\Y_{1}(\x)$ is summable over
$\bar{\I}(\x)$, Eq.\ (\ref{6.7}) and Lem.~\ref{5.2C}(iv) yield
\begin{equation}
\Y_{1}(\x) \text{ is } \bC^{0+} \text{ on } \bar{\I}(\x).
\label{6.11a}
\end{equation}

Next, using the Poincar\'{e}-Beltrami variant, one deduces the following 
equivalent of the Fredholm equation (\ref{6.7}):
\begin{equation}
\Y_{1}(\sigma) + \frac{1}{\pi i} \int_{\bar{\I}} d\sigma' \nu^{+}(\sigma')
\frac{\psi(\sigma')+W_{1}(\sigma')}{\sigma'-\sigma} = 0,
\label{6.11b}
\end{equation}
where
\begin{equation}
\psi(\sigma) := \frac{1}{\pi i} \int d\sigma' \nu^{+}(\sigma')^{-1}
\Y_{1}(\sigma') k_{21}(\sigma',\sigma).
\label{6.11c}
\end{equation}
From Lem.~\ref{5.2C}(iv) and (\ref{6.11a}),
\begin{equation}
\psi \text{ is } \bC^{0+} \text{ on } \bar{\I}(\x).
\end{equation}

Next, after replacing `$\sigma$' by `$\sigma''$' in Eq.\ (\ref{6.11b})
and then applying the operator
$$
\frac{1}{\pi i} \int_{\bar{\I}} d\sigma'' \nu^{+}(\sigma'')^{-1}
\frac{1}{\sigma''-\sigma},
$$
one finds that
\begin{equation}
\frac{1}{\pi i} \int_{\bar{\I}} d\sigma'' \nu^{+}(\sigma'')^{-1}
\frac{\Y_{1}(\sigma'')}{\sigma''-\sigma} + \psi(\sigma) + W_{1}(\sigma)
= 0,
\label{6.11e}
\end{equation}
from which equation one can derive the Alekseev-type equation (\ref{5.24e}).
\cheers

Let us summarize the results given by Thm.~\ref{5.1D} and Thm.~\ref{6.1B}
when $\bv \in K^{2+}$.

\begin{theorem}[Summary]
\label{6.2B} \mbox{ } \\
Suppose $\bv \in K^{2+}$, $\x \in D$, and $\F(\x)$ and
$\Y_{1}(\x)$ are $2 \times 2$ and $2 \times 1$ matrix functions,
respectively, such that
\begin{equation}
\dom{\F(\x)} = C - \bar{\I}(\x) \text{ and }
\dom{\Y_{1}(\x)} = \bar{\I}(\x).
\end{equation}
Then the following three statements are equivalent to one another:
\begin{romanlist}
\item 
The function $\F(\x)$ is the solution of the HHP corresponding to
$(\bv,\F^{M},\x)$, and $\Y_{1}(\x)$ is the function whose restriction
to $\I(\x)$ is defined by Eq.\ (\ref{5.25a}) and whose extension to 
$\bar{\I}(\x)$ is then defined by Eqs.\ (\ref{5.30b}) and (\ref{5.30a}).
[The existence and uniqueness of this extension is asserted by
Cor.~\ref{5.5C}.]
\item 
The function $\Y_{1}(\x)$ is $\bC^{0+}$ and its restriction to $\I(\x)$
is a solution of the Alekseev-type equation (\ref{5.24e});
and $\F(\x)$ is defined in terms of $\Y_{1}(\x)$ by Eq.\ (\ref{5.24b}).
\item 
The function $\Y_{1}(\x)$ is summable over $\bar{\I}(\x)$ and is a solution
of the Fredholm equation (\ref{6.7}) for all $\sigma \in \bar{\I}(\x)$.
\end{romanlist}
\end{theorem}

\proof
Directly from Thm.~\ref{5.1D} and Thm.~\ref{6.1B}.
\cheers

\begin{corollary}[Uniqueness of solutions]
\label{6.3B} \mbox{ } \\
When $\bv \in K^{2+}$, each of the solutions defined in (i), (ii) and~(iii)
of the preceding theorem is unique if it exists.
\end{corollary}

\proof
This follows from the preceding theorem and the uniqueness theorem
[Thm.~\ref{4.3D}(iv)] for the HHP.
\cheers

\setcounter{equation}{0}
\section{Existence and properties of the HHP solution $\F$ when
$\bv \in K^{2+}$\label{Sec_4}}

\setcounter{equation}{0}
\subsection{Homogeneous equations, theorems, etc.}

By considering a homogeneous version of the Fredholm equation (\ref{6.7}),
we found it possible to employ the Fredholm alternative theorem to establish
the existence of the solution of the HHP corresponding to $(\bv,\F^{M})$
when $\bv \in K^{2+}$.

\begin{definition}{Dfn.\ of \zip{HHP}}
The HHP that is defined as in Sec.~\ref{Sec_1} except that the condition
(2) is replaced by the condition
\begin{equation}
\F(\x,\infty) = 0 \text{ (\zip{HHP} condition)}
\label{6.14}
\end{equation}
will be called the \zip{HHP} {\em corresponding to} $(\bv,\F_{0},\x)$.
\end{definition}
Clearly, the $2 \times 2$ matrix function $\F(\x)$ with the domain
$C - \bar{\I}(\x)$ and the value $\F(\x,\tau)=0$ for all $\tau$ in
this domain is a solution of the \zip{HHP} corresponding to 
$(\bv,\F_{0},\bv)$.  It will be called the {\em zero solution}.

\begin{definition}{Dfn.\ of equation number with attached subscript `$0$'}
To each linear integral equation that occurs in these notes from Thm.~\ref{5.4A}
to Thm.~\ref{6.2B}, inclusive, and that has a term that is an integral whose
integrand involves `$\F$', `$\F^{\pm}$', `$\Y$' or `$\Y^{(i)}$' (or one of
their columns), there corresponds a homogeneous integral equation that will
be designated by the symbol that results when the subscript `$0$' is
attached to the equation number for the inhomogeneous integral equation.
\end{definition}

\begin{definition}{Dfn.\ of theorem label (etc.) with attached subscript
`$0$'}
When a {\em new} valid assertion results from subjecting a labelled
assertion to the following substitutions, that new valid assertion will
bear the same label with an attached subscript `$0$'.
\begin{arablist}
\item 
`HHP' $\rightarrow$ `\zip{HHP}'
\item 
$\F(\x,\infty)=I$' $\rightarrow$ `$\F(\x,\infty)=0$' 
in condition (2) of the HHP
\item 
each integral equation $\rightarrow$ the corresponding homogeneous
integral equation
\item 
each equation number for an integral equation $\rightarrow$ the same
equation number with an attached subscript `$0$'.
\end{arablist}
\end{definition}

\setcounter{equation}{0}
\subsection{Only a zero solution of homogeneous equation}

For our immediate purpose, we shall need the following explicit version of
Thm.~\zip{\ref{6.2B}}:

\begin{theorem}[Theorem \zip{\ref{6.2B}}]
\label{6.1C} \mbox{ } \\
Suppose $\bv \in K^{2+}$, $\x \in D$, and $\F(\x)$ and $\Y_{1}(\x)$
are $2 \times 2$ and $2 \times 1$ matrix functions, respectively, such that
\begin{equation}
\dom{\F(\x)} = C - \bar{\I}(\x) \text{ and }
\dom{\Y_{1}(\x)} = \bar{\I}(\x).
\end{equation}
Then the following three statements are equivalent to one another:
\begin{romanlist}
\item 
The function $\F(\x)$ is a solution of the \zip{HHP} corresponding
to $(\bv,\F^{M},\x)$; and $\Y_{1}(\x)$ is the continuous function
whose restriction to $\I(\x)$ is defined in terms of $\F^{\pm}(\x)$ by
Eq.\ (\ref{5.25a}), and whose existence and uniqueness are asserted by
Cor.~\zip{\ref{5.5C}}.
\item 
The function $\Y_{1}(\x)$ is $\bC^{0+}$ and its restriction to $\I(\x)$
is a solution of Eq.\ \zip{(\ref{5.24e})}; and $\F(\x)$ is defined
in terms of $\Y_{1}(\x)$ by Eq.\ \zip{(\ref{5.24b})}.
\item 
The function $\Y_{1}(\x)$ is summable over $\bar{\I}(\x)$ and is a 
solution of the homogeneous Fredholm integral equation \zip{(\ref{6.7})}
for all $\sigma \in \bar{\I}(\x)$.
\end{romanlist}
\end{theorem}

\proof
This theorem summarizes Thms.~\zip{\ref{5.1D}} and \zip{\ref{6.1B}} for
the case $\bv \in K^{2+}$.
\cheers

\begin{theorem}[Only a zero solution of \zip{HHP}]
\label{6.2C} \mbox{ } \\
For each $\bv \in K$, $\F_{0} \in \S_{\F}$ and $\x \in D$,
the only solution of the \zip{HHP} corresponding to $(\bv,\F_{0},\x)$
is its zero solution.
\end{theorem}

\proof
The proof will be given in four parts:
\begin{arablist}
\item 
From the hypothesis $\F_{0} \in \S_{\F}$,
\begin{equation}
\left[ \F_{0}(\x,\tau^{*})\right]^{\dagger} \A_{0}(\x,\tau)
\F_{0}(\x,\tau) = \A_{0}(\x_{0},\tau) \text{ for all }
\tau \in C-\bar{\I}(\x),
\label{6.15a}
\end{equation}
where
\begin{equation}
\A_{0}(\x,\tau) := (\tau-z)\Omega + \Omega h_{0}(\x) \Omega
\label{6.15b}
\end{equation}
and $h_{0}(\x)$ is computed from $\E_{0} \in \S_{\E}$ in the usual
way.\footnote{To prove Eq.\ (\ref{6.15a}), one first shows that Eq.\
(\ref{sdr}) is equivalent to $\A_{0}\Gamma_{0} = \frac{1}{2} \Omega dH_{0}
\Omega$ and then uses (\ref{linsys}) to show that the differential of the
left side of Eq.\ (\ref{6.15a}) vanishes.  The rest is obvious.} Since
\begin{equation}
h_{0}(\x_{0}) := h^{M}(\x_{0}) = \left( \begin{array}{cc}
\rho_{0}^{2} & 0 \\ 0 & 1
\end{array} \right)
\label{6.15c}
\end{equation}
in our gauge,
\begin{equation}
\A_{0}(\x_{0},\tau) = \A^{M}(\x_{0},\tau).
\label{6.15d}
\end{equation}
Equation (\ref{6.15a}) is clearly expressible in the alternative form
\begin{eqnarray}
\lefteqn{\F_{0}(\x,\tau) \left[\A^{M}(\x_{0},\tau)\right]^{-1} 
\left[\F_{0}(\x,\tau^{*})\right]^{\dagger} = 
\left[\A_{0}(\x,\tau)\right]^{-1} } \nonumber \\
& & \text{for all } \tau \in C-\bar{\I}(\x),
\label{6.15e}
\end{eqnarray}
since $[\F_{0}(\x,\tau)]^{-1}$ exists for all $\tau \in C-\bar{\I}(\x)$, and
\begin{equation}
\left[ \A_{0}(\x,\tau) \right]^{-1} =
\frac{\B_{0}(\x,\tau)}{\rho^{2}-(\tau-z)^{2}},
\label{6.15f}
\end{equation}
where
\begin{equation}
\B_{0}(\x,\tau) := h_{0}(\x) - (\tau-z) \Omega,
\label{6.15g}
\end{equation}
exists for all $\tau \in C - \{r,s\}$.

\item 
Next, condition (3) in the definition of the HHP (and the \zip{HHP}) that
is given in Sec.~\ref{Sec_1} asserts that $\F^{\pm}(\x)$ exist, and
Eq.\ (\ref{G3.17}) is expressible in the form
\begin{equation}
\F^{\pm}(\x,\sigma) = Y^{(i)}(\sigma) \F_{0}^{\pm}(\x,\sigma)
[v^{(i)}(\sigma)]^{-1} \text{ for each } i \in \{3,4\}
\text{ and } \sigma \in \I(\x).
\label{6.16a}
\end{equation}
From the definition of the group $K$,
\begin{equation}
[v^{(i)}(\sigma)]^{-1} [\A^{M}(\x_{0},\sigma)]^{-1}
[v^{(i)}(\sigma)^{\dagger}]^{-1} = \A^{M}(\x_{0},\sigma)^{-1}
\text{ for all } \sigma \in \I^{(i)} - \{r,s\}.
\label{6.16b}
\end{equation}
Therefore, from Eqs.\ (\ref{6.16a}), (\ref{6.16b}) and (\ref{6.15e}),
\begin{eqnarray}
\F^{\pm}(\x,\sigma) [\A^{M}(\x_{0},\sigma)]^{-1}
[\F^{\mp}(\x,\sigma)]^{\dagger} & = & Y(\x,\sigma) [\A_{0}(\x,\sigma)]^{-1}
Y(\x,\sigma)^{\dagger} \nonumber \\
& & \text{for all } \sigma \in \I(\x);
\label{6.16c}
\end{eqnarray}
or, equivalently, with the aid of Eqs.\ (\ref{6.15f}), (\ref{6.15g})
and (\ref{6.15c}),
\begin{eqnarray}
\lefteqn{\left[ \frac{\rho^{2}-(\sigma-z)^{2}}
{\rho_{0}^{2}-(\sigma-z_{0})^{2}} \right]
\F^{\pm}(\x,\sigma) \B^{M}(\sigma) [\F^{\mp}(\x,\sigma)]^{\dagger} = } 
\nonumber \\ 
& & Y(\x,\sigma) \B_{0}(\x,\sigma) Y(\x,\sigma)^{\dagger}
\text{ for all } \sigma \in \I(\x),
\label{6.16d}
\end{eqnarray}
where
\begin{equation}
\B^{M}(\tau) := \left( \begin{array}{cc}
\rho_{0}^{2} & -i(\tau-z_{0}) \\ i(\tau-z_{0}) & 1
\end{array} \right).
\label{6.16e}
\end{equation}

\item 
Next, let $Z(\x)$ denote the function with the (tentative) domain
$C - \bar{\I}(\x)$ and the values
\begin{eqnarray}
Z(\x,\tau) & := & Z(\x)(\tau) \nonumber \\
& := & \nu(\x,\tau)^{-1}
\F(\x,\tau) \B^{M}(\tau) 
[\nu(\x,\tau^{*})^{-1}\F(\x,\tau^{*})]^{\dagger} 
\nonumber \\
& & \text{for all } \tau \in C - \bar{\I}(\x),
\label{6.17a}
\end{eqnarray}
where note that
\begin{equation}
\nu(\x,\tau)^{-2} =
\frac{(\tau-r)(\tau-s)}{(\tau-r_{0})(\tau-s_{0})} =
\frac{(\tau-z)^{2}-\rho^{2}}{(\tau-z_{0})^{2}-\rho_{0}^{2}}.
\label{6.17b}
\end{equation}

We again appeal to the trilogy of elementary theorems due to Riemann and
Liouville.\footnote{See Refs.~\ref{acaa}, \ref{isohf} and~\ref{Liouville}.}
Using these, we shall define an extension of $Z(\x)$, and we shall let
$Z(\x)$ denote this extension as well.

From condition (1) in the definition of the HHP (and the \zip{HHP}),
and from Eqs.\ (\ref{6.17a}), (\ref{6.16e}) and (\ref{6.14}),
\begin{equation}
Z(\x,\tau) \text{ is a holomorphic function of } \tau 
\text{ throughout } C - \bar{\I}(\x),
\label{6.17c}
\end{equation}
and
\begin{equation}
Z(\x,\infty) = 0.
\label{6.17d}
\end{equation}
Let ($\Im{\zeta}>0$)
\begin{equation}
Z^{\pm}(\x,\sigma) := \lim_{\zeta \rightarrow 0} Z(\x,\sigma \pm \zeta)
\text{ for all } \sigma \in \I(\x),
\label{6.17e}
\end{equation}
which exist according to condition (3) in the definition of the HHP
(and the \zip{HHP}).  Then, from Eqs.\ (\ref{6.17a}), (\ref{6.17b})
and (\ref{6.16d}),
\begin{equation}
Z^{+}(\x,\sigma) = Z^{-}(\x,\sigma) 
= Y(\x,\sigma) \B_{0}(\x,\sigma) Y(\x,\sigma)^{\dagger}
\text{ for all } \sigma \in \I(\x).
\label{6.17f}
\end{equation}
The above equation permits us to define a single valued extension of
$Z(\x)$ to the domain $C - \{r,s,r_{0},s_{0}\}$ by letting
\begin{equation}
Z(\x,\sigma) := Z^{\pm}(\x,\sigma)
= Y(\x,\sigma) \B_{0}(\x,\sigma) Y(\x,\sigma)^{\dagger}
\text{ for all } \sigma \in \I(\x),
\label{6.17g}
\end{equation}
whereupon, from (\ref{6.17c}), (\ref{6.17g}) and the theorem on analytic
continuation across an arc,
\begin{equation}
Z(\x,\tau) \text{ is a holomorphic function of } \tau 
\text{ throughout } C -\{r,s,r_{0},s_{0}\}.
\label{6.17h}
\end{equation}

We next apply condition (4) in the definition of the HHP (and \zip{HHP}).
Since, according to condition (4), $\nu(\x)^{-1} \F(\x)$
and $Y(\x)$ are both bounded at $\x$, Eqs.\ (\ref{6.17a}) and (\ref{6.17g})
yield
\begin{equation}
\begin{array}{l}
\text{There exists a positive real number } M_{1}(\x) \text{ such that } \\
||Z(\x,\tau)|| < M_{1}(\x) \text{ as } \tau \rightarrow r \text{ and as } 
\tau \rightarrow s \\
\text{through any sequence of points in } C - \{r,s,r_{0},s_{0}\}.
\end{array}
\label{6.17i}
\end{equation}
Since $\F(\x)$ and $Y(\x)$ are both bounded at $\x_{0}$, Eqs.\
(\ref{6.17a}), (\ref{6.17b}) and (\ref{6.17g}) yield
\begin{equation}
\begin{array}{l}
\text{There exists a positive real number } M_{2}(\x) \text{ such that } \\
||(\tau-r_{0})(\tau-s_{0})Z(\x,\tau)|| < M_{2}(\x) 
\text{as } \tau \rightarrow r_{0} \text{ and as } \tau \rightarrow s_{0} \\
\text{through any sequence of points in } C - \{r,s,r_{0},s_{0}\}.
\end{array}
\label{6.17j}
\end{equation}
However, since $Y(\x)$ is bounded at $\x_{0}$, Eq.\ (\ref{6.17g}) yields

\begin{equation}
\begin{array}{l}
\text{There exists a positive real number } M_{3}(\x) \text{ such that } \\
||Z(\x,\sigma)|| < M_{3}(\x) \text{ as } \sigma \rightarrow r_{0}
\text{ and as } \sigma \rightarrow s_{0} \\ 
\text{through any sequence of points in } \I(\x).
\end{array}
\label{6.17k}
\end{equation}
The theorem on isolated singularities, together with statements (\ref{6.17h})
to (\ref{6.17k}), now informs us that
\begin{equation}
Z(\x) \text{ has a holomorphic extension [which we also denote by $Z(\x)$]
to C,}
\label{6.17l}
\end{equation}
whereupon Eq.\ (\ref{6.17d}) and the (generalized) theorem of Liouville yield
\begin{equation}
Z(\x,\tau) = 0 \text{ for all } \tau \in C.
\label{6.17m}
\end{equation}

\item 
Putting (\ref{6.17a}) and (\ref{6.17m}) together, one obtains
\begin{equation}
\F(\x,\sigma) \B^{M}(\sigma) \F(\x,\sigma)^{\dagger} = 0
\text{ for all } \sigma \in C - \bar{\I}(\x).
\label{6.18a}
\end{equation}
Note from Eq.\ (\ref{6.16e}), $\B^{M}(\sigma)$ is hermitian,
\begin{equation}
\begin{array}{rcl}
\tr{\B^{M}(\sigma)} & = & 1 + \rho_{0}^{2} \; \text{ and } \\
\det{\B^{M}(\sigma)} & = & (s_{0}-\sigma)(\sigma-r_{0}).
\end{array}
\end{equation}
Recall that $|r,r_{0}| < |s,s_{0}|$ for any type~A triple $\triple$;
and it is clear that
\begin{equation}
\B^{M}(\sigma) \text{ is hermitian and positive definite for all }
|r,r_{0}| < \sigma < |s,s_{0}|.
\end{equation}
Therefore, Eq.\ (\ref{6.18a}) implies
\begin{equation}
\F(\x,\sigma) = 0 \text{ for all } \sigma \text{ such that }
|r,r_{0}| < \sigma < |s,s_{0}|.
\end{equation}
However, $\F(\x,\tau)$ is a holomorphic function of $\tau$
throughout $C - \bar{\I}(\x)$, and this domain contains the open
interval between $|r,r_{0}|$ and $|s,s_{0}|$.  So,
\begin{equation}
\F(\x,\tau) = 0 \text{ for all } \tau \in C - \bar{\I}(\x).
\end{equation}
\end{arablist}
\cheers

\begin{theorem}[Only a zero solution of \zip{(\ref{6.7})}]
\label{6.3C} \mbox{ } \\
The only solution of the homogeneous Fredholm integral equation of the
second kind Eq.\ \zip{(\ref{6.7})} is its zero solution.
\end{theorem}

\proof
Let $\Y_{1}(\x)$, with domain $\bar{\I}(\x)$, denote a solution of
Eq.\ \zip{(\ref{6.7})}; and let $\F(\x)$, with domain $C -
\bar{\I}(\x)$, be defined in terms of $\Y_{1}(\x)$ by Eq.\ \zip{(\ref{5.24b})}.
Using Thm.~\ref{6.1C}, one obtains
\begin{equation}
\F(\x) \text{ is a solution of the \zip{HHP} corresponding to }
(\bv,\F^{M},\x),
\end{equation}
whereupon Thm.~\ref{6.2C} delivers
\begin{equation}
\F(\x,\tau) = 0 \text{ for all } \tau \in C - \bar{\I}(\x).
\end{equation}
It follows that
\begin{equation}
\F^{\pm}(\x,\sigma) = 0 \text{ for all } \sigma \in \I(\x),
\end{equation}
whereupon, from Thm.~\ref{6.2C}(i), Eq.\ (\ref{5.25a}) and the continuity
of $\Y_{1}(\x)$,
\begin{equation}
\Y_{1}(\x,\sigma) = 0 \text{ for all } \sigma \in \bar{\I}(\x).
\end{equation}
\cheers

\setcounter{equation}{0}
\subsection{Existence and uniqueness of HHP solution}

At this point, we note that Eq.\ (\ref{6.7}) is a regular Fredholm equation
in disguise when $\bv \in K^{2+}$.  In integrals such as those in 
Thm.~\ref{5.1B}, it is sometimes useful to introduce a new variable of
integration for the purpose of getting rid of the singularities of the
integrands at $\sigma' \in \{r,s,r_{0},s_{0}\}$.  This is especially 
important when one has to consider derivatives of the integrals with
respect to $r$ and $s$.  

\begin{definition}{Dfns.\ of $\Theta$, $\btheta(\x)$ and $\bsigma(\x)$}
Let $\Theta$ denote that union of arcs
\begin{equation}
\Theta := \left[0,\frac{\pi}{2}\right] + \left[\pi,\frac{3\pi}{2}\right]
\label{5.26a}
\end{equation}
whose assigned orientations are in the direction of increasing 
$\theta \in [0,\pi/2]$ and $\theta \in [\pi,3\pi/2]$.  For each 
$\x \in D$, let
\begin{equation}
\btheta(\x):\bar{\I}(\x) \rightarrow \Theta
\end{equation}
be a mapping such that
\begin{equation}
\btheta(\x)(\sigma) := \btheta(\x,\sigma),
\end{equation}
where 
\begin{equation}
\begin{array}{r}
0 \le \btheta(\x,\sigma) \le \frac{\pi}{2} \text{ and }
\cos[2\btheta(\x,\sigma)] := \frac{2\sigma-(r_{0}+r)}{r_{0}-r} \\[1ex]
\text{when } \sigma \in \bar{\I}^{(3)}(\x)
\end{array}
\end{equation}
and
\begin{equation}
\begin{array}{r}
\pi \le \btheta(\x,\sigma) \le \frac{3\pi}{2} \text{ and }
\cos[2\btheta(\x,\sigma)] := \frac{2\sigma-(s_{0}+s)}{s_{0}-s} \\[1ex]
\text{when } \sigma \in \bar{\I}^{(4)}(\x).
\end{array}
\end{equation}
Also let
\begin{equation}
\bsigma(\x):\Theta \rightarrow \bar{\I}(\x)
\end{equation}
be a mapping such that
\begin{equation}
\bsigma(\x)(\theta) := \bsigma(\x,\theta),
\end{equation}
where
\begin{eqnarray}
\bsigma(\x,\theta) & := & r_{0} \cos^{2}\theta + r \sin^{2}\theta
\text{ when } \theta \in \left[0,\frac{\pi}{2}\right] 
\label{5.26h}
\end{eqnarray}
and
\begin{eqnarray}
\bsigma(\x,\theta) & := & s_{0} \cos^{2}\theta + s \sin^{2}\theta
\text{ when } \theta \in \left[\pi,\frac{3\pi}{2}\right].
\label{5.26i}
\end{eqnarray}
\end{definition}
The mapping $\btheta(\x)$ is monotonic and is a continuous bijection
(one-to-one and onto) of $\I(\x)$ onto $\Theta$, and $\bsigma(\x)$
is its inverse mapping.  Moreover, $\bsigma(\x)$ is analytic [which
means that it has an analytic extension to an open subset of $R^{1}$].
Note, in particular, that 
\begin{eqnarray}
\sqrt{\frac{s-\bsigma(\x,\theta')}{s_{0}-\bsigma(\x,\theta')}}
\text{ is a real positive-valued analytic} \nonumber \\
\text{function of } (\x,\theta') \text{ on } D \times
\left[0,\frac{\pi}{2}\right]
\label{5.29a}
\end{eqnarray}
and
\begin{eqnarray}
\sqrt{\frac{\bsigma(\x,\theta')-r}{\bsigma(\x,\theta')-r_{0}}}
\text{ is a real positive-valued analytic} \nonumber \\
\text{function of } (\x,\theta') \text{ on } D \times 
\left[\pi,\frac{3\pi}{2}\right],
\label{5.29b}
\end{eqnarray}
since the left and right cuts are assumed not to overlap.

The following equation is equivalent to Eq.\ (\ref{6.7}) and has a 
$\bC^{0+}$ kernel and a $\bC^{1+}$ inhomogeneous term:
\begin{eqnarray}
\lefteqn{y_{1}(\x,\theta) - \frac{2}{\pi} \int_{\Theta} d\theta'
y_{1}(\x,\theta') \kappa_{21}(\x,\theta',\theta) } \nonumber \\
& = & u_{1}(\x,\theta) \text{ for all } \theta \in \Theta :=
[0,\pi/2] \cup [\pi,3\pi/2],
\label{6.20a}
\end{eqnarray}
where
\begin{eqnarray}
y_{1}(\x,\theta) & := & \Y_{1}(\x,\sigma(\x,\theta)),
\label{6.20b} \\
u_{1}(\x,\theta) & := & U_{1}(\x,\sigma(\x,\theta)),
\label{6.20c} \\
\kappa_{21}(\x,\theta',\theta) & := & q(\x,\theta')
K_{21}(\x,\sigma(\x,\theta'),\sigma(\x,\theta))
\end{eqnarray}
and
\begin{equation}
q(\x,\theta) := \left\{ \begin{array}{l}
(r_{0}-r)\cos^{2}\theta
\sqrt{\frac{s-\sigma(\x,\theta)}{s_{0}-\sigma(\x,\theta)}}
\text{ when } \theta \in [0,\pi/2], \\
(s_{0}-s)\cos^{2}\theta
\sqrt{\frac{\sigma(\x,\theta)-r}{\sigma(\x,\theta)-r_{0}}}
\text{ when } \theta \in [\pi,3\pi/2].
\end{array} \right.
\label{6.20e}
\end{equation}

Equations (\ref{6.2d}) and (\ref{6.2e}) are expressible in the following
forms, in which $\x$ and $\x_{0}$ are no longer suppressed:
\begin{eqnarray}
U_{1}(\x,\sigma) & = & W_{1}(\sigma) - \frac{2}{\pi} \int_{\Theta}
d\theta' p(\x,\theta')
\frac{W_{1}(\sigma(\x,\theta'))-W_{1}(\sigma)}{\sigma(\x,\theta')-\sigma}, 
\label{G5.3e} \\
K_{21}(\x,\sigma',\sigma) & = & k_{21}(\sigma',\sigma) - \frac{2}{\pi}
\int_{\Theta} d\theta'' p(\x,\theta'')
\frac{k_{21}(\sigma',\sigma(\x,\theta''))-k_{21}(\sigma',\sigma)}
{\sigma(\x,\theta'')-\sigma},
\label{G5.3f}
\end{eqnarray}
where
\begin{equation}
p(\x,\theta) := \left\{ \begin{array}{l}
(r_{0}-r)\sin^{2}\theta
\sqrt{\frac{s_{0}-\sigma(\x,\theta)}{s-\sigma(\x,\theta)}}
\text{ when } \theta \in [0,\pi/2], \\
(s_{0}-s)\sin^{2}\theta
\sqrt{\frac{\sigma(\x,\theta)-r_{0}}{\sigma(\x,\theta)-r}}
\text{ when } \theta \in [\pi,3\pi/2].
\end{array} \right.
\label{G5.3g}
\end{equation}

\begin{theorem}[Fredholm determinant not zero]
\label{6.4C} \mbox{ } \\
The Fredholm determinant corresponding to the kernel $\kappa_{21}(\x)$
is not zero.  Therefore, there exists exactly one solution of Eq.\
(\ref{6.20a}) for each given $\bv \in K^{2+}$ and $\x \in D$; or,
equivalently, there exists exactly one solution of Eq.\ (\ref{6.7})
for each given $\bv \in K^{2+}$ and $\x \in D$.
\end{theorem}

\proof
This follows from Thm.~\ref{6.3C} and the Fredholm alternative.
\cheers

Thus, in summation, we have the following theorem:

\begin{theorem}[Existence and uniqueness of HHP solution]
\label{Thm_13} \mbox{ } \\ \vspace{-3ex}
\begin{romanlist}
\item
If $\bv \in K^{2+}$, then the \zip{HHP} corresponding to 
$(\bv,\F^{M},\x)$ is equivalent to the homogeneous Fredholm
equation of the second kind that is obtained from Eq.\ (\ref{6.7}) by
deleting the term $U_{1}(\sigma)$, provided that the term $W_{1}(\sigma)$
is also deleted from the expression (\ref{5.24d}) for $\Y_{2}(\sigma)$. 
\item
For any given $\x \in D$, and $\bv \in K$, the \zip{HHP} corresponding
to $(v,\F^{M},\x)$ has the unique solution $\F(\x,\tau)=0$ for
all $\tau \in C - \I(\x)$.
\item
Therefore, if $\bv \in K^{2+}$, the only solution of the homogeneous
Fredholm equation is the zero solution.  Hence from the Fredholm alternative
theorem, the inhomogeneous Fredholm equation (\ref{6.7}) has exactly one
solution.  We conclude that there exists one and only one solution
of the HHP corresponding to $(\bv,\F^{M})$ when $\bv \in K^{2+}$.
\end{romanlist}
\end{theorem}

\proof
Directly from Thms.~\ref{6.1C}, \ref{6.2C}, \ref{6.3C} and \ref{6.4C}.
\cheers

\setcounter{equation}{0}
\subsection{The $2 \times 2$ matrix $H(\x)$ associated with each solution
of the HHP corresponding to $(\bv,\F_{0},\x)$ when $\bv \in K$}

\begin{theorem}[Properties of $H(\x)$ and $h(\x)$]
\label{7.1A} \mbox{ } \\
For each $\bv \in K$, $\F_{0} \in \S_{\F}$, $\x \in D$
and solution $\F(\x)$ of the HHP corresponding to
$(\bv,\F_{0},\x)$, there exists exactly one $2 \times 2$ matrix
$H(\x)$ such that
\begin{eqnarray}
\F(\x,\tau) & = & I + (2\tau)^{-1}\left[H(\x)-H^{M}(\x_{0})\right]\Omega
+ O(\tau^{-2}) \nonumber \\
& & \text{in at least one neighborhood of } \tau = \infty.
\label{7.1a}
\end{eqnarray}
Moreover,
\begin{eqnarray}
H(\x_{0}) & = & H^{M}(\x_{0}),
\label{7.1b} \\
H(\x)-H(\x)^{T} & = & 2z \Omega,
\label{7.1c} \\
h(\x) & := & -\Re{H(\x)} \text{ is symmetric,}
\label{7.1d}
\end{eqnarray}
and
\begin{equation}
h(\x_{0}) = \left( \begin{array}{cc}
\rho_{0}^{2} & 0 \\ 0 & 1
\end{array} \right).
\label{7.1e}
\end{equation}
\end{theorem}

\proof
From conditions (1) and (2) in the definition of the HHP, there exists
exactly one $2 \times 2$ matrix $B(\x)$ such that
\begin{eqnarray*}
\F(\x,\tau) & = & I + (2\tau)^{-1} B(\x) + O(\tau^{-2}) \\
& & \text{in at least one neighborhood of } \tau = \infty.
\end{eqnarray*}
Let
$$
H(\x) := H^{M}(\x_{0}) + B(\x) \Omega,
$$
whereupon statement (\ref{7.1a}) follows.  From Thm.~\ref{4.3D}(v)
[Eq.\ (\ref{4.30})], $B(\x_{0}) = 0$, whereupon Eq.\ (\ref{7.1b}) follows.

Next, from Thm.~\ref{4.3D}(iii),
\begin{eqnarray}
\det{\F(\x,\tau)} & = & \nu(\x,\tau) \nonumber \\
& = & 1 + (2\tau)^{-1} (r+s-r_{0}-s_{0}) + O(\tau^{-2}) \nonumber \\
& = & 1 + \tau^{-1} (z-z_{0}) + O(\tau^{-2}) \nonumber \\
& & \text{in at least one neighborhood of } \tau = \infty.
\label{7.2a}
\end{eqnarray}
Moreover, from Eq.\ (\ref{1.14c}),
\begin{equation}
H^{M}(\x_{0}) - [H^{M}(\x_{0})]^{T} = 2 z_{0} \Omega.
\label{7.2b}
\end{equation}
For any $2 \times 2$ matrix $M$, $M \Omega M^{T} = \Omega \det{M}$.  In
particular,
\begin{equation}
\F(\x,\tau) \Omega \F(\x,\tau)^{T} = \Omega
\nu(\x,\tau).
\label{7.2c}
\end{equation}
The next step is to consider Eq.\ (\ref{7.2c}) in at least one neighborhood
of $\tau = \infty$ for which the expansions given by Eqs.\ (\ref{7.1a}) and
(\ref{7.2a}) hold.  The reader can then easily deduce Eq.\ (\ref{7.1c}) by
using Eq.\ (\ref{7.2b}) and the relations $\Omega^{T} = - \Omega$ and
$\Omega^{2} = I$.

The statement (\ref{7.1d}) follows from Eq.\ (\ref{7.1c}) and the relation
$\Omega^{*} = - \Omega$.  Equation (\ref{7.1e}) is derived from Eqs.\
(\ref{7.1b}) and (\ref{1.14c}).
\cheers

\begin{theorem}[Quadratic relation]
\label{7.2A} \mbox{ } \\
For each $\bv \in K$, $\F_{0} \in \S_{\F}$, $\x \in D$
and solution $\F(\x)$ of the HHP corresponding to
$(\bv,\F_{0},\x)$, let $h(\x)$ be defined as in the preceding
theorem, and let
\begin{equation}
\A(\x,\tau) = (\tau-z) \Omega + \Omega h(\x) \Omega.
\label{7.3a}
\end{equation}
Then
\begin{equation}
\F^{\dagger}(\x,\tau) \A(\x,\tau) \F(\x,\tau) = \A(\x_{0},\tau)
\text{ for all } \tau \in [C - \bar{\I}(\x)] - \{\infty\},
\label{7.3b}
\end{equation}
where
\begin{equation}
\F^{\dagger}(\x,\tau) := [\F(\x,\tau^{*})]^{\dagger}
\text{ for all } \tau \in C - \bar{\I}(\x).
\end{equation}
\end{theorem}

\proof
Note that parts (1) and (2) in the proof of Thm.~\ref{6.2C} remain valid
here.  For the sake of convenience, we repeat below Eq.\ (\ref{6.16d})
from part (2) of that proof.
\begin{eqnarray}
[\nu(\x,\sigma)]^{-2} \F^{\pm}(\x,\sigma)
\B^{M}(\sigma) [\F^{\mp}(\x,\sigma)]^{\dagger} & = & Y(\x,\sigma)
\B_{0}(\x,\sigma) Y(\x,\sigma)^{\dagger} \nonumber \\
& & \text{for all } \sigma \in \I(\x),
\label{7.4a}
\end{eqnarray}
where
\begin{eqnarray}
\B^{M}(\tau) & := & \left( \begin{array}{cc}
\rho_{0}^{2} & -i(\tau-z_{0}) \\ i(\tau-z_{0}) & 1
\end{array} \right),
\label{7.4b} \\
\B_{0}(\x,\tau) & := & h_{0}(\x) - (\tau-z) \Omega \nonumber \\
& = & [\rho^{2}-(\tau-z)^{2}] \A_{0}(\x,\tau)^{-1}, \\
\nu(\x,\tau)^{-2} & = &
\frac{(\tau-r)(\tau-s)}{(\tau-r_{0})(\tau-s_{0})} \nonumber \\
& = & \frac{(\tau-z)^{2}-\rho^{2}}{(\tau-z_{0})^{2}-\rho_{0}^{2}}.
\label{7.4d}
\end{eqnarray}

Next, let $Z(\x)$ denote the function with the (tentative) domain
$[C - \bar{\I}(\x)] - \{\infty\}$ and the values
\begin{equation}
Z(\x,\tau) := \nu(\x,\tau)^{-1} \F(\x,\tau)
\B^{M}(\tau) \left[ \nu(\x,\tau^{*})^{-1} \F(\x,\tau^{*})
\right]^{\dagger}.
\label{7.5a}
\end{equation}
From conditions (1) and (2) in the definition of the HHP, and from Eqs.\
(\ref{7.4b}) and (\ref{7.5a}),
\begin{equation}
\begin{array}{l}
Z(\x,\tau) \text{ is a holomorphic function of } \tau \\
\text{throughout } [C - \bar{\I}(\x)] - \{\infty\} \\
\text{and has a simple pole at } \tau = \infty.
\end{array}
\label{7.5b}
\end{equation}
Note that Eq.\ (\ref{7.1e}) enables us to express (\ref{7.4b}) in the form
\begin{equation}
\B^{M}(\tau) = h(\x_{0}) - (\tau-z_{0}) \Omega.
\label{7.5c}
\end{equation}
Also, note that Eqs.\ (\ref{7.1c}) and (\ref{7.1d}) imply that
\begin{equation}
H(\x) + H(\x)^{\dagger} = - 2 h(\x) + 2 z \Omega
\label{7.5d}
\end{equation}
and that Eq.\ (\ref{7.4d}) yields
\begin{eqnarray}
\nu(\x,\tau)^{-2} & = & 1 + 2 \tau^{-1} (z_{0}-z) + 
O(\tau^{-2}) \nonumber \\
& & \text{in at least one neighborhood of } \tau = \infty.
\label{7.5e}
\end{eqnarray}
Upon using the relation $\nu(\x,\tau^{*})^{*} = 
\nu(\x,\tau)$ and upon inserting (\ref{7.1a}), (\ref{7.5c})
and (\ref{7.5e}) into the right side of Eq.\ (\ref{7.5a}), one obtains
the following result with the aid of Eqs.\ (\ref{7.1b}) and (\ref{7.4d}):
\begin{eqnarray}
Z(\x,\tau) & = & - (\tau-z) \Omega + h(\x) + O(\tau^{-1}) \nonumber \\
& & \text{in at least one neighborhood of } \tau = \infty.
\label{7.5f}
\end{eqnarray}

We again appeal to the trilogy of elementary theorems due to Riemann and
Liouville.\footnote{See Refs.~\ref{acaa}, \ref{isohf} and~\ref{Liouville}.}
We let $Z^{\pm}(\x,\sigma)$ be defined for all $\sigma \in \I(\x)$ by
Eq.\ (\ref{6.17e}), whereupon Eqs.\ (\ref{7.4a}) and (\ref{7.5a}) yield
\begin{equation}
Z^{+}(\x,\sigma) = Z^{-}(\x,\sigma) = Y(\x,\sigma) \B_{0}(\x,\sigma)
Y(\x,\sigma)^{\dagger} \text{ for all } \sigma \in \I(\x).
\end{equation}
The above equation permits us to define a single valued extension of $Z(\x)$
to the domain $C - \{r,s,r_{0},s_{0},\infty\}$ by letting
\begin{equation}
Z(\x,\sigma) := Z^{\pm}(\x,\sigma) = Y(\x,\sigma) \B_{0}(\x,\sigma)
Y(\x,\sigma)^{\dagger} \text{ for all } \sigma \in \I(\x),
\label{7.6b}
\end{equation}
whereupon (\ref{7.5b}), (\ref{7.6b}) and the theorem on analytic continuation
across an arc tell us that
\begin{equation}
\begin{array}{l}
Z(\x,\tau) \text{ is a holomorphic function of } \tau \text{throughout } \\
C - \{r,s,r_{0},s_{0},\infty\} \text{ and has a simple pole at } \tau = \infty.
\end{array}
\label{7.6c}
\end{equation}
We next use condition (4) in the definition of the HHP, and we obtain the
statements (\ref{6.17i}), (\ref{6.17j}) and (\ref{6.17k}) exactly as we did
in the proof of Thm.~\ref{6.2C}.  The theorem on isolated singularities, 
together with the statements (\ref{7.6c}), (\ref{6.17i}), (\ref{6.17j}) and
(\ref{6.17k}) now inform us that
\begin{equation}
\begin{array}{l}
Z(\x) \text{ has a holomorphic extension [which we also denote } \\
\text{by $Z(\x)$] to } C - \{\infty\} \text{ and has a simple pole at }
\tau = \infty,
\end{array}
\end{equation}
whereupon Eq.\ (\ref{7.5f}) and the theorem on entire functions that do not
have an essential singularity at $\tau = \infty$ yield
\begin{equation}
Z(\x,\tau) = - (\tau-z) \Omega + h(\x) \text{ for all } \tau \in 
C - \{\infty\}.
\label{7.6e}
\end{equation}

We are now close to completing our proof.  From Thm.~\ref{4.3D}(iii), Eqs.\ 
(\ref{7.5a}), (\ref{7.4b}) and (\ref{7.4d}),
\begin{equation}
\det{Z(\x,\tau)} = \rho^{2} - (\tau-z)^{2}.
\label{7.7a}
\end{equation}
Therefore, from Eqs.\ (\ref{7.3a}) and (\ref{7.6e}), the matrix
$-(\tau-z)\Omega+h(\x)$ is invertible when $\tau \notin \{r,s,\infty\}$,
and
\begin{equation}
\left[-(\tau-z)\Omega+h(\x)\right]^{-1} = 
\frac{\A(\x,\tau)}{\rho^{2}-(\tau-z)^{2}}.
\label{7.7b}
\end{equation}
[Above, we have used the fact that $M^{-1} = \Omega M^{T} \Omega/\det{M}$
for any invertible $2 \times 2$ matrix $M$.]

One then obtains from Eqs.\ (\ref{7.5a}), (\ref{7.5c}), (\ref{7.6e}) and
(\ref{7.7b}),
\begin{eqnarray*}
\lefteqn{\F(\x,\tau) [\A(\x_{0},\tau)]^{-1} \F^{\dagger}(\x,\tau)
= \A(\x,\tau)^{-1} } \nonumber \\
& & \text{for all } \tau \in [C - \bar{\I}(\x)] - \{\infty\},
\end{eqnarray*}
whereupon the conclusion (\ref{7.3b}) follows.
\cheers

\begin{theorem}[More properties of $h(\x)$]
\label{7.3A} \mbox{ } \\
Grant the same premises as in the preceding two theorems, and let $h(\x)$
be defined as before.  Then
\begin{equation}
\det{h(\x)} = \rho^{2}
\end{equation}
and
\begin{equation}
h(\x) \text{ is positive definite }
\end{equation}
as well as real and symetric.
\end{theorem}

\proof
Since $h(\x)$ is symmetric
$$
\det{[h(\x)-(\tau-z)\Omega]} = \det{h(\x)} - (\tau-z)^{2}.
$$
Therefore, Eq.\ (\ref{7.6e}) and (\ref{7.7a}) imply that $\det{h(\x)} =
\rho^{2}$.

From Eqs.\ (\ref{7.5a}), (\ref{7.6e}) and (\ref{7.4d}),
\begin{eqnarray}
Z(\x,\sigma) & = & \frac{(\sigma-r)(s-\sigma)}{(\sigma-r_{0})(s_{0}-\sigma)}
\F(\x,\sigma) \B^{M}(\sigma) \F(\x,\sigma)^{\dagger} \nonumber \\
& = & - (\sigma-z) \Omega + h(\x) \text{ for all } |r,r_{0}| < \sigma <
|s,s_{0}|.
\label{7.9a}
\end{eqnarray}
Equation (\ref{7.4b}) provides us with
\begin{equation}
\det{\B^{M}(\sigma)} = (s_{0}-\sigma)(\sigma-r_{0}) \text{ and }
\tr{\B^{M}(\sigma)} = 1 + \rho_{0}^{2}.
\end{equation}
Therefore,
$$
\frac{(\sigma-r)(s-\sigma)}{(\sigma-r_{0})(s_{0}-\sigma)} \B^{M}(\sigma)
$$
is a positive definite hermitian matrix when $|r,r_{0}| < \sigma < |s,s_{0}|$.
Therefore, the left side of Eq.\ (\ref{7.9a}) is a positive definite hermitian
matrix when $|r,r_{0}| < \sigma < |s,s_{0}|$ and must, therefore, have a real
positive trace when $|r,r_{0}| < \sigma < |s,s_{0}|$.  So,
\begin{equation}
\tr{[-(\sigma-z)\Omega+h(\x)]} = \tr{h(\x)} > 0;
\end{equation}
and, since the determinant of $h(\x)$ is also positive, $h(\x)$ is positive
definite.
\cheers

We caution the reader that the HHP solution $\F$ whose existence has been
proved in this section when $\bv \in K^{2+}$ is not necessarily a member of
$\S_{\F}$; and $H$ as defined by Eq.\ (\ref{7.1a}) is not necessarily a 
member of $\S_{H}$.  However, as we shall prove in Sec.~\ref{Sec_6},
$\F \in \S_{\F}$ and $H \in \S_{H}$ when $\bv \in K^{3}$.  To prepare for
this proof, we shall now investigate the differentiability of $\F$ and $H$
when $\bv \in K^{3}$.

\section{Derivatives of $\F$ and $H$ when $\bv \in K^{3}$\label{Sec_5}}

\setcounter{equation}{0}
\subsection{Fredholm equation solution $\Y_{1}$ corresponding to $\bv \in K^{3}$}

We again refer the reader to the mappings $\btheta(\x):\bar{\I}(\x)
\rightarrow \Theta$ and $\bsigma(\x):\Theta \rightarrow \bar{\I}(\x)$,
for we shall first be discussing the solution $y_{1}$ of the Fredholm
equation (\ref{6.20a}) with kernel $\kappa_{21}$ and inhomogeneous
term $u_{1}$ rather than the solution $\Y_{1}$ of the Fredholm equation
(\ref{6.7}) with kernel $K_{21}$ and inhomogeneous term $U_{1}$.

When $\bv \in K^{2+}$, the solution $y_{1}$ need not be differentiable. 
However, when $\bv \in K^{3}$, the kernel 
$K_{21}(\x,\sigma',\sigma)$ and the inhomogeneous term $U_{1}(\x,\sigma)$
in the Fredholm equation (\ref{6.7}) are $\bC^{1}$ and $\bC^{2}$ functions
of $(\x,\sigma',\sigma)$ and $(\x,\sigma)$, respectively; and the result
is a differentiable $y_{1}$ as we shall see in Thm.~\ref{7.3B}.
The following lemma is required for the proof of Thm.~\ref{7.3B}.

\begin{lemma}[Differentiability properties of $u_{1}$ and $\kappa_{21}$
when $\bv \in K^{3}$]
\label{7.1B} \mbox{ } \\
When $\bv \in K^{3}$, $u_{1}$ is $\bC^{2}$ and $\kappa_{21}$ is $\bC^{1}$.
Moreover, $\partial^{2}\kappa_{21}(\x,\theta',\theta)/\partial r \partial s$
exists and is a continuous function of $(\x,\theta',\theta)$ throughout
$D \times \Theta^{2}$ [whereupon, from a theorem of the calculus,
$\partial^{2}\kappa_{21}/\partial s \partial r$ also exists and is equal to
$\partial^{2}\kappa_{21}/\partial r \partial s$].
\end{lemma}

\proof
The proof will be given in three parts:
\begin{arablist}
\item 
From Eqs.\ (\ref{5.26h}) and (\ref{5.26i}), $\sigma(\x,\theta)$ is a real
analytic function of $\theta$ throughout $D \times \Theta$,
\begin{eqnarray}
\sigma(\x,\theta) & \in & |r,r_{0}| \text{ when } \theta \in [0,\pi/2], 
\label{7.12a} \\
\sigma(\x,\theta) & \in & |s,s_{0}| \text{ when } \theta \in [\pi,3\pi/2],
\label{7.12b}
\end{eqnarray}
Therefore,
\begin{equation}
W(\sigma(\x,\theta)) \text{ is a } \bC^{3} \text{ function of }
(\x,\theta) \text{ throughout } D \times \Theta
\label{7.12e}
\end{equation}
and
\begin{equation}
\begin{array}{r}
\lambda_{21}(\sigma(\x,\theta'),\sigma(\x,\theta''),\sigma(\x,\theta))
\text{ is a } \bC^{1} \text{ function of } \\
(\x,\theta',\theta'',\theta) \text{ throughout } D \times \Theta^{3}.
\end{array}
\label{7.12k}
\end{equation}

\item 
To prove that
\begin{eqnarray}
& & \frac{\partial^{2}\lambda_{21}\left(\sigma(\x,\theta'),\sigma(\x,\theta''),
\sigma(\x,\theta)\right)}{\partial r \partial s} \nonumber \\
& & \text{exists and is a continuous function of } \nonumber \\
& & (\x,\theta',\theta'',\theta) \text{ throughout } D \times \Theta^{3},
\label{7.13}
\end{eqnarray}
we consider three distinct cases, (a), (b) and (c):
\begin{letterlist}
\item 
\begin{equation}
(\theta'',\theta) \in [0,\pi/2] \times [\pi,3\pi/2] \text{ or }
(\theta'',\theta) \in [\pi,3\pi/2] \times [0,\pi/2].
\end{equation}
\item 
\begin{equation}
\begin{array}{l}
(\theta'',\theta) \in [0,\pi/2]^{2} \text{ and }
\theta' \in [\pi,3\pi/2], \text{ or } \\
(\theta'',\theta) \in [\pi,3\pi/2]^{2} \text{ and }
\theta' \in [0,\pi/2].
\end{array}
\label{7.14e}
\end{equation}
\item 
\begin{equation}
(\theta',\theta'',\theta) \in [0,\pi/2]^{3} \text{ or }
(\theta',\theta'',\theta) \in [\pi,3\pi/2]^{3}.
\end{equation}
\end{letterlist}

In case (a) and case (b), it is easily seen that the denominator
$\sigma(\x,\theta'')-\sigma(\x,\theta)$ is different from zero, and hence 
\begin{equation}
\lambda_{21}\left(\sigma(\x,\theta'),\sigma(\x,\theta''),\sigma(\x,\theta)\right)
\text{ is a } \bC^{2} \text{ function of } (\x,\theta',\theta'',\theta),
\end{equation}
from which the desired conclusion follows.

In case (3) we employ 
\begin{eqnarray}
\frac{\partial\sigma(\x,\theta)}{\partial s} & = & 0 \text{ when }
\theta \in [0,\pi/2], 
\label{7.12c} \\
\text{and} & & \nonumber \\
\frac{\partial\sigma(\x,\theta)}{\partial r} & = & 0 \text{ when }
\theta \in [\pi,3\pi/2],
\label{7.12d}
\end{eqnarray}
to show that the mixed second derivative of $\lambda_{21}$ exists and  
equals zero.

\item 
From Eqs.\ (\ref{6.20c}) to (\ref{G5.3f}), 
\begin{eqnarray}
u_{1}(\x,\theta) & = & W_{1}(\sigma(\x,\theta)) 
- \frac{2}{\pi} \int_{\Theta} d\theta' p(\x,\theta')
L_{1}\left(\sigma(\x,\theta'),\sigma(\x,\theta)\right) \nonumber \\
& & \text{for all } (\x,\theta) \in D \times \Theta
\label{7.15a} \\
\text{and} & & \nonumber \\
\kappa_{21}(\x,\theta',\theta) & = &
q(\x,\theta') \left[ k_{21}\left(\sigma(\x,\theta'),\sigma(\x,\theta)\right)
- \frac{2}{\pi} \int_{\Theta} d\theta'' p(\x,\theta'') 
\lambda_{21}\left(\sigma(\x,\theta'),\sigma(\x,\theta)\right) 
\right] \nonumber \\
& & \text{ for all } (\x,\theta',\theta) \in D \times \Theta^{2},
\label{7.15b}
\end{eqnarray}
where $p(\x,\theta)$ is defined by Eq.\ (\ref{G5.3g}), and $q(\x,\theta)$
is defined by Eq.\ (\ref{6.20e}).  From statements (\ref{5.29a}) and
(\ref{5.29b}),
\begin{equation}
\begin{array}{l}
p(\x,\theta) \text{ and } q(\x,\theta) \text{ are real analytic } \nonumber \\
\text{functions of } (\x,\theta) \text{ throughout } D \times \Theta.
\end{array}
\label{7.15c}
\end{equation}
From statements (\ref{7.12e}), (\ref{7.12k}), (\ref{7.13})
and (\ref{7.15c}), it is clear that the functions $u_{1}$ and $\kappa_{21}$
whose values are given by Eqs.\ (\ref{7.15a}) and (\ref{7.15b}), respectively,
satisfy the conclusions of our lemma.
\end{arablist}
\cheers

\begin{definition}{Dfn.\ of a function that is $\bC^{N_{1},\ldots,N_{L}}$
on $X \subset R^{L}$}
Suppose that $X$ is an open subset of $R^{L}$ or a closed or semi-closed
subinterval of $R^{L}$, $x=(x^{1},\ldots,x^{L})$ denotes any point in $X$,
$T$ is a topological space, $t$ denotes any point in $T$, and $N_{1},\ldots,
N_{L}$ are $L$ non-negative integers.  Suppose, furthermore, that
$F:(X \times T) \rightarrow C$ and that, for each $L$-tuple of integers
$(n_{1},\ldots,n_{L})$ such that $0 \le n_{k} \le N_{k}$ for all
$k=1,\ldots,L$,
\begin{equation}
\partial_{1 \cdots L}^{n_{1} \cdots n_{L}} F(x,t)
:= \left( \frac{\partial}{\partial x^{1}} \right)^{n_{1}} \cdots
\left( \frac{\partial}{\partial x^{k}} \right)^{n_{k}} \cdots
\left( \frac{\partial}{\partial x^{L}} \right)^{n_{L}} F(x,t)
\end{equation}
exists and is a continuous function of $(x,t)$ throughout $X \times T$.
[It is understood that $(\partial/\partial x^{k})^{0} = 1$.]  Then, we
shall say that {\em $F$ is $\bC^{N_{1},\ldots,N_{L}}$ on $X$}.  

Also, if $F:X \rightarrow C$ and $\partial_{1 \cdots L}^{n_{1} \cdots n_{L}}
F(x)$ exists and is a continuous function of $x$ throughout $X$ for each
choice of $(n_{1},\ldots,n_{L})$ that satisfies $0 \le n_{k} \le N_{k}$
for all $1 \le k \le L$, then we shall say that {\em $F$ is
$\bC^{N_{1},\ldots,N_{L}}$ on $X$}.
\end{definition}

Note:  If $F$ is $\bC^{N_{1},\ldots,N_{L}}$ on $X$, then a theorem of the
calculus tells us that, for each $(n_{1},\ldots,n_{L})$ satisfying
$0 \le n_{k} \le N_{k}$ for all $1 \le k \le L$, the existence and value
of $\partial_{1 \cdots L}^{n_{1} \cdots n_{L}} F$ are unchanged when the
operator factors $\partial/\partial x^{k}$ are subject to any permutation.

The following lemma is applicable to a broad class of Fredholm integral
equations and is clearly capable of further generalization in several
directions.  A $2 \times 2$ matrix version of the lemma for the case
$L=2$ was covered in a paper by the authors on the initial value problem
for colliding gravitational plane wave pairs.\footnote{I.~Hauser and
F.~J.~Ernst, J.\ Math.\ Phys.\ {\bf 32}, 198 (1991), Sec.~V.}  As regards
the current notes, the lemma will play a key role in the proof of
Thm.~\ref{7.3B}.

\begin{lemma}[Fredholm minor $M$ and determinant $\Delta$]
\label{7.2B} \mbox{ } \\
Let $X$, $x$ and $N_{k} \, (k=1,\ldots,L)$ be assigned the same meanings
as in the preceding definition; and let $Y$ denote a compact, oriented,
$m$-dimensional differentiable manifold, $y$ denote any point in $Y$, and
$dy$ denote a volume element at point $y$ (the value of a distinguished
non-zero $m$-form at $y$).  Suppose that $K:X \times (Y \times Y) \rightarrow
C$ and $K$ is $\bC^{N_{1},\ldots,N_{L}}$ on $X$.  Let us regard $K$ as an
$L$-parameter family of Fredholm kernels that is employed in Fredholm
integral equations of the form
\begin{equation}
f(x,y) - \int_{Y} dy' f(x,y') K(x,y',y) = g(x,y) \text{ for all }
(x,y) \in X \times Y,
\end{equation}
where $X$ is the parameter space.  Then, the corresponding Fredholm minor
$M$ and Fredholm determinant $\Delta$ are $\bC^{N_{1},\ldots,N_{L}}$
on $X$.
\end{lemma}

\proof
The Fredholm construction of $M$ and $\Delta$ are given by
\begin{eqnarray}
M(x,y',y) & = & \sum_{n=0}^{\infty} \frac{(-1)^{n}}{n!} M^{(n)}(x,y',y), \\
\Delta(x) & = & \sum_{n=0}^{\infty} \frac{(-1)^{n}}{n!} \Delta^{(n)}(x), 
\end{eqnarray}
where
\begin{eqnarray}
M^{(0)}(x,y',y) & := & K(x,y',y),
\label{7.18c} \\
M^{(n)}(x,y',y) & := & \int_{Y} dy_{1} \cdots \int_{Y} dy_{n}
D^{(n+1)}\left(x \left| \begin{array}{cc}
y & y_{1} \cdots y_{n} \\ y' & y_{1} \cdots y_{n} 
\end{array} \right. \right) \nonumber \\
& & \text{ for all } n > 0,
\label{7.18d} \\
\Delta^{(0)}(x) & := & 1, \\
\Delta^{(n+1)}(x) & := & \int_{Y} dy M^{(n)}(x,y,y) \text{ for all } n \ge 0,
\end{eqnarray}
and
\begin{eqnarray}
D^{(n)}\left(x \left| \begin{array}{c}
y_{1} \cdots y_{n} \\ y'_{1} \cdots y'_{n} 
\end{array} \right. \right) & := & \text{the determinant of that } n \times n
\nonumber \\
& & \text{matrix whose element in the $k$th } \nonumber \\
& & \text{row and $l$th column is } K(x,y'_{k},y_{l}).
\end{eqnarray}
In particular,
\begin{equation}
D^{(0)}\left( x \left| \begin{array}{c}
y \\ y' 
\end{array} \right. \right) := K(x,y',y).
\end{equation}

For each bounded and closed subspace $U$ of $X$, let
\begin{eqnarray}
||K_{u}|| & := & \sup\left\{\left| \partial_{1 \cdots L}^{n_{1} \cdots n_{L}}
K(x,y',y) \right|:(x,y',y) \in U \times Y^{2}, \right. \nonumber \\
& & \left. \text{and } 0 \le n_{k} \le N_{k} \text{ for all }
k=1,\ldots,L\right\}.
\end{eqnarray}
Also let
\begin{equation}
V := \int_{Y} dy.
\end{equation}
Then, from Eqs.\ (\ref{7.18c}) and (\ref{7.18d}), and from a generalization
of Hadamard's inequality that was formulated and proved by the authors in the
aforementioned paper on the initial value problem for colliding gravitational
plane wave pairs [see Thm.~7 in that paper],
\begin{eqnarray}
\left| \partial_{1 \cdots L}^{n_{1} \cdots n_{L}} M^{n}(x,y',y) \right|
& \le & V^{n} ||K_{U}||^{n+1} (n+1)^{N_{1}+\ldots+N_{L}+(n+1)/2}
\nonumber \\
& & \text{for all } (x,y',y) \in U \times Y^{2} \text{ and all } \nonumber \\
& & (n_{1},\ldots,n_{L}) \text{ such that } 0 \le n_{k} \le N_{k} \nonumber \\
& & \text{for each } k=1,\ldots,L.
\end{eqnarray}
It follows that, for each positive integer $N$,
\begin{eqnarray}
\sum_{n=0}^{N} \frac{1}{n!} \left| \partial_{1 \cdots L}^{n_{1} \cdots n_{L}}
M^{(n)}(x,y',y) \right| & \le & \sum_{n=0}^{N} \frac{V^{n}||K_{U}||^{n+1}}
{n!} (n+1)^{N_{1}+\ldots+N_{L}+(n+1)/2} \nonumber \\
& & \text{for all } (x,y',y) \in U \times Y^{2} \text{ and all } \nonumber \\
& & \text{choices (the usual) of } (n_{1},\ldots,n_{l}).
\label{7.19d}
\end{eqnarray}
The application of the ratio test to the series on the right side of the
above inequality (\ref{7.19d}) is straightforward and deomonstrates that
this series converges as $N \rightarrow \infty$.  Hence, from the comparison
test, the series on the left side of (\ref{7.19d}) converges for all
$(x,y',y) \in U \times Y^{2}$ and all choices of $(n_{1},\ldots,n_{L})$.
The theorems\footnote{See Sec.~2, Ch.~IV, of {\em Differential and Integral
Calculus} by R.~Courant (Interscience Publishers, Inc., 1936).\label{Courant}}
of the calculus on the continuity and term-by-term differentiability of a
uniformly convergent infinite series of functions then supply us with the 
following conclusions:
\begin{equation}
\begin{array}{l}
\text{For all choices of } (n_{1},\ldots,n_{L}) \text{ for which }
0 \le n_{k} \le N_{k} (1 \le k \le L), \nonumber \\
\partial_{1 \cdots L}^{n_{1} \cdots n_{L}} M(x,y',y) \text{ exists and is 
a continuous function of } (x,y',y)
\nonumber \\
\text{throughout } X \times Y^{2};
\end{array}
\end{equation}
and
\begin{eqnarray}
\partial_{1 \cdots L}^{n_{1} \cdots n_{L}} M(x,y',y) & = & 
\sum_{n=0}^{\infty} \frac{(-1)^{n}}{n!} 
\left[ \partial_{1 \cdots L}^{n_{1} \cdots n_{L}} M^{(n)}(x,y',y) \right],
\text{ and } \nonumber \\
& & \text{the infinite series converges absolutely } \nonumber \\
& & \text{and converges uniformly on each } \nonumber \\
& & \text{compact subspace of } X \times Y^{2}.
\end{eqnarray}

Hence, $M$ is $\bC^{N_{1},\ldots,N_{L}}$ on $X$.  The proof that $\Delta$
is also $\bC^{N_{1},\ldots,N_{L}}$ on $X$ is left for the reader.
\cheers

The following theorem concerns the solution $y_{1}(\x,\theta)$ of the
Fredholm equation (\ref{6.20a}) for all $(\x,\theta) \in D \times \Theta$.

\begin{theorem}[Differentiability properties of $y_{1}$ when $\bv \in K^{3}$]
\label{7.3B} \mbox{ } \\
If $\bv \in K^{3}$, then $y_{1}$ is $\bC^{1,1}$ on $D$; i.e., 
$\partial y_{1}(\x,\theta)/\partial r$, $\partial y_{1}(\x,\theta)/\partial s$
and $\partial^{2}y_{1}(\x,\theta)/\partial r \partial s$ exist and are
continuous functions of $(\x,\theta)$ throughout $D \times \Theta$.
\end{theorem}

\proof
Consider the inhomogenous Fredholm equation of the second kind (\ref{6.20a}).
According to Thm.~\ref{6.4C}, the Fredholm determinant for Eq.\ (\ref{6.20a})
is not zero for all choices of $\x \in D$.  Therefore, a unique solution
of the Fredholm equation exists and is given by
\begin{equation}
y_{1}(\x,\theta) = u_{1}(\x,\theta) + \frac{2}{\pi} \int_{\Theta} d\theta'
u_{1}(\x,\theta') R(\x,\theta',\theta)
\label{7.20a}
\end{equation}
for all $(\x,\theta) \in D \times \Theta$, where the resolvent kernel
$R(\x,\theta',\theta)$ is the following ratio of the Fredholm minor and
determinant:
\begin{equation}
R(\x,\theta',\theta) = \frac{M(\x,\theta',\theta)}{\Delta(\x)}.
\end{equation}
From Lem.~\ref{7.1B}, $\kappa_{21}$ is $\bC^{1}$.  Moreover, $\partial^{2}
\kappa_{21}(\x,\theta',\theta)/\partial r \partial s$ exists and is a
continuous function of $(\x,\theta',\theta)$ throughout $D \times
\Theta^{2}$.  Therefore,
\begin{equation}
\kappa_{21} \text{ is } \bC^{1,1} \text{ on } D.
\end{equation}
The preceding Lem.~\ref{7.2B} is now applied to the present case, for which
\begin{equation}
X=D, \quad Y=\Theta, \quad L=2, \quad m=1, \quad dy=2d\theta/\pi.
\end{equation}
Thereupon, one obtains
\begin{equation}
R \text{ is } \bC^{1,1} \text{ on } D.
\label{7.20e}
\end{equation}
Lemma~\ref{7.1B} also tells us that (amongst other things)
\begin{equation}
u_{1} \text{ is } \bC^{1,1} \text{ on } D.
\label{7.20f}
\end{equation}
Therefore, from Eq.\ (\ref{7.20a}), statements (\ref{7.20e}) and (\ref{7.20f}),
and the theorems\footnote{See Ref.~\ref{Courant}.} of the calculus on the
continuity and differentiability of an integral with respect to parameters,
\begin{equation}
y_{1} \text{ is } \bC^{1,1} \text{ on } D.
\end{equation}
\cheers

Note that, in terms of standard notation and terminology, $\lambda=1$ for
our particular Fredholm equation; and the statement that $\Delta(\x) \ne 0$
is equivalent to the statement that $1$ is not a characteristic value
(eigenvalue) of our kernel.

\setcounter{equation}{0}
\subsection{Concerning the partial derivatives of $\Y$, $\F$, $H$ and
$\F^{\pm}$ when $\bv \in K^{3}$}

\begin{definition}{Dfn.\ of $L^{(i)}(\sigma',\sigma)$ for each $\x \in D$
and $i \in \{3,4\}$}
For each $\sigma' \in \bar{\I}(\x)$ and $\sigma \in \I^{(i)}$, let
$$
L^{(i)}(\sigma',\sigma) := \frac{W(\sigma')-W^{(i)}(\sigma)}{\sigma'-\sigma}.
$$
\end{definition}

Employing the transformation defined by Eqs.\ (\ref{5.26a}) to (\ref{5.26i}),
the definition of $p(\x,\theta)$ by Eq.\ (\ref{G5.3g}), the definition of
$q(\x,\theta)$ by Eq.\ (\ref{6.20e}) and the definition of 
$L^{(i)}(\sigma',\sigma)$ that we just gave, one finds that Eqs.\ 
(\ref{5.30a}), (\ref{5.30b}) and (\ref{5.24b}) are expressible in the forms
(in which `$\x$' is no longer suppressed)
\begin{eqnarray}
\Y^{(i)}_{1}(\x,\sigma) & = & W^{(i)}_{1}(\sigma) - \frac{2}{\pi} \int_{\Theta}
d\theta' p(\x,\theta') y_{2}(\x,\theta') W_{1}^{T}(\sigma(\x,\theta')) J
L^{(i)}_{1}(\sigma(\x,\theta'),\sigma) \nonumber \\
& & \text{for all } \x \in D \text{ and } \sigma \in
\check{\I}^{(i)}(x^{7-i}) \nonumber \\
& & \text{[after the extension defined by (\ref{5.31a})],}
\label{7.21a} \\
\Y^{(i)}_{2}(\x,\sigma) & = & W^{(i)}_{2}(\sigma) + \frac{2}{\pi} \int_{\Theta}
d\theta' q(\x,\theta') y_{1}(\x,\theta') W_{2}^{T}(\sigma(\x,\theta')) J
L^{(i)}_{2}(\sigma(\x,\theta'),\sigma) \nonumber \\
& & \text{for all } \x \in D \text{ and } \sigma \in
\check{\I}^{(i)}(x^{7-i}) \nonumber \\
& & \text{[after the extension defined by (\ref{5.31a})],} 
\label{7.21b}
\end{eqnarray}
and
\begin{eqnarray}
\nu(\x,\tau)^{-1} \F(\x,\tau) & = & I - \frac{2}{\pi}
\int_{\Theta} d\theta' q(\x,\theta') y_{1}(\x,\theta')
\frac{W_{2}(\sigma(\x,\theta')) J}{\sigma(\x,\theta')-\tau} \nonumber \\
& & \text{for all } \x \in D \text{ and } \tau \in C - \bar{\I}(\x).
\label{7.22a}
\end{eqnarray}
Furthermore, from Eqs.\ (\ref{7.1a}), (\ref{7.2a}) and (\ref{7.22a}),
\begin{eqnarray}
H(\x) & = & H^{M}(\x_{0}) + 2(z-z_{0})\Omega - \frac{4i}{\pi} \int_{\Theta}
d\theta' q(\x,\theta') y_{1}(\x,\theta') W_{2}^{T}(\sigma(\x,\theta'))
\nonumber \\
& & \text{for all } \x \in D.
\end{eqnarray}

When proving the following theorem, one should bear in mind that
$\sigma(\x,\theta)$, $p(\x,\theta)$ and $q(\x,\theta)$ are analytic
functions of $(\x,\theta)$ throughout $D \times \Theta$.

\begin{theorem}[Differentiability properties of $\Y^{(i)}$, $\F$
and $H$ when $\bv \in K^{3}$]
\label{7.1D} \mbox{ } \\
If $\bv \in K^{3}$, then
\begin{equation}
\begin{array}{l}
\partial\Y^{(i)}(\x,\sigma)/\partial r, \;
\partial\Y^{(i)}(\x,\sigma)/\partial s, \;
\partial^{2}\Y^{(i)}(\x,\sigma)/\partial r \partial s, \\
\partial^{2}\Y^{(i)}(\x,\sigma)/\partial \sigma^{2}, \;
\partial^{2}\Y^{(i)}(\x,\sigma)/\partial r \partial \sigma, \text{ and }
\partial^{2}\Y^{(i)}(\x,\sigma)/\partial s \partial \sigma \\
\text{exist and are continuous functions of $(\x,\sigma)$ throughout } \\
\{(\x,\sigma):\x \in D, \sigma \in \check{\I}^{(i)}(x^{7-i})\}.
\end{array}
\label{7.23}
\end{equation}
Also, upon letting $\grave{\F}$ denote the restriction of $\F$ to
$$
\dom{\grave{\F}} := \{(\x,\tau):\x \in D, \tau \in 
C-\I(\x)-\{r,s,r_{0},s_{0}\}\},
$$
one has
\begin{equation}
\begin{array}{l}
\partial\grave{\F}(\x,\tau)/\partial r, \;
\partial\grave{\F}(\x,\tau)/\partial s, \text{ and }
\partial^{2}\grave{\F}(\x,\tau)/\partial r \partial s \\
\text{exist and are continuous functions of } (\x,\tau) \\
\text{throughout } \dom{\grave{\F}}; \text{ and, for each } \x \in D,
\text{ these } \\
\text{partial derivatives are holomorphic functions } \\
\text{of $\tau$ throughout } C - \I(\x)-\{r,s,r_{0},s_{0}\}.
\end{array}
\label{7.24a}
\end{equation}
Furthermore,
\begin{equation}
H \text{ is } \bC^{1,1} \text{ on } D.
\label{7.24b}
\end{equation}
\end{theorem}

\proof
From Thm.~\ref{7.3B}, statement (\ref{7.12e}) and the fact that $L^{(i)}$
is $\bC^{2}$, one concludes from Eq.\ (\ref{7.21b}) that
$$
\begin{array}{l}
\partial\Y^{(i)}_{2}(\x,\sigma)/\partial r, \quad 
\partial\Y^{(i)}_{2}(\x,\sigma)/\partial s, \\
\partial^{2}\Y^{(i)}_{2}(\x,\sigma)/\partial r \partial s, \quad
\partial^{2}\Y^{(i)}_{2}(\x,\sigma)/\partial \sigma^{2}, \\
\partial^{2}\Y^{(i)}_{2}(\x,\sigma)/\partial r \partial \sigma \text{ and }
\partial^{2}\Y^{(i)}_{2}(\x,\sigma)/\partial s \partial \sigma 
\end{array}
$$
exist and are continuous functions of $(\x,\sigma)$ throughout
$\{(\x,\sigma):\x \in D, \sigma \in \check{\I}^{(i)}(x^{7-i})\}$.
Then, from Eq.\ (\ref{7.21a}), one obtains like conclusions for
$\Y^{(i)}_{1}(\x,\sigma)$, whereupon the statement (\ref{7.23}) follows.

Statements (\ref{7.24a}) and (\ref{7.24b}) follow from Thm.~\ref{7.3B},
statement (\ref{7.12e}), the known differentiability and holomorphy
properties of $\nu(\x,\tau)^{-1}$ on $\dom{\grave{\F}}$, and the theorem 
on the holomorphy of functions given by Cauchy-type integrals.
\cheers

The following two lemmas will be used to prove Thm.~\ref{7.4D}.

\begin{lemma}[$d(\nu(\x,\tau)^{-1} \grave{\F}(\x,\tau))$]
\label{7.2D} \mbox{ } \\
If $\bv \in K^{3}$, then the first partial derivatives of 
\begin{equation}
\frac{\nu^{+}(\x,\sigma')^{-1} \Y_{1}(\x,\sigma') W_{2}^{T}(\sigma') J}
{\sigma'-\tau}
\label{7.25a}
\end{equation}
with respect to $r$ and with respect to $s$ are summable over $\sigma'
\in \bar{\I}(\x)$; and
\begin{eqnarray}
d\left[\nu(\x,\tau)^{-1} \grave{\F}(\x,\tau)\right] & = & 
- \frac{1}{\pi i} \int_{\bar{\I}} d\sigma'
\frac{d\left[\nu^{+}(\x,\sigma')^{-1}\Y_{1}(\x,\sigma')\right]
W_{2}^{T}(\sigma')J}{\sigma'-\tau} \nonumber \\
& & \text{for all } (\x,\tau) \in \dom{\grave{\F}}.
\label{7.25b}
\end{eqnarray}
\end{lemma}

\proof
We shall tacitly employ statements (\ref{7.23}) and (\ref{7.24a}) of
Thm.~\ref{7.1D} in some steps of this proof.  We shall supply the proof
only for 
$\partial[\nu(\x,\tau)^{-1}\grave{\F}(\x,\tau)]/\partial r$
and leave the proof for the partial derivative with respect to $s$ for
the reader.  The summability over $\bar{\I}(\x)$ of the partial derivative
with respect to $r$ of (\ref{7.25a}) is seen from the facts that
\begin{eqnarray}
\nu^{+}(\x,\sigma')^{-1} & = & \M^{+}(\sigma'-r)\M^{+}(\sigma'-s)
\left[\M^{+}(\sigma'-r_{0})\M^{+}(\sigma'-s_{0})\right]^{-1} \\
\text{and} & & \nonumber \\
\frac{\partial\nu^{+}(\x,\sigma')^{-1}}{\partial r} & = & 
-\frac{1}{2}\M^{+}(\sigma'-s) \left[\M^{+}(\sigma'-r)
\M^{+}(\sigma'-r_{0})\M^{+}(\sigma'-s_{0})\right]^{-1},
\end{eqnarray}
where 
$$
\M^{+}(\sigma) := \left\{ \begin{array}{rcl}
\sqrt{\sigma} & \text{if} & \sigma \ge 0, \\
i\sqrt{\sigma} & \text{if} & \sigma \le 0,
\end{array} \right.
$$
are both summable over $\bar{\I}(\x)$, and a summable function times a
continuous function over a bounded interval is summable.

In the proofs of this lemma and the next lemma, we shall employ the
shorthand notations
\begin{equation}
\begin{array}{rcl}
f(\x,\sigma') & := & \Y_{1}(\x,\sigma') W_{2}^{T}(\sigma') J, \\
g(\x,\sigma') & := & \nu(\x,\tau)^{-1} \F(\x,\tau),
\label{7.26a}
\end{array}
\end{equation}
whereupon Eq.\ (\ref{5.24b}) becomes
\begin{eqnarray}
g(\x,\tau) & = & I - \frac{1}{\pi i} \int_{\bar{\I}} d\sigma'
\nu^{+}(\x,\sigma')^{-1} \frac{f(\x,\sigma')}{\sigma'-\tau} \nonumber \\
& = & I - g_{1}(\x,\tau) - f(\x,r) g_{2}(\x,\tau),
\label{7.26b}
\end{eqnarray}
where
\begin{eqnarray}
g_{1}(\x,\tau) & := & \frac{1}{\pi i} \int_{\bar{\I}} d\sigma'
\nu^{+}(\x,\sigma')^{-1} \frac{f(\x,\sigma')-f(\x,r)}{\sigma'-\tau},
\label{7.26c} \\
\text{and} & & \nonumber \\
g_{2}(\x,\tau) & := & \frac{1}{\pi i} \int_{\bar{\I}} d\sigma'
\frac{\nu^{+}(\x,\sigma')^{-1}}{\sigma'-\tau}.
\label{7.26d}
\end{eqnarray}

We shall first deal with the term $f(\x,r)g_{2}(\x,\tau)$.
It is easy to show that 
\begin{equation}
g_{2}(\x,\tau) = \nu(\x,\tau)^{-1} - 1.
\label{7.26e}
\end{equation}
Therefore, for all $(\x,\tau) \in \dom{\grave{\F}}$,
\begin{equation}
\frac{\partial g_{2}(\x,\tau)}{\partial r} = - \frac{1}{2(\tau-r)}
\nu(\x,\tau)^{-1}.
\label{7.26f}
\end{equation}
Also, note that
\begin{eqnarray}
\frac{1}{\pi i} \int_{\bar{\I}} d\sigma' \frac{\partial
\nu^{+}(\x,\sigma')^{-1}/\partial r}{\sigma'-\tau} & = &
- \frac{1}{2\pi i} \int_{\bar{\I}} d\sigma'
\frac{\nu^{+}(\x,\sigma')^{-1}}{(\sigma'-r)(\sigma'-\tau)} \nonumber \\
& = & - \frac{\nu(\x,\tau)^{-1}}{2(\tau-r)}.
\label{7.26g}
\end{eqnarray}
So, from Eqs.\ (\ref{7.26d}), (\ref{7.26f}) and (\ref{7.26g}),
\begin{eqnarray}
\frac{\partial}{\partial r} \left[ f(\x,r)g_{2}(\x,\tau) \right] & = &
\frac{1}{\pi i} \int_{\bar{\I}} d\sigma'
\frac{\partial\left[\nu^{+}(\x,\sigma')^{-1}f(\x,r)\right]/\partial r}
{\sigma'-\tau} \nonumber \\
& & \text{for all } (\x,\tau) \in \dom{\grave{\F}}.
\label{7.26h}
\end{eqnarray}
That takes care of the term $f(\x,r)g_{2}(\x,\tau)$. 

We shall next deal with the term $g_{1}(\x,\tau)$.  From statement
(\ref{7.23}) in Thm.~\ref{7.1D} and from Eq.\ (\ref{7.26a}), one can
see that
\begin{equation}
\frac{\partial}{\partial r} \left\{ \M^{+}(\sigma'-r)\M^{+}(\sigma'-s)
\left[f(\x,\sigma')-f(\x,r)\right]\right\}
\end{equation}
exists and is a continuous function of $(\x,\sigma')$ throughout
$\{(\x,\sigma'):\x \in D, \sigma' \in \bar{\I}(\x)\}$.  [We leave
details for the reader.]  No loss of generality will be incurred if we
tentatively introduce a closed and bounded convex neighborhood $\N$ of
the point $\x_{0}$ in the space $D$, whereupon it is seen that
$$
\{(\x,\sigma'):\x \in \N, \sigma' \in \bar{\I}(\x)\}
$$
is a bounded closed subspace of $R^{3}$; and, therefore,
\begin{eqnarray}
\lefteqn{M(\N) := } \\
& & \sup \left\{ \left|\left| \frac{\partial}{\partial r}
\left\{ \M^{+}(\sigma'-r)\M^{+}(\sigma'-s) \left[f(\x,\sigma')-f(\x,r)\right]
\right\}\right|\right|:\x \in \N,\sigma' \in \bar{\I}(\x)\right\} \nonumber
\end{eqnarray}
is finite; and the integrand in the expression for $g_{1}(\x,\tau)$ that is
given by Eq.\ (\ref{7.26c}) satisfies
\begin{equation}
\left|\left|\frac{\partial}{\partial r} \left[\nu^{+}(\x,\sigma')^{-1}
\frac{f(\x,\sigma')-f(\x,r)}{\sigma'-\tau}\right]\right|\right| \le
\left[\sqrt{|\sigma'-r_{0}||\sigma'-s_{0}|}\;|\sigma'-\tau|\right]^{-1}
M(\N).
\end{equation}
Since the right side of the above inequality is summable over $\bar{\I}(\x)$
and is independent of $\x$, a well-known theorem\footnote{See
Ref.~\ref{McShane}, Sec.\ 39.} on differentiation of a Lebesgue integral
with respect to a parameter tells us that 
$\partial g_{1}(\x,\tau)/\partial r$ exists
(which, it happens, we already know) and is given by
\begin{equation}
\frac{\partial g_{1}(\x,\tau)}{\partial r} = \frac{1}{\pi i} \int_{\bar{\I}}
d\sigma' \frac{\frac{\partial}{\partial r} \left\{ \nu^{+}(\x,\sigma')^{-1}
\left[f(\x,\sigma')-f(\x,r)\right]\right\}}{\sigma'-\tau}
\label{7.27d}
\end{equation}
for all $\x \in \N$ and $\tau \in C - \I(\x)-\{r,s,r_{0},s_{0}\}$, where we
have used the
fact that the contribution to $\partial g_{1}(\x,\tau)/\partial r$ due to
differentiation of the integral with respect to the endpoint $r \in
\{a^{3},b^{3}\}$ of the integration interval $\bar{\I}^{(3)}(\x)$ vanishes,
because the integrand in Eq.\ (\ref{7.26c}) vanishes when $\sigma'=r$.

However, since $\N$ can always be chosen so that it covers any given point
in $D$, Eq.\ (\ref{7.27d}) holds for all $(\x,\tau) \in \dom{\grave{\F}}$;
and upon combining (\ref{7.27d}), (\ref{7.26h}) and (\ref{7.26b}), one
obtains
\begin{eqnarray}
\frac{\partial g(\x,\tau)}{\partial r} & = & - \frac{1}{\pi i} \int_{\bar{\I}}
d\sigma' \frac{\partial[\nu^{+}(\x,\sigma')^{-1}f(\x,\sigma')]/\partial r}
{\sigma'-\tau} \nonumber \\
& & \text{for all } (\x,\tau) \in \dom{\grave{\F}},
\end{eqnarray}
which is the coefficient of $dr$ in Eq.\ (\ref{7.25b}).
\cheers

Before we give the next lemma, note that application of the Plemelj relations
to Eq.\ (\ref{5.24b}) yields
\begin{eqnarray}
\frac{1}{2}[\F^{+}(\x,\sigma)+\F^{-}(\x,\sigma)] & = & 
- \Y_{1}(\x,\sigma) W_{2}^{T}(\sigma) J \nonumber \\
& & \text{for all } \x \in D \text{ and } \sigma \in \I(\x), 
\label{7.28a} \\
\text{and} & & \nonumber \\
\frac{1}{2}\nu^{+}(\x,\sigma)^{-1} [\F^{+}(\x,\sigma)-\F^{-}(\x,\sigma)]
& = & I - \frac{1}{\pi i} \int_{\bar{\I}} d\sigma' \nu^{+}(\x,\sigma')^{-1}
\frac{\Y_{1}(\x,\sigma')W_{2}^{T}(\sigma')J}{\sigma'-\sigma} \nonumber \\
& & \text{for all } \x \in D \text{ and } \sigma \in \I(\x).
\label{7.28b}
\end{eqnarray}

\begin{lemma}[Differentiability properties of $\F^{\pm}$ when $\bv \in K^{3}$]
\label{7.3D} \mbox{ } \\
As in the preceding lemma, suppose that $\bv \in K^{3}$ and $\F$ is
the solution of the HHP corresponding to $(\bv,\F^{M})$.  Then the
following three statements hold:
\begin{romanlist}
\item 
The partial derivatives $\partial \F^{\pm}(\x,\sigma)/\partial r$,
$\partial \F^{\pm}(\x,\sigma)/\partial s$ and 
$\partial^{2}\F^{\pm}(\x,\sigma)/\partial r \partial s$ exist and are
continuous functions of $(\x,\sigma)$ throughout 
$\{(\x,\sigma):\x \in D, \sigma \in \I(\x)\}$.
\item 
The $1$-form
\begin{equation}
\frac{d[\nu^{+}(\x,\sigma')^{-1} \Y_{1}(\x,\sigma')] W_{2}^{T}(\sigma') J}
{\sigma'-\sigma}
\label{7.29a}
\end{equation}
is, for each $\x \in D$ and $\sigma \in \I(\x)$, summable over
$\bar{\I}(\x)$ in the PV sense.
\item 
For all $\x \in D$ and $\sigma \in \I(\x)$,
\begin{eqnarray}
\lefteqn{d \left\{ \frac{1}{2}\nu^{+}(\x,\sigma)^{-1} [\F^{+}(\x,\sigma)
-\F^{-}(\x,\sigma)] \right\} = } \nonumber \\
& & -\frac{1}{\pi i} \int_{\bar{\I}} d\sigma'
\frac{d[\nu^{+}(\x,\sigma')^{-1}\Y_{1}(\x,\sigma')]W_{2}^{T}(\sigma')J}
{\sigma'-\sigma}.
\label{7.29b}
\end{eqnarray}
\end{romanlist}
\end{lemma}

\proofs
\begin{romanlist}
\item 
This follows from statement (\ref{7.23}), Eq.\ (\ref{7.28a}) and Eq.\ 
(\ref{5.25b}).
\cheers

The proofs of parts (ii) and (iii) will be supplied only for the
coefficients of $dr$ in Eqs.\ (\ref{7.29a}) and (\ref{7.29b}).  The
proofs for the coefficients of $ds$ are left to the reader.
\item 
As functions of $\sigma'$, $W_{2}^{T}(\sigma')$ is $\bC^{3}$,
$\Y_{1}(\x,\sigma')$ is $\bC^{2}$ and $\partial\Y_{1}(\x,\sigma')/\partial r$
is $\bC^{1}$ on $\bar{\I}(\x)$; and $\nu^{+}(\x,\sigma')^{-1}$ and
$\partial\nu^{+}(\x,\sigma')^{-1}/\partial r$ are summable over
$\bar{\I}(\x)$.  Therefore, for a sufficiently small $\epsilon > 0$,
\begin{equation}
\frac{\frac{\partial}{\partial r} \left[ \nu^{+}(\x,\sigma')^{-1}
\Y_{1}(\x,\sigma') \right] W_{2}^{T}(\sigma') J}{\sigma''-\sigma}
\label{7.30a}
\end{equation}
is summable over $\bar{\I}(\x) - ]\sigma-\epsilon,\sigma+\epsilon[$.
Moreover, since the numerator of (\ref{7.30a}) is a $\bC^{1}$ function of
$\sigma'$, it is well known that (\ref{7.30a}) is summable over
$[\sigma-\epsilon,\sigma+\epsilon]$ in the PV sense. 

Therefore, (\ref{7.30a}) is summable over $\bar{\I}(\x)$ in the PV sense.
\cheers
\item 
In terms of the shorthand notations (\ref{7.26a}), Eq.\ (\ref{7.28b}) is
expressible in the form
\begin{equation}
\frac{1}{2}[g^{+}(\x,\sigma)+g^{-}(\x,\sigma)] = I - \frac{1}{\pi i}
\int_{\bar{\I}} d\sigma' \nu^{+}(\x,\sigma')^{-1}
\frac{f(\x,\sigma')}{\sigma'-\sigma},
\label{7.30b}
\end{equation}
where Thm.~\ref{7.1D} furnishes the following properties of $f(\x,\sigma')$:
\begin{equation}
\begin{array}{l}
\partial f(\x,\sigma')/\partial r, \;
\partial f(\x,\sigma')/\partial s, \;
\partial^{2}f(\x,\sigma')/(\partial \sigma')^{2}, \\
\partial^{2}f(\x,\sigma')/\partial r \partial s, \; 
\partial^{2}f(\x,\sigma')/\partial r \partial \sigma' \text{ and } 
\partial^{2}f(\x,\sigma')/\partial s \partial \sigma' \\
\text{exist and are continuous functions of } (\x,\sigma') \\
\text{throughout } \{(\x,\sigma'):\x \in D, \sigma' \in \bar{\I}(\x)\}.
\end{array}
\label{7.30c}
\end{equation}
Let us introduce the additional shorthand notations
\begin{eqnarray}
f_{0}(\x,\sigma',\sigma) & := & 
\frac{f(\x,\sigma')-f(\x,\sigma)}{\sigma'-\sigma},
\label{7.30d} \\
f_{1}(\x,\sigma',\sigma) & := & f_{0}(\x,\sigma',\sigma)
-f_{0}(\x,r,\sigma), 
\label{7.30e} \\
g_{1}(\x,\sigma) & := & \frac{1}{\pi i} \int_{\bar{\I}} d\sigma'
\nu^{+}(\x,\sigma')^{-1} f_{1}(\x,\sigma',\sigma),
\label{7.30f} \\
g_{2}(\x,\sigma) & := & \frac{1}{\pi i} \int_{\bar{\I}} d\sigma'
\frac{\nu^{+}(\x,\sigma')^{-1}}{\sigma'-\sigma} \\
\text{and} & & \nonumber \\
g_{3}(\x,\sigma) & := & \frac{1}{\pi i} \int_{\bar{\I}} d\sigma'
\nu^{+}(\x,\sigma')^{-1}.
\end{eqnarray}
Then Eq.\ (\ref{7.30b}) is expressible in the form
\begin{equation}
\frac{1}{2}[g^{+}(\x,\sigma)+g^{-}(\x,\sigma)] = I - g_{1}(\x,\sigma)
- f(\x,\sigma) g_{2}(\x,\sigma) - f_{0}(\x,r,\sigma) g_{3}(\x,\sigma).
\label{7.30i}
\end{equation}

Let us first consider the above terms that contain $g_{2}$ and $g_{3}$.
A well-known formula yields
\begin{equation}
g_{2}(\x,\sigma) = -1,
\end{equation}
while the usual contour integration technique yields
\begin{equation}
g_{3}(\x,\sigma) = \frac{1}{2}(r+s-r_{0}-s_{0}).
\end{equation}
Therefore, by using 
\begin{equation}
\frac{1}{\pi i} \int_{\bar{\I}} d\sigma'' \nu^{+}(\sigma'')^{-1}
(\sigma''-\sigma)^{-1}(\sigma'-\sigma'')^{-1} = 0 \text{ for all }
\sigma \in \bar{\I}(\x)-\{r_{0},s_{0}\}.
\label{6.10d}
\end{equation}
and the fact that
$$
\partial\nu^{+}(\x,\sigma')^{-1}/\partial r = 
-\frac{1}{2}(\sigma'-r)^{-1} \nu^{+}(\x,\sigma')^{-1}, 
$$
the reader can prove that
\begin{eqnarray}
\frac{\partial g_{2}(\x,\sigma)}{\partial r} & = & 
\frac{1}{\pi i} \int_{\bar{\I}} d\sigma' 
\frac{\partial\nu^{+}(\x,\sigma')^{-1}/\partial r}{\sigma'-\sigma}
\\ \text{and} & & \nonumber \\
\frac{\partial g_{3}(\x,\sigma)}{\partial r} & = & 
\frac{1}{\pi i} \int_{\bar{\I}} d\sigma' 
\partial\nu^{+}(\x,\sigma')^{-1}/\partial r,
\end{eqnarray}
whereupon
\begin{eqnarray}
\frac{\partial[f(\x,\sigma)g_{2}(\x,\sigma)]}{\partial r} & = & \frac{1}{\pi i}
\int_{\bar{\I}} d\sigma'
\frac{\partial[f(\x,\sigma)\nu^{+}(\x,\sigma')^{-1}]/\partial r}
{\sigma'-\sigma},
\label{7.31e} \\
\text{and} & & \nonumber \\
\frac{\partial[f(\x,\sigma)g_{3}(\x,\sigma)]}{\partial r} & = & \frac{1}{\pi i}
\int_{\bar{\I}} d\sigma'
\frac{\partial}{\partial r} \left[f_{0}(\x,r,\sigma)\nu^{+}(\x,\sigma')^{-1}
\right].
\label{7.31f}
\end{eqnarray}
That completes the analysis of the terms in Eq.\ (\ref{7.30i}) that contain
$g_{2}$ and $g_{3}$.  

We next consider $g_{1}$.  From (\ref{7.30c}) to (\ref{7.30e}), one sees that
\begin{equation}
\begin{array}{l}
\text{For each } \sigma \in \I(\x), \partial f_{1}(\x,\sigma',\sigma)/\partial r
\text{ and } (\sigma'-r)^{-1} f_{1}(\x,\sigma',\sigma) \\
\text{exist and are continuous functions of } (\x,\sigma') \\
\text{throughout } \{(\x,\sigma'):\x \in D, \sigma' \in \bar{\I}(\x)\}.
\end{array}
\end{equation}
Therefore, as regards the integrand in the definition (\ref{7.30f}) of
$g_{1}(\x,\sigma)$, one readily deduces (by an argument similar to the one
used in the proof of the preceding lemma) that, corresponding to each closed
and bounded neighborhood $\N$ of the point $\x_{0}$ in the space $D$,
and each $\sigma \in \I(\x)$, there exists a positive real number
$M(\N,\sigma)$ such that
\begin{eqnarray}
\left|\left|\frac{\partial}{\partial r} \left[\nu^{+}(\x,\sigma')^{-1}
f_{1}(\x,\sigma',\sigma)\right]\right|\right| & \le &
\frac{M(\N,\sigma)}{\sqrt{|\sigma'-r_{0}||\sigma'-s_{0}|}} \nonumber \\
& & \text{for all } \x \in \N \text{ and } \sigma' \in \bar{\I}(\x)
- \{r_{0},s_{0}\}.
\end{eqnarray}
The remainder of the proof employs the same theorem on differentiation of a
Lebesgue integral with respect to a parameter that was used in the proof of
the preceding lemma.  The result is
\begin{equation}
\frac{\partial g_{1}(\x,\sigma)}{\partial r} = \frac{1}{\pi i} \int_{\bar{\I}}
d\sigma' \frac{\partial}{\partial r} \left[ \nu^{+}(\x,\sigma')^{-1}
f_{1}(\x,\sigma',\sigma) \right].
\label{7.32c}
\end{equation}
Upon combining the results given by Eqs.\ (\ref{7.31e}), (\ref{7.31f})
and (\ref{7.32c}), one obtains with the aid of Eqs.\ (\ref{7.30b}),
(\ref{7.30d}) to (\ref{7.30f}), and Eq.\ (\ref{7.30i}),
\begin{eqnarray}
\frac{\partial}{\partial r} \frac{1}{2} [g^{+}(\x,\sigma)+g^{-}(\x,\sigma)]
& = & -\frac{1}{\pi i} \int_{\bar{\I}} d\sigma'
\frac{\partial[\nu^{+}(\x,\sigma')^{-1}f(\x,\sigma')]/\partial r}
{\sigma'-\sigma} \nonumber \\
& & \text{for all } \x \in D \text{ and } \sigma \in \I(\x).
\end{eqnarray}
\cheers
\end{romanlist}

The point of the preceding two lemmas is the following crucial theorem.

\begin{theorem}[Limits of $d\F$ when $\bv \in K^{3}$]
\label{7.4D} \mbox{ } \\
Suppose $\bv \in K^{3}$ and $\F$ is the solution of the HHP
corresponding to $(\bv,\F^{M})$.  Then, the following three
statements hold:
\begin{romanlist}
\item 
For each $\x \in D$ and $\sigma \in \I(\x)$, $d\grave{\F}(\x,\sigma\pm\zeta)$
converges as $\zeta \rightarrow 0$ ($\Im{\zeta}>0$) and
\begin{equation}
\lim_{\zeta \rightarrow 0} d\grave{\F}(\x,\sigma\pm\zeta) = d\F^{\pm}(\x,\sigma).
\end{equation}
Note: The existences of $d\grave{\F}(\x,\tau)$ and $d\F^{\pm}(\x,\sigma)$ are
guaranteed by Thm.~\ref{7.1D} [statement (\ref{7.24a})] and by
Lem.~\ref{7.3D}(i), respectively.
\item 
$\F(\x,\tau)$ converges as $\tau \rightarrow r_{0}$ and as $\tau 
\rightarrow s_{0}$ [$\tau \in C - \bar{\I}(\x)$]; and
$\nu(\x,\tau)^{-1} \F(\x,\tau)$ converges as $\tau \rightarrow
r$ and as $\tau \rightarrow s$.
\item 
For each $i \in \{3,4\}$,
\begin{equation}
(\tau-x^{i}) \frac{\partial \grave{\F}(\x,\tau)}{\partial x^{i}}
\end{equation}
converges as $\tau \rightarrow r_{0}$ and as $\tau \rightarrow s_{0}$, while
\begin{equation}
\nu(\x,\tau)^{-1} (\tau-x^{i}) \frac{\partial \grave{\F}(\x,\tau)}
{\partial x^{i}}
\end{equation}
converges as $\tau \rightarrow r$ and as $\tau \rightarrow s$.
\end{romanlist}
\end{theorem}

\proofs
\begin{romanlist}
\item 
We shall prove statement (i) for the coefficient of $dr$ in
$d\grave{\F}(\x,\tau)$ and leave the proof for the coefficient of $ds$ to
the reader.

Employ the shorthand notation
\begin{equation}
f(\x,\sigma') := \Y_{1}(\x,\sigma') W_{2}^{T}(\sigma') J
\label{7.34a}
\end{equation}
in the integrand of Eq.\ (\ref{5.24b}), which then becomes
\begin{eqnarray}
\nu(\x,\tau)^{-1} \F(\x,\tau) & = & I - \frac{1}{\pi i}
\int_{\bar{\I}} d\sigma' \nu(\x,\sigma')^{-1}
\frac{f(\x,\sigma')}{\sigma'-\tau} \nonumber \\
& & \text{for all } (\x,\tau) \in \dom{\F},
\label{7.34b}
\end{eqnarray}
whereupon, from Eq.\ (\ref{7.25b}) in Lem.~\ref{7.2D}, and from Eq.\
(\ref{7.26g}), 
\begin{eqnarray}
\lefteqn{-\frac{\nu(\x,\tau)^{-1}}{2(\tau-r)} \F(\x,\tau)
+ \nu(\x,\tau)^{-1} \frac{\partial\F(\x,\tau)}{\partial r}
= } \nonumber \\
& & \mbox{ } - \Phi(\x,\tau) + \frac{\nu(\x,\tau)^{-1}}{2(\tau-r)}
f(\x,r) \nonumber \\
& & \text{for all } (\x,\tau) \in \dom{\grave{\F}},
\label{7.34c}
\end{eqnarray}
where
\begin{eqnarray}
\Phi(\x,\tau) & := & \frac{1}{\pi i} \int_{\bar{\I}} d\sigma'
\nu^{+}(\x,\sigma')^{-1} \frac{\phi(\x,\sigma')}{\sigma'-\tau}
\label{7.34d} \\
\text{and} & & \nonumber \\
\phi(\x,\sigma') & := & \frac{\partial f(\x,\sigma')}{\partial r}
- \frac{f(\x,\sigma')-f(\x,r)}{2(\sigma'-r)}.
\label{7.34e}
\end{eqnarray}
From Eq.\ (\ref{7.34e}) and the properties of $f(\x,\sigma')$ given by
statement (\ref{7.30c})
\begin{equation}
\begin{array}{l}
\partial\phi(\x,\sigma')/\partial\sigma' \text{ exists and is a continuous
function of } \\
(\x,\sigma')
\text{ throughout } \{(\x,\sigma'):\x \in D, \sigma' \in \bar{\I}(\x)\}.
\end{array}
\label{7.34f}
\end{equation}
Therefore, $\nu(\x,\sigma')^{-1} \phi(\x,\sigma')$ obeys a
H\"{o}lder condition of index $1$ on each closed subinterval of $\I(\x)$;
and it follows from the theorem in Sec.~16 in Muskhelishvili's 
treatise\footnote{See footnote \ref{Musk}.} that (\ref{7.34d}) satisfies
\begin{equation}
\Phi^{\pm}(\x,\sigma) := \lim_{\zeta \rightarrow 0} \Phi(\x,\sigma\pm\zeta)
\text{ exists for all } \sigma \in \I(\x).
\label{7.34g}
\end{equation}
Moreover, from the Plemelj relations [Eq.\ (17.2) in Sec.~17 of
Muskhelishvili's treatise],
\begin{equation}
\Phi^{\pm}(\x,\sigma) = \pm \nu^{+}(\x,\sigma)^{-1} \phi(\x,\sigma)
+ \frac{1}{\pi i} \int_{\bar{\I}} d\sigma' \nu^{+}(\x,\sigma')^{-1}
\frac{\phi(\x,\sigma')}{\sigma'-\sigma}.
\label{7.34h}
\end{equation}
[The existence of the above PV integral is demonstrated in Sec.~12 of 
Muskhelishvili's treatise.]  From Eq.\ (\ref{7.34c}), condition (3) in
the definition of the HHP [the one about the existence of $\F^{\pm}(\x)$]
and statement (\ref{7.34g}),
\begin{equation}
\lim_{\zeta \rightarrow 0} \frac{\partial\grave{\F}(\x,\sigma\pm\zeta)}
{\partial r} \text{ exists for each } \x \in D \text{ and }
\sigma \in \I(\x);
\label{7.34i}
\end{equation}
and, with the aid of Eqs.\ (\ref{7.28a}), (\ref{7.34a}) and (\ref{7.34h}),
\begin{equation}
\lim_{\zeta \rightarrow 0} \frac{1}{2} \left[
\frac{\partial\grave{\F}(\x,\sigma+\zeta)}{\partial r} + 
\frac{\partial\grave{\F}(\x,\sigma-\zeta)}{\partial r} \right] =
- \frac{\partial f(\x,\sigma)}{\partial r}
\end{equation}
and
\begin{eqnarray}
\lefteqn{\lim_{\zeta \rightarrow 0} \frac{1}{2} \frac{\partial}{\partial r}
\left[ \nu(\x,\sigma+\zeta)^{-1} \grave{\F}(\x,\sigma+\zeta)
+ \nu(\x,\sigma-\zeta)^{-1} \grave{\F}(\x,\sigma-\zeta) \right] }
\hspace{10em} \nonumber \\
& = & -\frac{1}{\pi i} \int_{\bar{\I}} d\sigma' \nu^{+}(\x,\sigma')^{-1}
\frac{\phi(\x,\sigma')}{\sigma'-\sigma}.
\label{7.34k}
\end{eqnarray}
However, from Eq.\ (\ref{6.10d}),
$$
\frac{1}{\pi i} \int_{\bar{\I}} d\sigma' \nu^{+}(\x,\sigma')^{-1}
\frac{f(\x,r)}{(\sigma'-r)(\sigma'-\sigma)} = 0.
$$
Therefore, from Eq.\ (\ref{7.34e}), Eq.\ (\ref{7.34k}) becomes
\begin{eqnarray}
\lefteqn{\lim_{\zeta \rightarrow 0} \frac{1}{2} \frac{\partial}{\partial r}
\left[ \nu(\x,\sigma+\zeta)^{-1} \grave{\F}(\x,\sigma+\zeta)
+ \nu(\x,\sigma-\zeta)^{-1} \grave{\F}(\x,\sigma-\zeta) \right] }
\hspace{10em} \nonumber \\
& = & -\frac{1}{\pi i} \int_{\bar{\I}} d\sigma' 
\frac{\partial[\nu^{+}(\x,\sigma')^{-1}f(\x,\sigma')]/\partial r}
{\sigma'-\sigma}.
\label{7.34l}
\end{eqnarray}

Next, from Eqs.\ (\ref{7.28a}) and (\ref{7.34a}),
\begin{equation}
\frac{1}{2} \left[ \frac{\partial\F^{+}(\x,\sigma)}{\partial r}
+ \frac{\partial\F^{-}(\x,\sigma)}{\partial r} \right] = 
- \frac{\partial f(\x,\sigma)}{\partial \sigma};
\label{7.35a}
\end{equation}
and, from Eq.\ (\ref{7.29b}) in Lem.~\ref{7.3D},
\begin{eqnarray}
\lefteqn{\frac{1}{2} \frac{\partial}{\partial r} \left\{
\nu^{+}(\x,\sigma)^{-1} [\F^{+}(\x,\sigma) -\F^{-}(\x,\sigma)] \right\} = } 
\nonumber \\
& & -\frac{1}{\pi i} \int_{\bar{\I}} d\sigma'
\frac{\partial[\nu^{+}(\x,\sigma')^{-1}f(\x,\sigma')]/\partial r}
{\sigma'-\sigma}.
\label{7.35b}
\end{eqnarray}
A comparison of the above Eqs.\ (\ref{7.35a}) and (\ref{7.35b}) with
Eqs. (\ref{7.35a}) and (\ref{7.34l}), together with the fact that
$$
\lim_{\zeta \rightarrow 0} \frac{\partial\nu(\x,\sigma\pm\zeta)^{-1}}
{\partial r} = \frac{\partial\nu^{\pm}(\x,\sigma)^{-1}}{\partial r},
$$
now yields
\begin{equation}
\lim_{\zeta \rightarrow 0} \frac{\partial\grave{\F}(\x,\sigma\pm\zeta)}
{\partial r} = \frac{\partial\F^{\pm}(\x,\sigma)}{\partial r}.
\label{7.35c}
\end{equation}
Statements (\ref{7.34i}) and (\ref{7.35c}) complete the proof of part (i)
of our theorem for $\partial\F(\x,\tau)/\partial r$.
\cheers

\item 
Since
\begin{equation}
\nu^{+}(\x,\sigma)^{-1} = \frac{\M^{+}(\sigma-r)\M^{+}(\sigma-s)}
{\M^{+}(\sigma-r_{0})\M^{+}(\sigma-s_{0})},
\label{7.36a}
\end{equation}
one has 
\begin{equation}
\nu^{+}(\x,\sigma)^{-1}f(\x,\sigma)=0 \text{ when } \sigma=r
\text{ and when } \sigma=s.
\end{equation}
Therefore, from statement $1^{0}$ in Sec.~29 of Muskhelishvili's treatise,
and from our Eq.\ (\ref{7.34b}),
\begin{equation}
\nu(\x,\tau)^{-1}\F(\x,\tau) \text{ converges as }
\tau \rightarrow r \text{ and as } \tau \rightarrow s \;
[\tau \in C - \bar{\I}(\x)].
\label{7.36c}
\end{equation}
Furthermore, from Eqs.\ (\ref{7.26d}), (\ref{7.26e}) and (\ref{7.34b}),
\begin{eqnarray}
\lefteqn{\F(\x,\tau) = \nu(\x,\tau) I 
+ [\nu(\x,\tau)-1]f(\x,r_{0}) } \nonumber \\
& & \mbox{ } - \frac{\nu(\x,\tau)}{\pi i} \int_{\bar{\I}}
d\sigma' \nu^{+}(\x,\sigma')^{-1} \left[
\frac{f(\x,\sigma')-f(\x,r_{0})}{\sigma'-r} \right]. 
\label{7.36d}
\end{eqnarray}
From statement (\ref{7.30c}), $\partial f(\x,\sigma')/\partial \sigma'$
exists and is a continuous function of $\sigma'$ throughout $\bar{\I}(\x)$.
Therefore, as one can see from Eq.\ (\ref{7.36a}),
\begin{equation}
\nu^{+}(\x,\sigma)^{-1} [f(\x,\sigma')-f(\x,r_{0})] = 0
\text{ when } \sigma = r_{0};
\end{equation}
and it then follows from Eq.\ (\ref{7.36d}) and the same statement $1^{0}$
in Sec.~29 of Muskhelishvili that was used before that
\begin{equation}
\F(\x,\tau) \text{ converges [to $-f(\x,r_{0})$] as } \tau \rightarrow
r_{0}.
\label{7.36f}
\end{equation}
Similarly, one proves that
\begin{equation}
\F(\x,\tau) \text{ converges [to $-f(\x,s_{0})$] as } \tau \rightarrow
s_{0}.
\label{7.36g}
\end{equation}
Statements (\ref{7.36c}), (\ref{7.36f}) and (\ref{7.36g}) together 
constitute part (ii) of our theorem.
\cheers

\item 
We shall prove this part of our theorem for $i=3$, and the proof for $i=4$
is left to the reader.

We start with the definition (\ref{7.34d}) of $\Phi(\x,\tau)$.  The proof
that we have just given for part (ii) of this theorem is also applicable
to $\Phi(\x,\tau)$.  Specifically, the proof of part (ii) remains valid
if one makes all of the following substitutions in its wording and
equations:
\begin{eqnarray*}
f(\x,\sigma') & \rightarrow & \phi(\x,\sigma'), \\
\text{Eq.\ (\ref{7.34b})} & \rightarrow & \text{Eq.\ (\ref{7.34d})}, \\
\nu(\x_{0},\x,\tau)\F(\x,\tau) & \rightarrow & \Phi(\x,\tau), \\
\text{statement (\ref{7.30c})} & \rightarrow & 
\text{condition (\ref{7.34f})}.
\end{eqnarray*}
Therefore, the conclusion of part (ii) of our theorem remains valid if one
makes the substitution `$\nu(\x,\tau)^{-1}\F(\x,\tau)$'
$\rightarrow$ `$\Phi(\x,\tau)$'.  So, for all $(\x,\tau) \in \dom{\F}$,
\begin{equation}
\begin{array}{l}
\Phi(\x,\tau) \text{ converges as } \tau \rightarrow r \text{ and as }
\tau \rightarrow s, \text{ and } \\
\nu(\x,\tau) \Phi(\x,\tau) \text{ converges as }
\tau \rightarrow r_{0} \text{ and as } \tau \rightarrow s_{0}.
\end{array}
\label{7.37}
\end{equation}
When the above statement (\ref{7.37}) is applied to Eq.\ (\ref{7.34c}),
one obtains the statement in part (iii) of our theorem for the case $i=3$.
\cheers
\end{romanlist}

Note: The meanings that we assigned above to `$f(\x,\sigma')$',
`$\phi(\x,\sigma')$' and `$\Phi(\x,\tau)$' will not be used in the
remainder of these notes.  They were temporary devices for the purpose
of abbreviating the proofs of the preceding theorem and two lemmas.

\section{Proof of the generalized Geroch conjecture\label{Sec_6}}

\setcounter{equation}{0}
\subsection{Generalized Abel transforms of the initial data and the
identification of the sets $\S_{\F}^{\Box}$ and $\S_{\E}^{\Box}$}

In Sec.~\ref{Sec_1}A, we introduced a linear system $\F_{HE}$ for the Ernst
equation that is related to $\F = \F_{KC}$ by Eqs.\ (\ref{G2.14a}) to 
(\ref{Uref}).  It will now be useful to introduce one more linear system
$\tilde{\F}_{HE}$ such that
\begin{equation}
\tilde{\F}_{HE}(\x,\tau) := P^{M}(\x_{0},\tau) \F_{HE}(\x,\tau)
P^{M}(\x_{0},\tau)^{-1},
\label{tF_HE}
\end{equation}
whereupon Eq.\ (\ref{G2.14a}) and the fact that
\begin{equation}
\F^{M}(\x,\tau) = P^{M}(\x,\tau) P^{M}(\x_{0},\tau)^{-1}
\end{equation}
yields
\begin{equation}
\F(\x,\tau) = A(\x) \F^{M}(\x,\tau) \tilde{\F}_{HE}(\x,\tau),
\end{equation}
where
\begin{equation}
A := \frac{1}{\sqrt{h_{22}}} \left( \begin{array}{cc}
1 & h_{12} \\ 0 & h_{22}
\end{array} \right).
\end{equation}
Note that
\begin{equation}
h = A \, h^{M} \, A^{T}, \quad h^{M} = \left( \begin{array}{cc}
\rho^{2} & 0 \\ 0 & 1
\end{array} \right).
\end{equation}
Therefore, from Eqs.\ (\ref{6.15a}) to (\ref{6.15d}), and the fact that
$A^{T} \Omega A = (\det{A}) \Omega = \Omega$,
\begin{equation}
\left[\tilde{\F}_{HE}(\x,\tau^{*})\right]^{\dagger} \A^{M}(\x_{0},\tau)
\tilde{\F}_{HE}(\x,\tau) = \A^{M}(\x_{0},\tau).
\label{tquad}
\end{equation}
Obviously, $d\tilde{\F}_{HE} = \tilde{\Gamma}_{HE} \tilde{\F}_{HE}$, where
\begin{equation}
\tilde{\Gamma}_{HE}(\x,\tau) = P^{M}(\x_{0},\tau) \Gamma_{HE}(\x,\tau)
P^{M}(\x_{0},\tau)^{-1},
\end{equation}
and $\Gamma_{HE}$ is given by Eq.\ (\ref{GammaHE}).  Note that 
$\tilde{\Gamma}_{HE}$
can be obtained by making the following substitutions in $\Gamma_{HE}$:
\begin{eqnarray}
J & \rightarrow & \tilde{J}(\tau) := P^{M}(\x_{0},\tau) J 
P^{M}(\x_{0},\tau)^{-1} = \left( \begin{array}{cc}
-i & 2(\tau-z_{0}) \\ 0 & i
\end{array} \right), \\
\N(\tau) & \rightarrow & \tilde{\N}(\tau) := P^{M}(\x_{0},\tau) \N(\tau)
P^{M}(\x_{0},\tau)^{-1} = \left( \begin{array}{cc}
-\tau+z_{0} & -i\rho_{0}^{2} \\ -i & \tau-z_{0}
\end{array} \right),
\label{tN}
\end{eqnarray}
where
\begin{equation}
J := i\sigma_{2} \text{ and } \N(\tau) := \mu(\x_{0},\tau) \sigma_{3}.
\end{equation}

The properties of $\tilde{\F}_{HE}$ can be deduced from those of $\F_{HE}$.
For example, consider the {\em generalized Abel transforms}\footnote{In
the Weyl case the $\alpha^{(i)}$ are easily expressed in terms of Abel
transforms.} (our term)
\begin{equation}
\begin{array}{rcl}
\alpha^{(3)}(\sigma) & := & \F_{HE}^{+}((\sigma,s_{0}),\sigma)^{-1}
\text{ for } \sigma \in \I^{(3)} \text{ and } \\
\alpha^{(4)}(\sigma) & := & \F_{HE}^{+}((r_{0},\sigma),\sigma)^{-1}
\text{ for } \sigma \in \I^{(4)}
\end{array}
\end{equation}
of the initial data functions
\begin{equation}
\begin{array}{rcl}
\E^{(3)}(r) & = & \E(r,s_{0}) \text{ for } r \in \I^{(3)} \text{ and } \\
\E^{(4)}(s) & = & \E(r_{0},s) \text{ for } s \in \I^{(4)}.
\end{array}
\end{equation}
Analysis\footnote{For details, see our {\em Magnum Opus} (gr-qc/9903104).}
yields
\begin{equation}
\alpha^{(i)} = I \alpha_{0}^{(i)} + J \alpha_{1}^{(i)}
+ \N^{+}(\sigma) [I \alpha_{2}^{(i)} + J \alpha_{3}^{(i)}],
\end{equation}
where $\N^{+}(\sigma) = \mu^{+}(\x_{0},\sigma) \sigma_{3}$,
\begin{eqnarray}
& & \alpha_{k}^{(i)} : \I^{(i)} \rightarrow R^{1} \; (k=0,1,2,3),
\label{alp1} \\
& & \alpha_{k}^{(i)} \text{ is } H(1/2) \text{ on each closed
subinterval of } \I^{(i)},
\end{eqnarray}
\begin{equation}
\alpha_{k}^{(i)} \text{ is } \bC^{n-1} \text{ if } \E^{(i)} 
\text{ is } \bC^{n} \text{ and } \alpha_{k}^{(i)} \text{ is analytic if } 
\E^{(i)} \text{ is analytic},
\end{equation}
and
\begin{equation}
\det{\alpha^{(i)}} = [\alpha_{0}^{(i)}]^{2} + [\alpha_{1}^{(i)}]^{2}
+ (\sigma-r_{0})(\sigma-s_{0}) \left\{ [\alpha_{2}^{(i)}]^{2} +
[\alpha_{3}^{(i)}]^{2} \right\} = 1.
\label{alp4}
\end{equation}
Instead of $\alpha^{(3)}$ and $\alpha^{(4)}$, we shall be employing 
\begin{equation}
\begin{array}{rcl}
V^{(3)}(\sigma) & := & \tilde{\F}_{HE}^{+}((\sigma,s_{0}),\sigma)^{-1}
\text{ for } \sigma \in \I^{(3)} \text{ and } \\
V^{(4)}(\sigma) & := & \tilde{\F}_{HE}^{+}((r_{0},\sigma),\sigma)^{-1}
\text{ for } \sigma \in \I^{(4)},
\end{array}
\label{Vs}
\end{equation}
whose pertinent properties are easily deduced from those of $\alpha^{(3)}$
and $\alpha^{(4)}$ by using Eq.\ (\ref{tF_HE}).  For example,
\begin{equation}
V^{(i)} = I \alpha_{0}^{(i)} + \tilde{J}(\sigma) \alpha_{1}^{(i)}
+ \tilde{\N}(\sigma) [I \alpha_{2}^{(i)} + \tilde{J}(\sigma) 
\alpha_{3}^{(i)}].
\end{equation}
Furthermore, with the aid of Eq.\ (\ref{tquad}) and the definitions
of $K$ and $K^{\Box}$ by Eqs.\ (\ref{gKbox}) to (\ref{gK}), one readily
deduces from Eqs.\ (\ref{alp1}) to (\ref{alp4}) that
\begin{equation}
\bV \in K \text{ where } \bV := (V^{(3)},V^{(4)}),
\end{equation}
and
\begin{eqnarray}
& & \bV \in K^{n-1} \text{ if } \E^{(3)} \text{ and } \E^{(4)} 
\text{ are } \bC^{n}, \nonumber \\
& & \bV \in K^{\infty} \text{ if } \E^{(3)} \text{ and } \E^{(4)} 
\text{ are } \bC^{\infty} \text{ and } \nonumber \\
& & \bV \in K^{an} \text{ if } \E^{(3)} \text{ and } \E^{(4)} 
\text{ are } \bC^{an}.
\end{eqnarray}

Defining
\begin{eqnarray}
\S_{\bV} & := & \text{ the set of all ordered pairs } \bV = (V^{(3)},V^{(4)}), 
\nonumber \\
& & \text{ where $V^{(i)}$ is a $2 \times 2$ matrix function with the} 
\nonumber \\
& & \text{ domain $\I^{(i)}$ and there exists } \F \in \S_{\F}
\label{G3.1} \\
& & \text{ such that Eqs.\ (\ref{Vs}) hold,} \nonumber
\end{eqnarray}
\begin{eqnarray}
B(\I^{(i)}) & := & \text{ the multiplicative group of all }
\nonumber \\ & &
\exp \left( \tilde{J} \bvarphi^{(i)} \right) = I \cos{\bvarphi^{(i)}}
+ \tilde{J} \sin{\bvarphi^{(i)}} \\ & &
\text{ such that $\bvarphi^{(i)}$ is any real-valued function} 
\nonumber \\ & &
\text{ that has the domain $\I^{(i)}$ and is $H(1/2)$} 
\nonumber \\ & &
\text{ on every closed subinterval of $\I^{(i)}$,} 
\nonumber
\end{eqnarray}
it will turn out to be possible to identify the sets $\S_{\F}^{\Box}$
involved in the generalized Geroch conjecture in terms of the more
fundamental sets
\begin{equation}
\S_{\bV}^{\Box} := \{\bV \in \S_{\bV}: \text{ there exists }
\bw \in B(\I^{(3)}) \times B(\I^{(4)}) \text{ for which } \bV\bw \in k^{\Box}\},
\label{G3.8b}
\end{equation}
where 
\begin{equation}
k^{\Box} = k \cap K^{\Box}, \quad
k := \{\bV\bw: \bV \in \S_{\bV}, \bw \in B(\I^{(3)}) \times B(\I^{(4)})\},
\end{equation}
and, for any members $\bv = (v^{(3)},v^{(4)})$ and 
$\bv' = (v^{(3)\prime},v^{(4)\prime})$ of $K$, 
$$
\bv\bv' := (v^{(3)}v^{(3)\prime},v^{(4)}v^{(4)\prime}).
$$
Specifically, we let
\begin{equation}
\S_{\F}^{\Box} := \text{ the set of all } \F \in \S_{\F} 
\text{ for which } \bV \in \S_{\bV}^{\Box}.
\end{equation}

Having defined $\S_{\F}^{\Box}$, we can easily identify the remaining
important sets.  Thus,
\begin{equation}
\S_{\E}^{\Box} := \text{ the set of all } \E \in \S_{\E}
\text{ for which } \F \in \S_{\F}^{\Box},
\end{equation}
with a like definition of $\S_{H}^{\Box}$.

We leave the proof of the following theorem, which actually motivated how
we formulated our HHP corresponding to $(\bv,\F_{0})$, to the reader:
\begin{abc}
\begin{theorem}[Motivation]
\label{4.1C}
\mbox{ } \\
For all $\bv \in K$ and for all $\S_{\F}$ members $\F$ and 
$\F_{0}$ whose corresponding $\S_{\bV}$ members are $\bV$ and
$\bV_{0}$, respectively, the following statements (i) and (ii) are
equivalent to one another:
\begin{romanlist}
\item 
There exists $\bw \in B(\I^{(3)}) \times B(\I^{(4)})$ such that 
\begin{equation}
\bv = \bV \bw \bV_{0}^{-1}.
\end{equation}

\item 
For each $\x \in D$, $i \in \{3,4\}$ and $\sigma \in \I^{(i)}(\x)$,
\begin{equation}
\F^{+}(\x,\sigma) v^{(i)}(\sigma) [\F_{0}^{+}(\x,\sigma)]^{-1}
= 
\F^{-}(\x,\sigma) v^{(i)}(\sigma) [\F_{0}^{-}(\x,\sigma)]^{-1}.
\label{4.16b}
\end{equation}
\end{romanlist}
Moreover, if $\E^{(i)}$ and $\E_{0}^{(i)}$ are $\bC^{n_{i}}$ (resp.\ 
analytic) and $w^{(i)}$ is $\bC^{n_{i}-1}$ (resp.\ analytic), then the
function of $\sigma$ that equals each side of Eq.\ (\ref{4.16b}) has
a $\bC^{n_{i}-1}$ (resp.\ analytic) extension $Y^{(i)}(\x)$
to the interval
\begin{equation}
\dom{Y^{(i)}(\x)} = \check{\I}^{(i)}(x^{7-i})
\end{equation}
and, if $\bv \in K^{\Box}$ and $\F_{0} \in \S_{\F}^{\Box}$,
then $\bV \in \S_{\bV}^{\Box}$ and $\F \in \S_{\F}^{\Box}$.
\end{theorem}
\end{abc}

\begin{theorem}[Relation of $\F_{0}$ and $\bV_{0}$]
\label{4.1D} \mbox{ } \\
For each $\F_{0} \in \S_{\F}$ whose corresponding member of
$\S_{\bV}$ is $\bV_{0}$, and for each $\bw \in B(\I^{(3)}) \times B(\I^{(4)})$,
$\F_{0}$ is a solution of the HHP corresponding to $(\bV_{0}\bw,
\F^{M})$.
\end{theorem}

\begin{abc}
\proof
For each $\x \in D$ and $i \in \{3,4\}$,
\begin{arablist}
\item 
Thm.~\ref{Thm_3}(i) states that $\F_{0}(\x)$ is holomorphic
throughout its domain $C - \bar{\I}(\x)$,

\item 
Thm.~\ref{3.7G} states that $\F^{\pm}(\x)$ exist and, from Thm.~\ref{4.1C}
and the fact that $\bV^{M} = (I,I)$,
\begin{eqnarray}
Y_{0}^{(i)}(\x,\sigma) & := & \F_{0}^{+}(\x,\sigma) V_{0}^{(i)}(\sigma)
w^{(i)}(\sigma) [\F^{M+}(\x,\sigma)]^{-1} \nonumber \\
& = & \F_{0}^{-}(\x,\sigma) V_{0}^{(i)}(\sigma)
w^{(i)}(\sigma) [\F^{M-}(\x,\sigma)]^{-1} \nonumber \\
& & \text{ for all } \sigma \in \I^{(i)}(\x);
\label{4.24}
\end{eqnarray}
and Thm.~\ref{3.7G} and Thm.~\ref{Thm_3}(iii) imply that $\F_{0}(\x)$
is bounded at $\x_{0}$ and $\nu(\x)^{-1} \F_{0}(\x)$ is bounded
at $\x$, while the function $Y_{0}(\x)$ whose domain is $\I(\x)$ and whose 
values are given by $Y_{0}^{(i)}(\x,\sigma)$ at each $\sigma \in \I^{(i)}(\x)$
satisfies the condition
\begin{equation}
Y_{0}(\x) \text{ is bounded at $\x$ and at $\x_{0}$.}
\label{4.25}
\end{equation}
Thus, $\F_{0}$ is a solution of the HHP corresponding to
$(\bV\bw,\F^{M})$.
\end{arablist}
\cheers
\end{abc}

\begin{theorem}[Reduction theorem]
\label{4.2D}
\mbox{ } \\
For each $\x \in D$ and $2 \times 2$ matrix function $\F(\x)$
with the domain $C - \bar{\I}(\x)$, for each $\bv \in K$ and $\F_{0}
\in \S_{\F}$ whose corresponding member of $\S_{\bV}$ is $\bV_{0}$,
and for each $\bw \in B(\I^{(3)}) \times B(\I^{(4)})$, the following two 
statements are equivalent to one another:
\begin{arablist}
\item 
The function $\F(\x)$ is a solution of the HHP corresponding to
$(\bv,\F_{0},\x)$.
\item
The function $\F(\x)$ is a solution of the HHP corresponding to
$(\bv\bV_{0}\bw,\F^{M},\x)$.
\end{arablist}
\end{theorem}

\begin{abc}
\proof
Suppose that statement (i) is true.  Then $\F(\x)$ satisfies all
four conditions (1) through (4) in the definition of the HHP corresponding
to $(\bv,\F_{0},\x)$.  In particular, from conditions (3) and (4),
\begin{eqnarray}
Y^{(i)}(\x,\sigma) & := & \F^{+}(\x,\sigma) v^{(i)}(\sigma)
[\F^{+}_{0}(\x,\sigma)]^{-1} \nonumber \\
& = & \F^{-}(\x,\sigma) v^{(i)}(\sigma)
[\F^{-}_{0}(\x,\sigma)]^{-1} \nonumber \\
& & \text{ for all } i \in \{3,4\} \text{ and } \sigma \in \I^{(i)}(\x);
\label{4.26}
\end{eqnarray}
and
\begin{eqnarray}
Y(\x) \text{ is bounded at $\x$ and at $\x_{0}$.}
\label{4.27}
\end{eqnarray}
So, from the preceding Thm.~\ref{4.1D} and Eqs.\ (\ref{4.24}) and
(\ref{4.26}),
\begin{eqnarray}
X^{(i)}(\x,\sigma) & := & \F^{+}(\x,\sigma) u^{(i)}(\sigma)
[\F^{M+}(\x,\sigma)]^{-1} \nonumber \\
& = & \F_{0}^{-}(\x,\sigma) u^{(i)}(\sigma)
[\F^{M-}(\x,\sigma)]^{-1} \nonumber \\
& & \text{ for all } i \in \{3,4\} \text{ and } \sigma \in \I^{(i)}(\x),
\label{4.28a}
\end{eqnarray}
where 
\begin{equation}
\bu := \bv(\bV_{0}\bw)
\label{4.28b}
\end{equation}
(which is a member of $K$, since $\S_{\bV} \subset K$ and $B \subset K$);
and, furthermore,
\begin{equation}
X(\x) = Y(\x) Y_{0}(\x)
\label{4.28c}
\end{equation}
and, from (\ref{4.25}), (\ref{4.27}) and (\ref{4.28c}),
\begin{equation}
X(\x) \text{ is bounded at $\x$ and $\x_{0}$.}
\label{4.28d}
\end{equation}
Therefore, we have proved that statement (ii) is true if statement (i)
is true.

Next, suppose statement (ii) is true.  Then $\F(\x)$ satisfies
all four conditions in the definition of the HHP corresponding to
$(\bu,\F^{M},\x)$, where $\bu$ is defined by Eq.\ (\ref{4.28b}).
In particular, from conditions (3) and (4), Eq.\ (\ref{4.28a}) and
the statement (\ref{4.28d}) hold.  Since $\det{V_{0}^{(i)}} =
\det{w^{(i)}} = 1$ and since $\det{\F_{0}(\x)} = \det{\F^{M}(\x)}
= \nu(\x)$ [Thm.~\ref{Thm_3}(iii)], Eq.\ (\ref{4.24}) yields
$\det{Y^{(i)}(\x)} = 1$.  Therefore, both sides of Eq.\ (\ref{4.24})
are invertible, and
\begin{eqnarray}
[Y_{0}^{(i)}(\x,\sigma)]^{-1} & = & \F^{M+}(\x,\sigma)
[V_{0}^{(i)}(\sigma) w^{(i)}(\sigma)]^{-1}
[\F_{0}^{+}(\x,\sigma)]^{-1} \nonumber \\
& = & \F^{M-}(\x,\sigma)
[V_{0}^{(i)}(\sigma) w^{(i)}(\sigma)]^{-1}
[\F_{0}^{-}(\x,\sigma)]^{-1} \nonumber \\
& & \text{ for all } i \in \{3,4\} \text{ and } \sigma \in \I^{(i)}(\x);
\label{4.29a}
\end{eqnarray}
and, from (\ref{4.25}),
\begin{equation}
Y_{0}(\x)^{-1} \text{ is bounded at $\x$ and at $\x_{0}$.}
\label{4.29b}
\end{equation}
So, by multiplying both sides of Eq.\ (\ref{4.28a}) by the corresponding
sides of Eq.\ (\ref{4.29a}), and then using (\ref{4.28b}), (\ref{4.28d})
and (\ref{4.29b}), we establish that $\F$ is a solution of the
HHP corresponding to $(\bv,\F_{0},\x)$.
\cheers
\end{abc}

\setcounter{equation}{0}
\subsection{The HHP solution $\F$ is a member of $\S_{\F}^{\Box}$
when $\bv \in K^{\Box}$ and $\Box$ is $n \ge 3$, $n+ \; (n \ge 3)$,
$\infty$ or `an'}

\begin{theorem}[$\partial\grave{\F}/\partial x^{i} = \Gamma_{i} 
\grave{\F}$]
\label{7.1E} \mbox{ } \\
When $\bv \in K^{3}$, $\F$ is the solution of the HHP corresponding
to $(\bv,\F^{M})$ and $H$ is the function defined by Eq.\ (\ref{7.1a})
in Thm.~\ref{7.1A}, then [from Thm.~\ref{7.1D}] $d\grave{\F}(\x,\tau)$
and $dH(\x)$ exist; and, for each $i \in \{3,4\}$,
\begin{equation}
\frac{\partial \grave{\F}(\x,\tau)}{\partial x^{i}} = \Gamma_{i}(\x,\tau)
\grave{\F}(\x,\tau) \text{ for all } (\x,\tau) \in \dom{\grave{\F}},
\end{equation}
where
\begin{equation}
\Gamma_{i}(\x,\tau) := \frac{1}{2(\tau-x^{i})} \frac{\partial H(\x)}
{\partial x^{i}} \Omega.
\end{equation}
\end{theorem}

\proof
From Thm.~\ref{4.3D}(ii), $\F(\x,\tau)^{-1}$ exists for all
$(\x,\tau) \in \dom{\F}$; and, for the continuous extension of $Y$
that is defined by Cor.~\ref{5.5C} (also, see the beginning of
Sec.~\ref{Sec_4}F) and Eq.\ (\ref{5.22b}), $Y(\x,\sigma)^{-1}$ exists
for all $\x \in D$ and $\sigma \in \bar{\I}(\x)$.  From Thm.~\ref{7.1D},
$d\grave{\F}(\x,\tau)$, $dY(\x,\sigma)$ and $dH(\x)$ exist and are continuous
functions of $(\x,\tau)$, $(\x,\sigma)$ and $\x$ throughout $\dom{\grave{\F}}$,
$\dom{Y} := \{(\x,\sigma):\x \in D, \sigma \in \bar{\I}(\x)\}$
and $D$, respectively; and, for each $\x \in D$, $d\grave{\F}(\x,\tau)$
is a holomorphic function of $\tau$ throughout $C - \I(\x)
-\{r,s,r_{0},s_{0}\}$.  It
then follows, with the aid of conditions (1) through (3) in the definition
of the HHP, Eq.\ (\ref{7.1a}) in Thm.~\ref{7.1A}, and Thm.~\ref{7.4D}(i)
that, for each $\x \in D$,
\begin{eqnarray}
Z_{i}(\x,\tau) & := & (\tau-x^{i}) \frac{\partial \grave{\F}(\x,\tau)}
{\partial x^{i}} \grave{\F}(\x,\tau)^{-1} \text{ exists and is a holomorphic }
\nonumber \\
& & \text{function of $\tau$ throughout } C-\I(\x)-\{r,s,r_{0},s_{0}\} 
\label{7.39a} \\
Z_{i}(\x,\tau) & = & \frac{1}{2} \frac{\partial H(\x)}{\partial x^{i}}
\Omega + O(\tau^{-1}) \text{ in at least } \nonumber \\
& & \text{one neighborhood of } \tau = \infty,
\label{7.39b} \\
Z_{i}^{\pm}(\x,\sigma) & & \text{exists for each } \sigma \in \I(\x) \\
\text{and} & & \nonumber \\
Z_{i}^{+}(\x,\sigma) & = & Z_{i}^{-}(\x,\sigma) \nonumber \\
& = & (\sigma-x^{i}) \frac{\partial Y(\x,\sigma)}{\partial x^{i}}
Y(\x,\sigma)^{-1} + Y(\x,\sigma) \frac{1}{2} \frac{\partial H^{M}(\x)}
{\partial x^{i}} \Omega Y(\x,\sigma)^{-1} \nonumber \\
& & \text{for all } \sigma \in \bar{\I}(\x),
\label{7.39d}
\end{eqnarray}
where we have used the fact that the defining equation in condition (3)
for the HHP corresponding to $(\bv,\F^{M},\x)$ is expressible in
the form
\begin{equation}
\F^{\pm}(\x,\sigma) = Y^{(j)}(\x,\sigma) \F^{M\pm}(\x,\sigma)
[v^{(j)}(\sigma)]^{-1} \text{ for all } \sigma \in \I^{(j)}(\x);
\label{7.39e}
\end{equation}
and we have used the fact that, since $\F^{M} \in \S_{\F}$,
\begin{equation}
\frac{\partial\grave{\F}^{M}(\x,\tau)}{\partial x^{j}} = \Gamma_{i}^{M}(\x,\tau)
\grave{\F}^{M}(\x,\tau) \text{ for all } \tau \in C - \I(\x) -
\{r,s,r_{0},s_{0}\}.
\label{7.39f}
\end{equation}

We next define a continuous extension of $Z_{i}(\x)$ [which we also denote
by $Z_{i}(\x)$] to the domain $C - \{r,s,r_{0},s_{0}\}$ by letting
\begin{equation}
Z_{i}(\x,\sigma) := Z_{i}^{\pm}(\x,\sigma).
\end{equation}
Then, from the statement (\ref{7.39a}) and the theorem of Riemann that we
have already used in a different context, 
\begin{equation}
Z_{i}(\x,\tau) \text{ is a holomorphic function of } \tau 
\text{ throughout } C - \{r,s,r_{0},s_{0}\}.
\label{7.40b}
\end{equation}
However, from Eq.\ (\ref{7.39a}) and Thms.~\ref{7.4D}(ii) and~(iii),
\begin{equation}
\begin{array}{l}
\nu(\x,\tau) Z_{i}(\x,\tau) \text{ converges as }
\tau \rightarrow r_{0} \text{ and as } \tau \rightarrow s_{0},
\nonumber \\
\text{and } \nu(\x,\tau)^{-1} Z_{i}(\x,\tau)
\text{ converges as } \tau \rightarrow r \text{ and as } 
\tau \rightarrow s.
\end{array}
\label{7.40c}
\end{equation}
Also, from Eq.\ (\ref{7.39d}) and the continuity on $\bar{\I}(\x)$
of $dY(\x,\sigma)$ and $Y(\x,\sigma)^{-1} = \Omega Y(\x,\sigma)^{T} \Omega$,
\begin{equation}
Z_{i}(\x,\sigma) \text{ converges as } \sigma \rightarrow r_{0},
\sigma \rightarrow s_{0}, \sigma \rightarrow r, \sigma \rightarrow s.
\label{7.40d}
\end{equation}
Combining (\ref{7.40b}), (\ref{7.40c}) and (\ref{7.40d}), one obtains,
by reasoning that should now be familiar to us, $Z_{i}(\x,\tau) = 
Z_{i}(\x,\infty)$, whereupon the conclusion of our theorem follows from
Eqs.\ (\ref{7.39a}) and (\ref{7.39b}).
\cheers

\begin{corollary}[$d\grave{\F}=\Gamma\grave{\F}$]
\label{7.2E} \mbox{ } \\
For each $(\x,\tau) \in \dom{\grave{\F}}$,
\begin{equation}
d\grave{\F}(\x,\tau) = \Gamma(\x,\tau) \grave{\F}(\x,\tau),
\label{7.41a}
\end{equation}
where
\begin{eqnarray}
\Gamma(\x,\tau) & := & \frac{1}{2} (\tau-z+\rho\star)^{-1} dH(\x) \Omega
\nonumber \\
& = & \sum_{i} dx^{i} \Gamma_{i}(\x,\tau).
\label{7.41b}
\end{eqnarray}
\end{corollary}

\proof
Obvious.
\cheers

\begin{theorem}[$\A \Gamma = \frac{1}{2} \Omega dH \Omega$]
\label{7.3E} \mbox{ } \\
Suppose $\bv \in K^{3}$ and $\F$ is the solution of the HHP corresponding
to $(\bv,\F^{M})$.  Then
\begin{equation}
\A \Gamma = \frac{1}{2} \Omega dH \Omega,
\label{7.42}
\end{equation}
where $H$, $\A$ and $\Gamma$ are defined by Eqs.\ (\ref{7.1a}), (\ref{7.3a})
and (\ref{7.41b}), respectively.
\end{theorem}

\proof
The proof will be given in three parts:
\begin{arablist}
\item 
For each $H' \in \S_{H}$, note that
\begin{equation}
\Re{H'} = -h'
\end{equation}
and that the defining differential equation for $\Im{H'}$ in terms of
$\Re{H'}$ is expressible in the form 
\begin{equation}
h' \Omega d(\Re{H'}) = \rho \star (i \Im{H'}).
\label{7.43b}
\end{equation}
Recall that $h'$ is symmetric and $\det{h'} = \rho^{2}$.  So,
\begin{equation}
(h'\Omega)^{2} = \rho^{2} I.
\label{7.43c}
\end{equation}
From Eq.\ (\ref{7.43c}), Eq.\ (\ref{7.43b}) is equivalent to the equation
\begin{equation}
h'\Omega d(i\Im{H'}) = \rho \star d(\Re{H'});
\end{equation}
and, therefore, Eq.\ (\ref{7.43b}) is equivalent to the equation
\begin{equation}
h'\Omega dH' = \rho \star dH'.
\label{7.43c'}
\end{equation}
Furthermore, the above Eq.\ (\ref{7.43c'}) yields
$$
\A' \Gamma' =
[(\tau-z)\Omega+\Omega h' \Omega] \frac{1}{2} (\tau-z+\rho\star)^{-1}
dH' \Omega = \frac{1}{2} (\tau-z+\rho\star)^{-1}
[(\tau-z) \Omega dH' \Omega + \rho \star \Omega dH' \Omega].
$$
So, Eq.\ (\ref{7.43c'}) implies
\begin{equation}
\A' \Gamma' = \frac{1}{2} \Omega dH' \Omega
\text{ for each } H' \in \S_{H}.
\label{7.43d}
\end{equation}
The reader should have no difficulty in proving that, conversely, Eq.\ 
(\ref{7.43d}) implies (\ref{7.43c'}).  We shall use the above result
later in our proof.

\item 
We now obtain a second result that we shall need for the proof.  From
Eq.\ (\ref{7.3b}) in Thm.~\ref{7.2A},
\begin{equation}
[\F^{\mp}(\x,\sigma)]^{\dagger} \A(\x,\sigma) \F^{\pm}(\x,\sigma) =
\A(\x_{0},\sigma) \text{ for all } \sigma \in \I(\x).
\label{7.44a}
\end{equation}
Now, recall that $\A(\x_{0},\sigma) = \A^{M}(\x_{0},\sigma)$ in our
gauge.  Also, recall that
\begin{equation}
[\F'(\x,\tau^{*})]^{\dagger} \A'(\x,\tau) \F'(\x,\tau) =
\A^{M}(\x_{0},\sigma) \text{ for all } \F' \in \S_{\F}.
\label{7.44b}
\end{equation}
Therefore, we obtain the following result by using Eqs.\ (\ref{7.39e})
[condition (3) in the definition of the HHP corresponding to
$(\bv,\F^{M},\x)$], (\ref{7.44a}), (\ref{4.2}) 
and (\ref{7.44b}) [for $\F' = \F^{M}$]:
$$
Y^{\dagger}(\x,\sigma) \A(\x,\sigma) Y(\x,\sigma) = \A^{M}(\x,\sigma)
\text{ for all } \sigma \in \I(\x).
$$
However, recall that $Y(\x,\sigma)$ is now a continuous function of 
$\sigma$ throughout $\bar{\I}(\x)$.  Therefore,
\begin{equation}
Y^{\dagger}(\x,\sigma) \A(\x,\sigma) Y(\x,\sigma) = \A^{M}(\x,\sigma)
\text{ for all } \sigma \in \bar{\I}(\x).
\label{7.44c}
\end{equation}
We shall use this result below.

\item 
From the definition of $\A$ and $\Gamma$, each component of $\A(\x,\tau)
\Gamma(\x,\tau)$ is a holomorphic function of $\tau$ throughout
$C - \{r,s\}$ and has no essential singularity at $\tau = r$ and at
$\tau = s$.  In fact, if there are any singularities at these points,
they are simple poles.  That much is obvious.

From Eqs.\ (\ref{7.41a}), (\ref{7.39e}) and (\ref{7.39f}),
\begin{eqnarray}
\A(\x,\sigma) \Gamma(\x,\sigma) & = & \A(\x,\sigma) [d\F^{\pm}(\x,\sigma)]
[\F^{\pm}(\x,\sigma)]^{-1} \nonumber \\
& = & \A(\x,\sigma) \left[dY(\x,\sigma) + Y(\x,\sigma) \Gamma^{M}(\x,\sigma)
\right] [Y(\x,\sigma)]^{-1} \nonumber \\
& & \text{for all } \sigma \in \I(\x).
\end{eqnarray}
The above equation becomes, after using Eq.\ (\ref{7.44c}),
$$
\A(\x,\sigma) \Gamma(\x,\sigma) = \left\{ \A(\x,\sigma) dY(\x,\sigma)
+ [Y(\x,\sigma)^{\dagger}]^{-1} \A^{M}(\x,\sigma) \Gamma^{M}(\x,\sigma)
\right\} [Y(\x,\sigma)]^{-1},
$$
which becomes, after using Eq.\ (\ref{7.43d}) with $H' = H^{M}$,
\begin{equation}
\A(\x,\sigma) \Gamma(\x,\sigma) = \left\{ \A(\x,\sigma) dY(\x,\sigma)
+ [Y(\x,\sigma)^{\dagger}]^{-1} \frac{1}{2} \Omega dH^{M}(\x) \Omega 
\right\} [Y(\x,\sigma)]^{-1}.
\end{equation}
From Thm.~\ref{7.1D} and the fact that $\det{Y(\x,\sigma)} = 1$, the
right side of the above equation is a continuous function of $\sigma$
throughout $\bar{\I}(\x)$.  Therefore, $\A(\x,\tau) \Gamma(\x,\tau)$
is extendable to a holomorphic function of $\tau$ throughout $C$; and
it follows that
$$
\A(\x,\tau) \Gamma(\x,\tau) = [\A(\x,\tau)\Gamma(\x,\tau)]_{\tau=\infty}
= \frac{1}{2} \Omega dH(\x) \Omega.
$$
\end{arablist}
\cheers

\begin{corollary}[$h \Omega dH = \rho \star dH$]
\label{7.4E} \mbox{ } \\
Suppose $H$ is defined as in the preceding theorem.  Then
\begin{equation}
h \Omega dH = \rho \star dH.
\label{7.46}
\end{equation}
\end{corollary}

\proof
Multiply both sides of Eq.\ (\ref{7.42}) through by $2\Omega(\tau-z+\rho\star)$
on the left, and by $\Omega$ on the right; and then set $\tau=z$.
\cheers

\begin{theorem}[HHP solution $\F \in \S_{\F}^{\Box}$]
\label{7.5E} \mbox{ } \\ \vspace{-3ex}
\begin{romanlist}
\item 
For each $\bv \in K^{\Box}$, where $\Box$ is $n \ge 3$, $n+$ ($n \ge 3$),
$\infty$ or `an', and, for each $\F_{0} \in \S_{\F}^{\Box}$,
there exists exactly one solution $\F$ of the HHP corresponding to
$(\bv,\F_{0})$.

\item 
Let $H$ be defined in terms of $\F$ by Eq.\ (\ref{7.1a}), and let
$\E := H_{22}$.  Then $\E \in \S_{\E}$, and $H$ is identical with the
unique member of $\S_{H}$ that is constructed from $\E$ in the usual way.

\item 
Furthermore, $\F$ is identical with the member of $\S_{\F}$ that
is defined in terms of $H$ in Sec.~\ref{Sec_1}  [and whose existence and
uniqueness for a given $H \in \S_{H}$ is asserted in Thm.~\ref{Thm_3}.]

\item 
Finally, let $\F_{HE}$ be 
defined in terms of $\F_{KC}=\F$ by Eq.\ (\ref{G2.14a}), and let $\bV$
denote the member of $\S_{\bV}$ that is defined in terms of $\tilde{\F}_{HE}$
by Eq.\ (\ref{Vs}).  Then $\bV \in \S_{\bV}^{\Box}$ and, therefore
(by definition), $\E \in \S_{\E}^{\Box}$, $H \in \S_{H}^{\Box}$ and
$\F \in \S_{\F}^{\Box}$.
\end{romanlist}
\end{theorem}

\proofs
\begin{romanlist}
\item 
Let $\bV_{0}$ denote the member of $\S_{\bV}$ that corresponds to
$\F_{0}$.  Since $\F_{0} \in \S_{\F}^{\Box}$, there
exists (by definition of $\S_{\F}^{\Box}$) $\bw \in B(\I^{3}) \times B(\I^{(4)})$
such that
\begin{equation}
\bV_{0} \bw \in k^{\Box} \subset K^{\Box};
\end{equation}
and, since $K^{\Box}$ is a group,
\begin{equation}
\bv \bV_{0} \bw \in K^{\Box}.
\label{7.47b}
\end{equation}
From Thm.~\ref{Thm_13}, there exists exactly one solution $\F$ of the
HHP corresponding to $(\bv\bV_{0}\bw,\F^{M})$; and, from Thm. \ref{4.2D},
it then follows that $\F$ is also a solution of the HHP corresponding to
$(\bv,\F_{0})$.  Finally, from Thm. \ref{4.3D}(iv), there is no other
solution of the HHP corresponding to $(\bv,\F_{0})$.
\cheers

\item 
From the premises of this theorem, $\bv \in K^{3}$.  Therefore, from 
statement (\ref{7.24b}) in Thm.~\ref{7.1D},
\begin{equation}
dH \text{ exists and is continuous }
\end{equation}
throughout $D$; and since
\begin{eqnarray}
(d^{2}H)(\x) & = & dr ds \left[ \frac{\partial^{2}H(\x)}{\partial r \partial s}
- \frac{\partial^{2}H(\x)}{\partial s \partial r} \right] \\
\text{and} & & \nonumber \\
(d\star dH)(\x) & = & -dr ds \left[
\frac{\partial^{2}H(\x)}{\partial r \partial s}
+ \frac{\partial^{2}H(\x)}{\partial s \partial r} \right], 
\end{eqnarray}
it is true that 
\begin{equation}
d^{2}H \text{ exists and vanishes }
\end{equation}
and
\begin{equation}
d\star dH \text{ exists and is continuous }
\end{equation}
throughout $D$.  Also, Eq.\ (\ref{7.46}) in Cor.~\ref{7.4E} asserts that
\begin{equation}
\rho \star dH = h \Omega dH,
\label{7.48d}
\end{equation}
where we recall from Eq.\ (\ref{7.1b}) in Thm.~\ref{7.1A} that
\begin{equation}
h := - \Re{H} = h^{T}
\label{7.48e}
\end{equation}
and, from Thm.~\ref{7.3A},
\begin{equation}
\det{h} = \rho^{2} \text{ and } f := \Re{\E} = -g_{22} < 0,
\label{7.48f}
\end{equation}
where $g_{ab}$ denotes the element of $h$ in the $a$th row and $b$th
column.  Since $\star\star=1$, Eq.\ (\ref{7.48d}) is equivalent to the
equation 
\begin{equation}
\rho dH = h \Omega \star dH
\label{7.48g}
\end{equation}
from which we obtain
\begin{equation}
\rho dH^{\dagger} \Omega dH = dH^{\dagger} \Omega h \Omega \star dH.
\label{7.48h}
\end{equation}
Upon taking the hermitian conjugates of the terms in the above equation,
and upon noting that $\Omega^{\dagger} = \Omega$, $h^{\dagger} = h$,
\begin{equation}
(\omega\eta)^{T} = - \eta^{T} \omega^{T} \text{ and }
\omega \star \eta = - (\star \omega) \eta \text{ for any }
n \times n \text{ matrix $1$-forms, }
\end{equation}
one obtains
\begin{equation}
-\rho dH^{\dagger} \Omega dH = dH^{\dagger} \Omega h \Omega \star dH.
\label{7.48j}
\end{equation}
From Eqs.\ (\ref{7.48j}) and (\ref{7.48h}),
\begin{equation}
dH^{\dagger} \Omega dH = 0.
\label{7.48k}
\end{equation}
[The above result (\ref{7.48k}) was first obtained by the authors in a
paper\footnote{I.~Hauser and F.~J.~Ernst, J.\ Math.\ Phys.\ {\bf 21},
1116-1140 (1980).  See Eq.\ (37).} which introduced an abstract geometric
definition of the Kinnersley potential $H$ and which derived other 
properties of $H$ that we shall not need in these notes.]

We next consider the $(2,2)$ matrix elements of Eqs.\ (\ref{7.48d}) and
(\ref{7.48k}).  With the aid of Eqs.\ (\ref{7.48e}) and (\ref{7.48f}), 
one obtains
\begin{equation}
\rho \star d\E = i(g_{12} d\E + f dH_{12})
\label{7.49a}
\end{equation}
and 
\begin{equation}
dH_{12}^{*} d\E - d\E^{*} dH_{12} = 0.
\label{7.49b}
\end{equation}
From Eq.\ (\ref{7.49a}),
\begin{eqnarray*}
f d(\rho \star d\E) - \rho d\E \star d\E & = & 
if (dg_{12} d\E + df dH_{12} - d\E dH_{12}) \\
& = & if \left[ -\frac{1}{2}(dH_{12}+dH_{12}^{*})d\E
+ \frac{1}{2}(d\E + d\E^{*}) dH_{12} - d\E dH_{12} \right] \\
& = & \frac{if}{2} (- dH_{12}^{*} d\E + d\E^{*} dH_{12}).
\end{eqnarray*}
Therefore, from Eq.\ (\ref{7.49b}),
\begin{equation}
f d(\rho \star d\E) - \rho d\E \star d\E = 0.
\label{7.49c}
\end{equation}
Furthermore, from Eqs.\ (\ref{1.12c}), (\ref{7.1b}) and (\ref{7.48f}),
\begin{equation}
\E(\x_{0}) = -1 \text{ and } \Re{\E} < 0.
\label{7.49d}
\end{equation}
Therefore,
\begin{equation}
\E \in \S_{\E},
\end{equation}
since $\E$ satisfies the Ernst equation (\ref{7.49c}) and the requisite
gauge conditions (\ref{7.49d})

Next, let
\begin{equation}
\chi := \Im{\E} \text{ and } \omega := g_{12}/g_{22}.
\end{equation}
Then, by taking the imaginary parts of the terms in Eq.\ (\ref{7.49a}),
one deduces
\begin{equation}
d\omega = \rho f^{-2} \star d\chi.
\label{7.50b}
\end{equation}
Furthermore, Eqs.\ (\ref{7.48e}) and (\ref{7.48f}) enable us to express
$h$ in the form
\begin{equation}
h = A \left( \begin{array}{cc}
\rho^{2} & 0 \\ 0 & 1
\end{array} \right) A^{T},
\end{equation}
where
\begin{equation}
A := \left( \begin{array}{cc}
1 & \omega \\ 0 & 1
\end{array} \right) \left( \begin{array}{cc}
1/\sqrt{-f} & 0 \\ 0 & \sqrt{-f}
\end{array} \right).
\end{equation}
Finally, the imaginary parts of the terms in Eq.\ (\ref{7.48g}) give us
\begin{equation}
\rho d(\Im{H}) = - h J \star dh.
\label{7.50e}
\end{equation}
A comparison of Eqs.\ (\ref{7.1b}), (\ref{7.1d}) and (\ref{7.50b}) to
(\ref{7.50e}) with the definition of $\S_{H}$ that is given in
Sec.~\ref{Sec_1} demonstrates that $H$ is precisely that member of $\S_{H}$
that is computed from $\E$ in the usual way.
\cheers

\item 
From statement (\ref{7.24a}) in Thm.~\ref{7.1D},
\begin{equation}
d\grave{\F}(\x,\tau) \text{ exists for all } \x \in D \text{ and }
\tau \in C - \I(\x) - \{r,s,r_{0},s_{0}\};
\label{7.51a}
\end{equation}
and, from Cor.~\ref{7.2E},
\begin{equation}
d\grave{\F}(\x,\tau) = \Gamma(\x,\tau) \grave{\F}(\x,\tau) \text{ for all }
\x \in D \text{ and } \tau \in C - \I(\x)-\{r,s,r_{0},s_{0}\}.
\end{equation}
Furthermore, from Thm. \ref{4.3D}(v),
\begin{equation}
\F(\x_{0},\tau) = I \text{ for all } \tau \in C.
\end{equation}
Finally, consider that $\bar{\I}(\x) = \bar{\I}^{(3)}(\x)$ when $s=s_{0}$,
and $\bar{\I}(\x) = \bar{\I}^{(4)}(\x)$ when $r=r_{0}$; and, from condition 
(1) in the definition of the HHP, $\F(\x,\tau)$ is a holomorphic
function of $\tau$ throughout $C - \bar{I}(\x)$.  Therefore,
\begin{equation}
\begin{array}{l}
\F((r,s_{0}),\tau) \text{ and } \F((r_{0},s),\tau) \\
\text{are continuous functions of $\tau$ at } \\
\text{$\tau=s_{0}$ and at $\tau=r_{0}$, respectively. }
\end{array}
\label{7.51d}
\end{equation}
From the above statements (\ref{7.51a}) to (\ref{7.51d}) and from the
definition of $\S_{\F}$ in Sec.~\ref{Sec_1}, it follows that
$\F \in \S_{\F}$.
\cheers

\item 
From condition (3) in the definition of the HHP and from Thm. \ref{4.1C},
there exists $\bw' \in B(\I^{(3)}) \times B(\I^{(4)})$ such that
\begin{equation}
\bv = \bV \bw' \bV_{0}^{-1}.
\label{7.52a}
\end{equation}
Therefore,
\begin{equation}
\bV = \bv \bV_{0} (\bw')^{-1}.
\end{equation}
However, from the proof of part (i) this theorem [see Eq.\ (\ref{7.47b})],
there then exists $\bw \in B(\I^{(3)}) \times B(\I^{(4)})$ such that
\begin{equation}
\bV(\bw' \bw) = \bv \bV_{0} \bw \in K^{\Box}.
\end{equation}
Therefore, since $\bw' \bw \in B(\I^{(3)}) \times B(\I^{(4)})$, it follows from the
definition of $\S_{\bV}^{\Box}$ given by Eq.\ (\ref{G3.8b}) that
$\bV \in \S_{\bV}^{\Box}$.  Hence, by definition, $\E \in \S_{\E}^{\Box}$,
$H \in \S_{H}^{\Box}$ and $\F \in \S_{\F}^{\Box}$.
\cheers
\end{romanlist}

\begin{corollary}[$k^{\Box} = K^{\Box}$]
\label{7.6E} \mbox{ } \\
Suppose that $\Box$ is $n \ge 3$, $n+$ ($n \ge 3$), $\infty$ or `an'.
Then
\begin{equation}
k^{\Box} := \{ \bV \bw:\bV \in \S_{\bV}, \bw \in B(\I^{(3)}) \times B(\I^{(4)}),
\bV \bw \in K^{\Box}\} = K^{\Box}.
\end{equation}
\end{corollary}

\proof
From its definition
\begin{equation}
k^{\Box} \subset K^{\Box}.
\label{7.54a}
\end{equation}

Now, suppose $\bv \in K^{\Box}$.  Since
\begin{equation}
\bV^{M} = (\delta^{(3)},\delta^{(4)}),
\label{7.54b}
\end{equation}
where
\begin{equation}
\delta^{(i)}(\sigma) := I \text{ for all } \sigma \in \I^{(i)},
\label{7.54c}
\end{equation}
we know that
\begin{equation}
\F^{M} \in \S_{\F}^{an} \subset \S_{\F}^{\Box}.
\label{7.54d}
\end{equation}
Therefore, from the preceding theorem, there exists $\F \in
\S_{\F}^{\Box}$ such that $\F$ is the solution of the HHP
corresponding to $(\bv,\F^{M})$; and, if $\bV$ denotes the member
of $\S_{\bV}^{\Box}$ that corresponds to $\F$, Eq.\ (\ref{7.52a})
in the proof of the preceding theorem informs us that $\bw' \in
B(\I^{(3)}) \times B(\I^{(4)})$ exists such that $\bv = \bV \bw'$.  So $\bv \in
k^{\Box}$.

We have thus proved that
\begin{equation}
K^{\Box} \subset k^{\Box},
\end{equation}
whereupon (\ref{7.54a}) and (\ref{7.54b}) furnish us with the conclusion
$k^{\Box} = K^{\Box}$.
\cheers

\setcounter{equation}{0}
\subsection{The generalized Geroch group $\K^{\Box}$}

\begin{definition}{Dfn.\ of $Z^{(i)}$}
Let $Z^{(i)}$ denote the subgroup of $K(\x_{0},\I^{(i)})$ that is given by
\begin{equation}
Z^{(i)} := \{\delta^{(i)},-\delta^{(i)}\},
\end{equation}
where $\delta^{(i)}$ is defined by Eq.\ (\ref{7.54c}).
\end{definition}

\begin{theorem}[Center of $K$]
\label{7.1F} \mbox{ } \\
The center of $K(\x_{0},\I^{(i)})$ is $Z^{(i)}$.  Hence the center of
$K$ is $Z^{(3)} \times Z^{(4)}$.
\end{theorem}

\proof
Left for the reader.  Hint:  See the proof of Lem.~\ref{7.2F}(i).
\cheers

\begin{definition}{Dfn.\ of $[\bv]$ for each $\bv \in K^{3}$}
For each $\bv \in K^{3}$, let $[\bv]$ denote the function such that
\begin{equation}
\dom{[\bv]} := \S_{\F}^{3}
\end{equation}
and, for each $\F_{0} \in \S_{\F}^{3}$,
\begin{equation}
[\bv](\F_{0}) := \text{ the solution of the HHP corresponding to }
(\bv,\F_{0}).
\end{equation}
Note that the existence of $[\bv]$ is guaranteed by Thm. \ref{4.2D}
and Thm. \ref{Thm_13}(iii).
\end{definition}

\begin{definition}{Dfn.\ of $\K^{\Box}\s$ when $\Box$ is $n \ge 3$,
$n+$ ($n \ge 3$), $\infty$ or `an'}
Let 
\begin{equation}
\K^{\Box} := \{[\bv]:\bv \in K^{\Box}\}.
\end{equation}
\end{definition}

The following lemma concerns arbitrary members $\bv$ and $\bv'$ of $K$,
and arbitrary members $\F_{0}$ and $\F$ of $\S_{\F}$.
Therefore, the lemma could have been given as a theorem in
Sec.~\ref{Sec_1}.  However, we have saved it for now, because the
lemma is directly applicable in the proof of the next theorem.

\begin{lemma}[Properties of $K$]
\label{7.2F} \mbox{ } \\ \vspace{-3ex}
\begin{romanlist}
\item 
Suppose that $\bv \in K$, $\F_{0} \in \S_{\F}$ and $\F
\in \S_{\F}$.  Then $\F$ is the solution of the HHP corresponding
to $(\bv,\F_{0})$ if and only if $\bV^{-1} \bv \bV_{0} \in 
B(\I^{(3)}) \times B(\I^{(4)})$, where $\bV_{0}$ and $\bV$ are the 
members of $\S_{\bV}$
corresponding to $\F_{0}$ and $\F$, respectively.

In particular, the solution of the HHP corresponding to $(\bv,\F_{0})$
is $\F_{0}$ if and only if $\bV_{0}^{-1} \bv \bV_{0} \in
B(\I^{(3)}) \times B(\I^{(4)})$; and the solution of the HHP corresponding to
$(\bv,\F^{M})$ is $\F^{M}$ if and only if $\bv \in
B(\I^{(3)}) \times B(\I^{(4)})$.

\item 
In addition to the premises of part (i) of this lemma, suppose that
$\bv' \in K$.  Thereupon, if $\F$ is the solution of the HHP
corresponding to $(\bv,\F_{0})$, and $\F'$ is the solution
of the HHP corresponding to $(\bv',\F)$, then $\F'$ is the
solution of the HHP corresponding to $(\bv' \bv,\F_{0})$.

If $\F$ is the solution of the HHP corresponding to $(\bv,\F_{0})$,
then $\F_{0}$ is the solution of the HHP corresponding to 
$(\bv^{-1},\F)$.
\end{romanlist}
\end{lemma}

\proofs
\begin{romanlist}
\item 
This theorem follows from Thm. \ref{4.1C} and the properties of members of
$\S_{\F}$ that are given in Thm. \ref{Thm_3} [specifically, the
properties $\F(\x,\infty)=I$ and (iv)] and Thm. \ref{3.7G}.  The
reader can easily fill in the details of the proof.
\cheers

\item 
This follows from the obvious facts that the equations
\begin{eqnarray*}
Y^{(i)}(\x,\sigma) & := & \F^{+}(\x,\sigma) v^{(i)}(\sigma)
[\F_{0}^{+}(\x,\sigma)]^{-1} \\
& = & \F^{-}(\x,\sigma) v^{(i)}(\sigma)
[\F_{0}^{-}(\x,\sigma)]^{-1}
\end{eqnarray*}
and
\begin{eqnarray*}
Y^{\prime(i)}(\x,\sigma) & := & \F^{\prime +}(\x,\sigma) v^{\prime(i)}(\sigma)
[\F^{+}(\x,\sigma)]^{-1} \\
& = & \F^{\prime -}(\x,\sigma) v^{\prime(i)}(\sigma)
[\F^{-}(\x,\sigma)]^{-1}
\end{eqnarray*}
imply
\begin{eqnarray*}
Y^{\prime(i)}(\x,\sigma) Y^{(i)}(\x,\sigma) & = & 
\F^{\prime +}(\x,\sigma) v^{\prime(i)}(\sigma) v^{(i)}(\sigma)
[\F_{0}^{+}(\x,\sigma)]^{-1} \\
& = & \F^{\prime -}(\x,\sigma) v^{\prime(i)}(\sigma) v^{(i)}(\sigma)
[\F_{0}^{-}(\x,\sigma)]^{-1}
\end{eqnarray*}
and
\begin{eqnarray*}
[Y^{(i)}(\x,\sigma)]^{-1} & = & \F_{0}^{+}(\x,\sigma)
[v^{(i)}(\sigma)]^{-1} [F^{+}(\x,\sigma)]^{-1} \\
& = & \F_{0}^{+}(\x,\sigma)
[v^{(i)}(\sigma)]^{-1} [F^{+}(\x,\sigma)]^{-1}
\end{eqnarray*}
for all $i \in \{3,4\}$ and $\sigma \in \I^{(i)}$.
\cheers
\end{romanlist}

Finally, we prove the following generalized Geroch conjecture:
\begin{theorem} 
\mbox{ } \\ \vspace{-3ex}
\begin{romanlist}
\item 
The mapping $[\bv]$ is the identity map on $\S_{\F}^{\Box}$ iff
$\bv \in Z^{(3)} \times Z^{(4)}$.
\item 
The set $\K^{\Box}$ is a group of permutations of $\S_{\F}^{\Box}$
such that the mapping $\bv \rightarrow [\bv]$ is a homomorphism of
$K^{\Box}$ onto $\K^{\Box}$; and the mapping $\{\bv\bw: \bw \in
Z^{(3)} \times Z^{(4)}\} \rightarrow [\bv]$ is an isomorphism [viz,
the isomorphism of $K^{\Box}/(Z^{(3)} \times Z^{(4)})$ onto $\K^{\Box}$].
\item 
The group $\K^{\Box}$ is transitive [i.e., for each $\F_{0},
\F \in \S_{\F}^{\Box}$ there exists at least one element of
$\K^{\Box}$ that transforms $\F_{0}$ into $\F$]. 
\end{romanlist}
\end{theorem}

\proofs
\begin{romanlist}
\item 
The statement that $[\bv]$ is the identity mapping on $\S_{\F}^{\Box}$
means that each $\F_{0} \in \S_{\F}$ is the solution of the
HHP corresponding to $(\bv,\F_{0})$; and, from Lem.~\ref{7.2F}(i),
this is equivalent to the following statement:
\begin{equation}
\text{For each } \bV_{0} \in \S_{\bV}^{\Box},
\bV_{0}^{-1} \bv \bV_{0} \in B(\I^{(3)}) \times B(\I^{(4)}).
\label{7.58a}
\end{equation}
Since $k^{\Box}=K^{\Box}$ (Cor.~\ref{7.6E}), each $\bv' \in K^{\Box}$ is
also a member of $k^{\Box}$, and this means that there exist $\bV' \in
\S_{\bV}$ and $\bw' \in B(\I^{(3)}) \times B(\I^{(4)})$ such that $\bv' = \bV' \bw'$.
Therefore, from statement (\ref{7.58a}),
$$
\begin{array}{l}
\text{For each } \bv' \in K^{\Box}, \text{ there exists }
\bw' \in B(\I^{(3)}) \times B(\I^{(4)}) \\
\text{such that } \bw' (\bv')^{-1} \bv \bV' (\bw')^{-1} \in
B(\I^{(3)}) \times B(\I^{(4)}).
\end{array}
$$
So, since $B(\I^{(3)}) \times B(\I^{(4)})$ is a group,
\begin{equation}
\text{For each } \bv' \in K^{\Box}, (\bv')^{-1} \bv \bv' \in
B(\I^{(3)}) \times B(\I^{(4)}).
\label{7.58b}
\end{equation}
In particular, since $\bV^{M} \in K^{\Box}$ [see Eqs.\ (\ref{7.54b}) to
(\ref{7.54d})] and $(\bV^{M})^{-1} \bv \bV^{M} = \bv$,
\begin{equation}
\bv \in B(\I^{(3)}) \times B(\I^{(4)}).
\end{equation}
Therefore, there exist
\begin{equation}
\alpha_{0}^{(i)}:\I^{(i)} \rightarrow R^{1} \text{ and }
\alpha_{1}^{(i)}:\I^{(i)} \rightarrow R^{1}
\end{equation}
such that
\begin{equation}
v^{(i)} = I \alpha_{0}^{(i)} + \tilde{J} \alpha_{1}^{(i)};
\label{7.58e}
\end{equation}
and, since $\bv \in K^{\Box}$ and $\det{v^{(i)}} = 1$,
\begin{equation}
\alpha_{0}^{(i)} \text{ and } \alpha_{1}^{(i)} \text{ are } \bC^{\Box}
\end{equation}
and
\begin{equation}
(\alpha_{0})^{2} + (\alpha_{1})^{2} = 1.
\label{7.58g}
\end{equation}

Also [see Eq.\ (\ref{tN})], the function whose domain is $\I^{(i)}$
and whose values are given by
$$
u^{(i)}(\sigma) = \exp{\tilde{\N}(\sigma)}
$$
is a member of $K^{an}(\I^{(i)}) \subset K^{\Box}(\I^{(i)})$, where
$K^{\Box} = K^{\Box}(\I^{(3)}) \times K^{\Box}(I^{(4)})$.  Upon
letting $\bv' = (u^{(3)},u^{(4)})$ in Eq.\ (\ref{7.58b}), and upon
using Eq.\ (\ref{7.58e}), one obtains
$$
I \alpha_{0}^{(i)} + J \alpha_{1}^{(i)} [u^{(i)}]^{2} \in B(\I^{(i)});
$$
and this is true if and only if
\begin{equation}
\alpha_{1}^{(i)}(\sigma) \sinh[2\tilde{\N}(\sigma)] = 0
\text{ for all } \sigma \in \I^{(i)}.
\label{7.58h}
\end{equation}
However, $\alpha_{1}^{(i)}$ is continuous.  Therefore, the condition
(\ref{7.58h}) can hold if and only if $\alpha_{1}^{(i)}$ is identically
zero, whereupon (\ref{7.58e}) and (\ref{7.58g}) yield $v^{(i)} = \pm
\delta^{(i)}$.  Hence $\bv \in Z^{(3)} \times Z^{(4)}$ is a necessary
and sufficient condition for $[\bv] = $ the identity map on
$\S_{\F}^{\Box}$.
\cheers

\item 
Suppose $\bV \in K^{\Box}$ and suppose $\F \in \S_{\F}^{\Box}$
and the corresponding member of $\S_{\bV}^{\Box}$ is $\bV$.  From the
definition of $\S_{\F}^{\Box}$ and Cor.~\ref{7.6E}, there exists
$\bw \in B(\I^{(3)}) \times B(\I^{(4)})$ such that 
$$
\bV \bw \in k^{\Box} = K^{\Box}.
$$
Therefore, since $\bv \in K^{\Box}$ and $K^{\Box}$ is a group,
$$
\bv^{-1} \bV \bw \in K^{\Box} = k^{\Box}.
$$
Therefore, from the definition of $k^{\Box}$, there exist $\bV_{0}
\in \S_{\bV}$ and $\bw' \in B(\I^{(3)}) \times B(\I^{(4)})$ such that
$$
\bv^{-1} \bV \bw = \bV_{0} \bw'.
$$
So, since $B(\I^{(3)}) \times B(\I^{(4)})$ is a group,
$$
\bV^{-1} \bv \bV_{0} \in B(\I^{(3)}) \times B(\I^{(4)}).
$$
It then follows from Lem.~\ref{7.2F}(i) that $\F$ is the solution
of the HHP corresponding to $(\bv,\F_{0})$, where $\F_{0}$
is the member of $\S_{\F}^{\Box}$ that corresponds to $\bV_{0}$.

We have thus shown that every member $\F$ of $\S_{\F}^{\Box}$
is in the range of $[\bv]$; i.e., 
\begin{equation}
[\bv] \text{ is a mapping of } \S_{\F}^{\Box} \text{ onto }
\S_{\F}^{\Box}.
\label{7.59a}
\end{equation}

Next, suppose $\F_{0}$ and $\F'_{0}$ are members of 
$\S_{\F}^{\Box}$ such that
$$
\F := [\bv](\F_{0}) = [\bv](\F'_{0}).
$$
Then, $\F$ is the solution of the HHP's corresponding to
$(\bv,\F_{0})$ and to $(\bv,\F'_{0})$, whereupon
Lem.~\ref{7.2F}(ii) informs us that $\F'_{0}$ is the solution
of the HHP corresponding to $(\bv^{-1}\bv,\F_{0})$.  Hence,
$\F'_{0} = \F_{0}$.

We have thus shown that $[\bv]$ is one-to-one.  Upon combining this result
with (\ref{7.59a}), we obtain
\begin{equation}
\begin{array}{l}
\text{For each } \bv \in K^{\Box}, [\bv] \text{ is a permutation of }
\S_{\F}^{\Box} \\
\text{\{ i.e., $[\bv]$ is a one-to-one mapping of $\S_{\F}^{\Box}$
onto $\S_{\F}^{\Box}$ \}. }
\end{array}
\end{equation}

Furthermore, the reader can easily show from Lem.~\ref{7.2F}(ii) that, if
\begin{equation}
[\bv'] \circ [\bv] := \text{ the composition of the mappings }
[\bv'] \text{ and } [\bv],
\end{equation}
then
\begin{equation}
[\bv'] \circ [\bv] = [\bv' \bv].
\end{equation}
Lemma \ref{7.2F}(ii) also yields
\begin{equation}
[\bv]^{-1} = [\bv^{-1}].
\end{equation}
Therefore, since $K^{\Box}$ is a group, $\K^{\Box}$ is a group with
respect to composition of mappings.

The remainder of the proof is straightforward and is left to the reader.
\cheers

\item 
Let $\F_{0}$ and $\F$ be any members of $\S_{\F}^{\Box}$
such that the corresponding members of $\S_{\bV}^{\Box}$ are $\bV_{0}$
and $\bV$, respectively.  By definition of $\S_{\bV}^{\Box}$, there exist
members $\bw_{0}$ and $\bw$ of the group $B(\I^{(i)}) \times B(\I^{(4)})$ such that
$$
\bV_{0} \bw_{0} \text{ and } \bV \bw \text{ are members of }
k^{\Box} = K^{\Box}.
$$
Then, from Lem.~\ref{7.2F}(i), $\F$ is the solution of the HHP
corresponding to $(\bv,\F_{0})$, where
$$
\bv := \bV \bw (\bw_{0})^{-1} \bV_{0}^{-1},
$$
and is clearly a member of $K^{\Box}$.  So, for each $\F_{0}
\in \S_{\F}^{\Box}$ and $\F \in \S_{\F}^{\Box}$,
there exists $[\bv] \in \K^{\Box}$ such that $\F = [\bv](\F_{0})$;
and that is what is meant by the statement that $\K^{\Box}$ is
transitive.
\cheers
\end{romanlist}

As a final note, the K--C subgroup of $\K^{3}$ is
$$
\left\{ [\bv] \in \K^{an}:v^{(3)} \text{ and } v^{(4)}
\text{ have equal analytic extensions to the domain } ]r_{1},s_{1}[
\right\}.
$$

\section*{Acknowledgement}
Research supported in part by grants PHY-93-07762, PHY-96-01043 and
PHY-98-00091 from the National Science Foundation to FJE Enterprises.  

\end{document}